\documentclass[letterpaper, 12pt]{article}
\pdfoutput=1
\usepackage{cite}
\usepackage{silence}
\usepackage{amsmath}
\usepackage{amssymb}
\usepackage{amsfonts}

\usepackage{comment}
\usepackage{slashed}

\usepackage{hyperref}
\hypersetup{bookmarksnumbered}

\newcommand{\Tr}{\operatorname{Tr}}

\renewcommand{\title}[1]{\vbox{\center\LARGE{#1}}\vspace{5mm}}
\renewcommand{\author}[1]{\vbox{\center#1}\vspace{5mm}}
\newcommand{\address}[1]{\vbox{\center\em#1}}

\numberwithin{equation}{section}
\allowdisplaybreaks

\usepackage{amsfonts,amscd,mathrsfs,amsmath,amsthm,amssymb}
\usepackage{yfonts}
\usepackage{mathtools}
\usepackage[shortlabels]{enumitem}
\usepackage{graphicx}
\usepackage{float}
\usepackage{bm}
\usepackage{bbm}
\usepackage{lipsum}
\usepackage{caption}
\usepackage{subcaption}

\PassOptionsToPackage{hyphens}{url}\usepackage{hyperref}
\usepackage{cite}
\usepackage[capitalise]{cleveref}

\usepackage{physics}
\usepackage{slashed}
\usepackage{tensor}

\hypersetup{
    colorlinks=true,
    linkcolor=red,
    filecolor=black,
    citecolor=red,
}

    \numberwithin{equation}{section}

    \allowdisplaybreaks

    \newcommand{\bfit}[1]{\textcolor{black}{\textit{\textbf{#1}}}}
    
    \DeclareMathAlphabet{\mathsfit}{T1}{\sfdefault}{\mddefault}{\sldefault}
    \SetMathAlphabet{\mathsfit}{bold}{T1}{\sfdefault}{\bfdefault}{\sldefault}

    \newcommand{\Z}{\mathbb{Z}}	
    \newcommand{\R}{\mathbb{R}} 
    \newcommand{\C}{\mathbb{C}} 

    \renewcommand{\H}{\mathcal{H}} 
    
    \newcommand{\sket}[2]{{\ket{\smqty{#1 \\ #2}}}}
    \newcommand{\sbra}[2]{{\bra{\smqty{#1 \\ #2}}}}
    
    \newcommand{\SOket}[3]{{\ket*{\smqty{#1\hfill \\#2 \, #3}}}}

    \renewcommand{\S}{\mathsf{S}}
    \renewcommand{\L}{\mathsf{L}}
    \newcommand{\G}{\mathsf{G}}
    \renewcommand{\P}{\mathsf{P}}
    
    \newcommand{\monopole}{{}_{j}Y^\ell_m}
    \newcommand{\monopoleY}[3]{{}_{#1}Y^{#2}_{#3}}

    \newcommand{\ds}{\hspace{0.05em}}

\begin{document}

\pagenumbering{roman} 
\begin{titlepage}

\begin{center}

\hfill UTWI-15-2022 \\
\hfill \\
\vskip 2.25cm

\title{Quantum Error Correction in the \\ Lowest Landau Level}

\author{Yale Fan,$^\ast$ Willy Fischler,$^\ast$ and Eric Kubischta$^\dag$}

\address{\small $^\ast$Department of Physics, University of Texas at Austin, \\ Austin, TX 78712, USA}

\address{\small $^\dag$Department of Mathematics, University of Maryland, \\ College Park, MD 20742, USA}

\vspace{5 mm}
\vbox{\center\tt yalefan@gmail.com, fischler@physics.utexas.edu, erickub@umd.edu}

\end{center}

\vskip 1cm

\abstract
{
We develop finite-dimensional versions of the quantum error-correcting codes proposed by Albert, Covey, and Preskill (ACP) for continuous-variable quantum computation on configuration spaces with nonabelian symmetry groups.  Our codes can be realized by a charged particle in a Landau level on a spherical geometry---in contrast to the planar Landau level realization of the qudit codes of Gottesman, Kitaev, and Preskill (GKP)---or more generally by spin coherent states.  Our quantum error-correction scheme is inherently approximate, and the encoded states may be easier to prepare than those of GKP or ACP.
}

\end{titlepage}
\pagenumbering{arabic} 

\setcounter{tocdepth}{2} 
\tableofcontents

\newpage

\section{Introduction}

What analog information lacks in robustness, it makes up for in expressiveness.  This can be said of both classical and quantum information.  Continuous-variable quantum information processing represents a promising approach to quantum computation that is especially suitable for optical platforms \cite{Braunstein, Lloyd}.  As with analog classical computation, the implementation of error correction and fault tolerance in the continuous setting presents an enduring challenge.  Twenty years ago, Gottesman, Kitaev, and Preskill (GKP) put forth a prototypical scheme for robustly encoding quantum information in continuous-variable quantum systems \cite{Gottesman:2000di}.  The simplest quantum error-correcting codes of GKP encode an $n$-level system in the infinite-dimensional Hilbert space of a single bosonic mode (harmonic oscillator): $\mathbb{C}^n\subset L^2(\mathbb{R})$.  Such codes can correct sufficiently small errors that act simultaneously on any pair of canonically conjugate variables on the physical Hilbert space.

GKP codes are mathematically elegant, and they have stimulated an enormous amount of interest.  However, the requisite codewords are challenging to prepare, leading to a long time lag between their theoretical proposal and their experimental implementation \cite{Puri}.  It is natural to ask whether there exist finite-dimensional, or ``discretized,'' versions of GKP codes that retain their desirable property of protecting against small diffusive errors.  Such codes would encode a logical qudit in a larger, but still finite-dimensional, physical qudit.  GKP themselves gave an affirmative answer to this question and proposed an appealing physical realization thereof: an electron on a periodically identified patch of the plane in a uniform background magnetic field, i.e., in a toroidal Landau level \cite{Haldane-Rezayi, Stockholm}.  Rymarz et al.\ \cite{PhysRevX.11.011032} have since proposed a practical implementation of the requisite ``grid states'' for this setup.

GKP codes lend themselves to a variety of generalizations, such as to multiple modes \cite{Conrad2022gottesmankitaev, vanLoock} or to planar rotors \cite{Englert}, as already anticipated by GKP \cite{Gottesman:2000di}.  The key requirement is that the configuration space be an abelian group, such as the group of translations of the real line.  Recently, Albert, Covey, and Preskill (ACP) \cite{PhysRevX.10.031050} took substantial steps toward generalizing GKP codes to \emph{nonabelian} group or coset configuration spaces: $\mathbb{C}^n\subset L^2(G), L^2(G/H)$.  In doing so, they brought tools from representation theory and harmonic analysis to bear on the problem of designing GKP-like codes.  In particular, ACP develop schemes for encoding a qudit in the rotational states of a rigid body.  Examples include the rotational degrees of freedom of a generic molecule (with configuration space $SO(3)$) and a diatomic molecule with a symmetry axis (whose configuration space is $SO(3)/SO(2)\cong S^2$).  The associated codes are known as \emph{molecular codes} and \emph{linear rotor codes}, respectively.

Despite the advances in continuous-variable quantum error correction afforded by ACP, a simple and natural question about these codes has remained unanswered: do there exist finite-dimensional versions of ACP codes that retain their desirable properties?  Such codes would embed a logical qudit in a \emph{finite-dimensional} quantum system whose configuration space has a nonabelian symmetry group.  We show that the answer is yes, and we propose an intuitive physical realization via a spherical Landau level \cite{Haldane:1983xm, Greiter}.  The (re)appearance of Landau levels in this context both establishes a similar ``discretization'' for ACP codes as for GKP codes, and suggests a role for quantum Hall physics in quantum computation beyond the already well-appreciated applications to topological quantum computation \cite{Nayak}.  However, our scheme is not limited to Landau levels, and can be realized more generally by spin coherent states.

We begin by reviewing continuous-variable GKP codes and their generalizations due to ACP in Section \ref{QEC}.  We emphasize that all of these codes, as well as their finite-dimensional predecessors, can be described in the unified framework of Calderbank-Shor-Steane (CSS) codes.  Our main results are presented in Section \ref{LLLcodes}.  After recalling how the qudit codes of GKP can be realized in a planar Landau level with periodic boundary conditions, we construct finite-dimensional analogues of the linear rotor codes of ACP.  Specifically, we show how to encode a qudit in a spherical Landau level in a way that protects against certain rotational errors.  We also review the physics of the lowest Landau level (LLL) as necessary, most notably the fact that the phase space and the configuration space coincide.  The spherical phase space has a different algebraic structure than the torus, which necessitates a different encoding scheme than for GKP qudit codes.  Importantly, angular momentum errors are energetically suppressed in the LLL limit, which simplifies our task to that of designing codes that protect only against rotational errors.  More abstractly, our codes rely on using coherent states for the group $SU(2)$ to construct ``smeared'' versions of the linear rotor codewords of ACP.  In Section \ref{discussion}, we highlight a few open questions and comment on possible extensions of our codes.  Technical formulas are given in the appendices, where we spell out our conventions for Landau levels (Appendix \ref{LLs}), derive useful properties of spin coherent states (Appendix \ref{spincoherentstates}), elaborate on the approximation error induced by smearing position eigenstates to spin coherent states (Appendix \ref{approxQEC}), and present some additional codes beyond those considered in the main text (Appendix \ref{morecodes}).

\section{Quantum Error Correction} \label{QEC}

A quantum error-correcting code (QECC) is an embedding (or partial isometry) of a logical Hilbert space into a larger physical Hilbert space: $\H_\text{logical}\hookrightarrow \H_\text{phys}$.  The aim is to perform computations on states in $\H_\text{logical}$ while using the extra degrees of freedom from $\H_\text{phys}$ to monitor the states for errors.  If $\H_\text{phys}$ is finite-dimensional, then we call the code \bfit{finite}; otherwise, we call the code \bfit{continuous}.

\subsection{Finite Qubit Codes}

Consider first a finite QECC that encodes $k$ qubits into $n$ qubits: $\C^{2^k} \hookrightarrow \C^{2^n}$. The standard notion of performance in this setting is given by the \emph{code distance} $\delta$. In particular, consider the Pauli group $\P_n$ on $n$ qubits, consisting of $n$-fold tensor products of the Pauli matrices $I$, $X$, $Y$, and $Z$ along with an overall phase of $\pm 1$ or $\pm i$. The \emph{weight} of an operator $Q \in \P_n$ is the number of tensor factors in $Q$ that differ from $I$. Then a code with distance $\delta$ can correct all errors from $\P_n$ with weight $\lfloor (\delta-1)/2 \rfloor$ or less. The standard notation for such a code is $[[n, k, \delta]]$.

The \bfit{stabilizer codes} (also called \emph{additive codes}) \cite{GottesmanThesis, preskillnotes, NielsenChuang, Gottesman2009} form the richest class of finite qubit QECCs. Given an abelian subgroup $\mathbf{S} \subset \P_n$ such that $-I \not\in \mathbf{S}$, a stabilizer code is defined as the linear subspace of $\C^{2^n}$ stabilized by $\mathbf{S}$ (i.e., the common $+1$ eigenspace of the generators of $\mathbf{S}$). If $\mathbf{S}$ has $r$ generators, then the code comprises $k = n - r$ logical qubits. If $Z(\mathbf{S})$ is the centralizer of $\mathbf{S}$ in $\P_n$, then the distance of a stabilizer code is the smallest weight of the operators in $Z(\mathbf{S})\setminus \mathbf{S}$. Such a code can correct a set of Pauli errors $\mathcal{E}$ if for any two errors $E_1, E_2\in \mathcal{E}$, we have $E_1^\dag E_2\not\in Z(\mathbf{S})\setminus \mathbf{S}$. The $r$ generators of the stabilizer $\mathbf{S}$ are called \bfit{check operators}, being what we physically measure to diagnose errors. The operators in $Z(\mathbf{S})\setminus \mathbf{S}$ are called \bfit{logical operators} because they preserve the code subspace (but not the individual states within it). In fact, $Z(\mathbf{S})/\mathbf{S} \cong \P_k$, the logical Pauli group.

A special class of stabilizer codes, the \bfit{CSS codes} (cf.\ \cite{GottesmanThesis}), will underlie the entirety of our presentation. These are stabilizer codes for which there exists a choice of generators such that each generator is either an ``$X$-type operator'' (consisting of tensor products of only $X$ and $I$) or a ``$Z$-type operator'' (consisting of tensor products of only $Z$ and $I$). One way to construct CSS codes is to consider a nested sequence of groups
\begin{equation}
\S \subset \L \subset \G
\end{equation}
where $\G$ is the group of all $X$-type errors, $\S$ is the subgroup of all $X$-type check operators (which leaves all states in the code subspace invariant), and $\L$ is the subgroup of all $X$-type logical operators (which acts transitively on a basis for the code subspace). A logical ``position basis'' for the code is found by taking a uniform superposition of the operators from each coset in $\L/\S$ and applying the sum to some fiducial position ket. The dimension of the code is the number of cosets $|\L/\S|$.

A dual description of CSS codes is afforded by the sequence of groups
\begin{equation}
\widehat{\S} \subset \widehat{\L} \subset \widehat{\G}.
\end{equation}
Here, $\widehat{\G}$---thought of as the group of all $Z$-type errors---is the Pontryagin dual of the \emph{abelian} group $\G$, i.e., the group of all complex unitary irreps of $\G$.\footnote{For an $n$-letter qubit code, $\G$ is isomorphic to $\Z_2^n$, so it is indeed abelian and the Pontryagin dual makes sense.} On the other hand, $\widehat{\S}$ and $\widehat{\L}$ do \emph{not} denote the Pontryagin duals of $\S$ and $\L$: rather, $\widehat{\S}\equiv \L^\perp$ and $\widehat{\L}\equiv \S^\perp$, where $\L^\perp$ is the subgroup of irreps in $\widehat{\G}$ such that $\L$ is represented trivially and similarly for $\S^\perp$.\footnote{In other words, $\L^\perp$ is the collection of all trivial restricted (subduced) irreps of $\L$ in $\G$, or the collection of all irreps in $\widehat{\G}$ that \emph{branch} to the trivial irrep of $\L$.} To be sure, $\smash{\widehat{\S}}$ is a subgroup of $Z$-type operators that leaves the code invariant (the $Z$-type check operators), while $\smash{\widehat{\L}}$ is a subgroup of $Z$-type operators that acts transitively on a basis for the code (the $Z$-type logical operators). A logical ``momentum basis'' for the code is found by taking a uniform superposition of the operators from each coset in $\smash{\widehat{\L}}/\smash{\widehat{\S}}$ and applying the sum to some fiducial momentum ket. The stabilizer of a CSS code is generated as $\mathbf{S} = \langle\S, \widehat{\S}\rangle$, and the logical operators for a CSS code are generated as $\langle\L, \widehat{\L}\rangle$.

\subsection{Finite GKP Codes} \label{finiteGKP}

The preceding discussion generalizes readily to other finite dimensions. Suppose that the physical Hilbert space is $\C^N$. We denote the ``position basis'' by $|x\rangle$ and the ``momentum basis'' by $|p\rangle$ for $x, p\in \Z_N$. The two bases are related by a Fourier transform:
\begin{equation}
|p\rangle = \frac{1}{\sqrt{N}}\sum_{x=0}^{N-1} \omega^{xp}|x\rangle, \qquad |x\rangle = \frac{1}{\sqrt{N}}\sum_{p=0}^{N-1} \omega^{-xp}|p\rangle,
\end{equation}
where $\omega\equiv e^{2\pi i/N}$. Sylvester's clock and shift matrices generalize the Pauli matrices $Z$ and $X$ to $N$ dimensions, respectively. They act on the aforementioned bases as
\begin{alignat}{2}
X|x\rangle &= |x + 1\rangle, \qquad & Z|x\rangle &= \omega^x|x\rangle, \label{actionZbasis} \\
X|p\rangle &= \omega^{-p}|p\rangle, \qquad & Z|p\rangle &= |p + 1\rangle, \label{actionXbasis}
\end{alignat}
with addition understood modulo $N$, and they satisfy the commutation relation 
\begin{equation}
ZX = \omega XZ.
\label{genPauliCommRel}
\end{equation}
The generalized Pauli group $\P_N$ consists of all operators $\omega^c X^a Z^b$ where $a,b,c \in \Z_N$ (here, we assume $N > 2$). The operators $X^a Z^b$ form a complete and orthonormal set of operators on $\C^N$. For CSS codes, an $X$-type operator is any operator in $\P_N$ with $b = 0$, and a $Z$-type operator is any operator in $\P_N$ with $a = 0$.\footnote{Since $\G$ is isomorphic to $\Z_N$, an abelian group, the Pontryagin dual $\widehat{\G}$ still makes sense.}

Now consider a code that embeds a $K$-level system into an $N$-level system for general $K < N$: $\C^K \hookrightarrow \C^N$.  We can use the CSS subgroup chain $\S \subset \L \subset \G$, where $\G\cong \Z_N$ is the collection of all $X$-type operators (``position errors'') $X^a$.  A finite cyclic group of order $N$ has exactly one subgroup for each divisor of $N$.  Suppose that $N = Kr_1 r_2$ and that we choose $\S = \langle X^{Kr_1}\rangle\cong \Z_{r_2}$ and $\L = \langle X^{r_1}\rangle\cong \Z_{Kr_2}$.  The dual groups can be shown to be $\widehat{\S} = \linebreak[0] \langle Z^{Kr_2}\rangle\cong \Z_{r_1}$ and $\widehat{\L} = \linebreak[0] \langle Z^{r_2}\rangle\cong \Z_{Kr_1}$.  It follows that the stabilizer $\langle\S, \smash{\widehat{\S}}\rangle$ is generated by $X^{Kr_1}$ and $Z^{Kr_2}$, while the logical operators $\langle\L, \smash{\widehat{\L}}\rangle$ are generated by $\overline{X} = X^{r_1}$ and $\overline{Z} = Z^{r_2}$.

The logical position and momentum kets, which both form bases for the $K$-dimensional code subspace, are as follows:
\begin{equation}
|\overline{x} = j\rangle = \frac{1}{\sqrt{r_2}}\sum_{n=0}^{r_2 - 1} |x = (Kn + j)r_1\rangle, \qquad |\overline{p} = j\rangle = \frac{1}{\sqrt{r_1}}\sum_{n=0}^{r_1 - 1} |p = (Kn + j)r_2\rangle,
\end{equation}
where $j\in \mathbb{Z}_K$.  These two sets of codewords are related by a Fourier transform within the code subspace:
\begin{equation}
|\overline{p}\rangle = \frac{1}{\sqrt{K}}\sum_{x=0}^{K-1} (\omega^{r_1 r_2})^{xp}|\overline{x}\rangle, \qquad |\overline{x}\rangle = \frac{1}{\sqrt{K}}\sum_{p=0}^{K-1} (\omega^{r_1 r_2})^{-xp}|\overline{p}\rangle.
\end{equation}
Indeed, the logical operators satisfy
\begin{equation}
\overline{Z}\overline{X} = \omega^{r_1 r_2}\overline{X}\overline{Z} = e^{2\pi i/K}\overline{X}\overline{Z}.
\label{logicalcommutator}
\end{equation}
The logical position kets are spaced a distance $r_1$ apart and the logical momentum kets are spaced a distance $r_2$ apart, which suggests that we can correct all errors $X^a$ and $Z^b$ with $|a|$ and $|b|$ sufficiently small relative to $r_1$ and $r_2$, respectively.

To formally deduce the properties of this code, it is easier to examine the stabilizer rather than the states.  A general Pauli operator $X^a Z^b$ satisfies the following commutation relations with the stabilizer generators $Z^{Kr_2}$ and $X^{Kr_1}$:
\begin{align}
Z^{Kr_2}(X^a Z^b) &= e^{2\pi ia/r_1}(X^a Z^b)Z^{Kr_2}, \\
X^{Kr_1}(X^a Z^b) &= e^{-2\pi ib/r_2}(X^a Z^b)X^{Kr_1}.
\end{align}
In particular, for any logical state $|\overline{\psi}\rangle$ in the code subspace, we have
\begin{align}
Z^{Kr_2}(X^a Z^b|\overline{\psi}\rangle) &= e^{2\pi ia/r_1}X^a Z^b|\overline{\psi}\rangle, \\
X^{Kr_1}(X^a Z^b|\overline{\psi}\rangle) &= e^{-2\pi ib/r_2}X^a Z^b|\overline{\psi}\rangle.
\end{align}
The operators $Z^{Kr_2}$ and $X^{Kr_1}$ commute, so they can be measured simultaneously.\footnote{To perform a projective measurement of an observable (which projects onto eigenspaces), all one needs is a spectral theorem.  Hence all normal operators, of which Hermitian and unitary operators are examples, can be regarded as observables.  The operators $X$ and $Z$ are unitary, but not (in general) Hermitian.  Their logarithms are Hermitian, but not single-valued on the torus, hence not good observables.}  Measuring $Z^{Kr_2}$ determines $a$ modulo $r_1$ and measuring $X^{Kr_1}$ determines $b$ modulo $r_2$, so this code can correct all errors of the form $|\overline{\psi}\rangle\mapsto X^a Z^b|\overline{\psi}\rangle$ with\footnote{A general error takes $|\overline{\psi}\rangle$ to a superposition of terms $X^a Z^b|\overline{\psi}\rangle$.  These measurements collapse the per\-turbed state to a single such term in the superposition, which can then be corrected (assuming sufficiently small $|a|, |b|$) to recover the original state $|\overline{\psi}\rangle$.  If errors act locally, then errors with large $|a|, |b|$ are unlikely to occur.}
\begin{equation}
|a| < \frac{r_1}{2}, \qquad |b| < \frac{r_2}{2}.
\end{equation}
Assuming that both $r_1$ and $r_2$ are odd, there are $r_1 r_2 = N/K$ such errors.  This is then a ``perfect'' code in that the $N$-dimensional physical Hilbert space contains exactly $r_1 r_2$ copies of the $K$-dimensional code space, one for each possible error.  In addition, we see that $r_1$ and $r_2$ control the number of correctable position and momentum shifts, respectively. If we want them to be equal, $r_1 = r_2$, then $N/K$ must be a perfect square.

While the above construction is quite general, we call the resulting codes \bfit{finite GKP codes} because in a certain limit, they approach the GKP codes on $L^2(\R)$ that we discuss next.  We also refer to them as ``qudit codes,'' as they embed a qudit in a larger qudit.

\subsection{Continuous Codes: GKP} \label{GKPcodes}

We now consider the continuous codes constructed by GKP \cite{Gottesman:2000di} that embed a $d$-level qudit into the infinite-dimensional Hilbert space of a harmonic oscillator: $\C^d \hookrightarrow L^2(\R)$.

The Pauli group on $L^2(\R)$, which we denote by $\P$, can be obtained as the $d\to \infty$ limit of the single-qudit generalized Pauli group $\P_d$. $\P$ is also called the Heisenberg-Weyl group over $\R$. It is generated by ``$X$-type'' displacement operators $X_a = e^{-iap}$, ``$Z$-type'' displacement operators $Z_b = e^{ibx}$, and an overall phase $e^{ic}$ for $a, b, c\in \R$, where $x$ and $p$ are conjugate position and momentum operators satisfying $[x, p] = i$ (we set $\hbar = 1$ for convenience). $X_a$ is a translation operator in the position basis ($X_a|x\rangle = |x + a\rangle$), while $Z_b$ is a translation operator in the momentum basis ($Z_b|p\rangle = |p + b\rangle$). These operators obey the commutation relation $X_a Z_b = e^{-iab}Z_b X_a$. The $X_a Z_b$ form a complete set of operators on $L^2(\R)$.

A \bfit{GKP code} is simply the result of applying the CSS subgroup chain construction to the configuration space $\R$. Consider a chain $\S \subset \L \subset \G$, where $\G\cong \R$ is the collection of all $X$-type translation operators. We suppose further that $\S$ and $\L$ are \emph{discrete} subgroups, so our only choices are $\beta\Z$ for some $\beta > 0$. We choose $\S = \langle X_{d\alpha}\rangle\cong d\alpha\Z$ and $\L = \langle X_\alpha\rangle\cong \alpha\Z$ for some $\alpha > 0$. The $d$ cosets in the quotient group $\L/\S$ take the form $\S_j\cong j\alpha + d\alpha\Z$ for $j = 0, \ldots, d - 1$. To get a logical position basis $|\overline{x}\rangle$ for the $d$-dimensional code subspace, we take a uniform sum of the operators in each $\S_j$ and apply them to a fixed position ket, say $|x = 0\rangle$. The resulting codewords are infinite superpositions of position eigenstates:
\begin{equation}
|\overline{x} = j\rangle = \sum_{k\in \Z} |x = \alpha(j + dk)\rangle, \qquad j = 0, \ldots, d - 1.
\end{equation}
Now consider the dual description $\widehat{\S} \subset \widehat{\L} \subset \widehat{\G}$, where $\widehat{\G}\cong \R$ is the collection of all $Z$-type translation operators.\footnote{To be sure, $\R$ is abelian but also locally compact, so the Pontryagin dual still makes sense.} It is not hard to see that $\widehat{\S} = \L^\perp = \langle Z_{2\pi/\alpha}\rangle\cong \tfrac{2\pi}{\alpha}\Z$ and $\widehat{\L} = \linebreak[0] \S^\perp = \linebreak[0] \langle Z_{2\pi/d\alpha}\rangle\cong \tfrac{2\pi}{d\alpha}\Z$. The $d$ cosets in $\smash{\widehat{\L}}/\smash{\widehat{\S}}$ are $\smash{\widehat{\S}_j}\cong \tfrac{2\pi j}{d\alpha} + \tfrac{2\pi}{\alpha}\Z$ for $j = 0, \ldots, d - 1$. We get a logical momentum basis $|\overline{p}\rangle$ by taking a uniform sum of the operators in each $\smash{\widehat{\S}_j}$ and applying them to a fixed momentum ket, say $|p = 0\rangle$. The resulting codewords are infinite superpositions of momentum eigenstates:
\begin{equation}
|\overline{p} = j\rangle = \sum_{k\in \Z} \left|p = \frac{2\pi}{d\alpha}(j + dk)\right\rangle, \qquad j = 0, \ldots, d - 1.
\end{equation}
By Poisson resummation, we obtain the Fourier transform relations
\begin{equation}
|\overline{p}\rangle = \frac{1}{\sqrt{d}}\sum_{x=0}^{d-1} \omega_d^{xp}|\overline{x}\rangle, \qquad |\overline{x}\rangle = \frac{1}{\sqrt{d}}\sum_{p=0}^{d-1} \omega_d^{-xp}|\overline{p}\rangle,
\end{equation}
where $\omega_d\equiv e^{2\pi i/d}$. Of course, the exact codewords are non-normalizable (i.e., distributions) and cannot be realistically constructed. However, one can construct approximate codewords that belong to $L^2(\R)$ by smearing out the infinite peaks and by modulating the would-be Dirac comb by an overall Gaussian envelope.

The stabilizer subgroup of this code is generated by $\S$ and $\widehat{\S}$, so the corresponding check operators are $X_{d\alpha}$ and $Z_{2\pi/\alpha}$. The logical group for this code is generated by $\L$ and $\widehat{\L}$, so the logical $X$ and $Z$ operators are $\overline{X} = X_\alpha$ and $\overline{Z} = Z_{2\pi/d\alpha}$. These logical operators satisfy
\begin{equation}
\overline{Z}\overline{X} = \omega_d\overline{X}\overline{Z},
\end{equation}
and they act as follows on the logical position and momentum bases:
\begin{alignat}{2}
\overline{X}|\overline{x}\rangle &= |\overline{x + 1}\rangle, \qquad & \overline{Z}|\overline{x}\rangle &= \omega_d^x|\overline{x}\rangle, \\
\overline{X}|\overline{p}\rangle &= \omega_d^{-p}|\overline{p}\rangle, \qquad & \overline{Z}|\overline{p}\rangle &= |\overline{p + 1}\rangle.
\end{alignat}
Note that the logical group is generated by the ``$d^\text{th}$ roots'' of the check operators.

Because operators in the continuous Pauli group $\P$ can no longer be decomposed into tensor products, the notion of code distance from the finite case must be amended. However, the same basic idea still applies. The distance of a finite code measures the minimum weight of any logical operation. In the present case, distance is an even more natural notion because $X_a$ and $Z_b$ are generated by the Lie algebra operators $p$ and $x$, and the Lie algebra can be thought of as the tangent space at the identity. Thus the distance of a code is dictated by \emph{both} $|a|$ and $|b|$. A good code should protect against Pauli errors $X_a$ and $Z_b$ where $|a|$ and $|b|$ are as large as possible. Here, we are tacitly assuming that errors act locally, which is true for closed systems but is true for open systems only in certain circumstances. For example, it holds over sufficiently short time intervals or when the oscillator is only weakly coupled to its environment.

In position space, the logical codewords are offset from each other by $\alpha$, so we can correct position shifts with $|a| < \alpha/2$. In momentum space, the codewords are set apart by $2\pi/d\alpha$, so we can correct momentum shifts with $|b| < \pi/d\alpha$. We can correct both types of errors simultaneously because the corresponding check operators commute. In fact, this conclusion follows directly from examining the stabilizer. A generic error operator $X_a Z_b$ satisfies the following commutation relations with the stabilizer generators $\overline{Z}{}^d$ and $\overline{X}{}^d$:
\begin{align}
\overline{Z}{}^d(X_a Z_b) &= e^{2\pi ia/\alpha}(X_a Z_b)\overline{Z}{}^d, \\
\overline{X}{}^d(X_a Z_b) &= e^{-ibd\alpha}(X_a Z_b)\overline{X}{}^d.
\end{align}
After $X_a Z_b$ acts on an encoded state,\footnote{More precisely, an arbitrary noise channel $\mathcal{E}$ acting on the state $\rho$ of an oscillator can be expanded as
\begin{equation}
\mathcal{E}(\rho) = \int da\, db\, da'\, db'\, C(a, b; a', b')X_a Z_b\rho Z_{b'}^\dag X_{a'}^\dag,
\end{equation}
where the coefficients $C$ are such that $\mathcal{E}$ is completely positive and trace-preserving \cite{Gottesman:2000di}.  If $\rho$ is a state in the code subspace and if $C$ has support only on sufficiently small values of its arguments, then the GKP code can recover $\rho$ from $\mathcal{E}(\rho)$.} measuring $\overline{Z}{}^d$ and $\overline{X}{}^d$ simultaneously will reveal $a$ modulo $\alpha\Z$ and $b$ modulo $\tfrac{2\pi}{d\alpha}\Z$. Thus we can correct all combined shifts with
\begin{equation}
|a| < \frac{\alpha}{2}, \qquad |b| < \frac{\pi}{d\alpha}
\end{equation}
(so that $|ab| < \frac{\pi}{2d}$). The key insight of GKP here is that while one cannot measure $x$ and $p$ simultaneously, one can measure both of them ``up to some lattice'' without violating the uncertainty principle. Note also that we left $\alpha$ as an adjustable parameter. The larger $\alpha$ is, the more position errors we can correct but the fewer momentum errors we can correct, and vice versa. When $\alpha = \sqrt{2\pi/d}$, we correct position and momentum errors equally well.

This single-oscillator GKP code easily extends to the configuration space $\R^n$. The Pauli group there is the Heisenberg-Weyl group on $\R^n$, so $a$ and $b$ become vector-valued.

Finally, to see that the finite GKP codes really do approach these continuous codes on $\R$, note that as $N\to\infty$, the $X$ and $Z$ in \eqref{genPauliCommRel} limit to infinite-dimensional displacement operators.  We may formally write $X\to e^{-i\epsilon p}$ and $Z\to e^{2\pi i\epsilon x}$ where $\epsilon = 1/\sqrt{N}$.  Setting $r_1 = \alpha/\epsilon$ and $r_2 = 1/K\alpha\epsilon$ for some $\alpha > 0$ then shows that as $N\to\infty$, the logical operators $\overline{X} = X^{r_1}$ and $\overline{Z} = Z^{r_2}$ approach $X_\alpha$ and $Z_{2\pi/d\alpha}$ with $d = K$.\footnote{In a different limit, the finite GKP codes approach GKP codes on the circle $S^1$ \cite{Gottesman:2000di}.}  In brief, a (finite) GKP code is a CSS-type stabilizer code for which the stabilizer is a discrete abelian subgroup of the generalized Pauli group or the continuous Heisenberg-Weyl group.

\subsection{Continuous Codes: Beyond GKP}

For GKP codes and their immediate analogues, the configuration space can be identified as an \emph{abelian} Lie group: $\R$, $\R^n$, or $S^1 = U(1)$.  Elements of these groups label position kets: $\{|g\rangle : g \in \G\}$.  On the other hand, momentum kets are labeled by elements of the Pontryagin dual $\widehat{\G}$, where $\widehat{\R} = \R$, $\widehat{\R^n} = \R^n$, and $\widehat{U(1)} = \Z$.  For example, the momentum eigenstates of a planar rotor with $\G = U(1)$ are $\{|\ell\rangle : \ell \in \Z\}$.

Can GKP codes generalize to configuration spaces described by \emph{nonabelian} Lie groups?  The main problem is that Pontryagin duality only works for locally compact abelian groups, so it is unclear what the ``momentum'' space should be in this setting.  This problem was tackled and solved by ACP in \cite{PhysRevX.10.031050}.  Given a nonabelian Lie group $\G$ as our configuration space, the quantum Hilbert space is $L^2(\G)$.  The Peter-Weyl theorem suggests that we look at the matrix coefficients of $\G$.  Let $Z_{mn}^\ell(g)$ denote the $(m, n)$ matrix element of the $\ell^\text{th}$ irrep matrix for the element $g\in \G$.  The collection of all matrix coefficients is again denoted by $\widehat{\G}$, but unlike in the abelian case, it is no longer a group.  We write momentum kets as $\SOket{\ell}{m}{n}$.  They are related to the position kets $|g\rangle$ via
\begin{equation}
\SOket{\ell}{m}{n} = \int_\G dg\, \sqrt{\frac{d_\ell}{|\G|}} Z_{mn}^\ell(g) |g\rangle, \qquad |g\rangle = \sum_{\ell, m, n \in \widehat{\G}} \sqrt{\frac{d_\ell}{|\G|}} Z_{mn}^\ell(g)^\ast \SOket{\ell}{m}{n},
\end{equation}
where $d_\ell$ is the dimension of the $\ell^\text{th}$ irrep of $\G$ and $|\G|$ is the volume of $\G$ with respect to the Haar measure.

Because $\G$ is nonabelian, there exist two natural kinds of $X$-type operators on $L^2(\G)$, generated by the left- and right-regular actions of $\G$ on itself:
\begin{equation}
\overrightarrow{X}_h|g\rangle = |hg\rangle, \qquad \overleftarrow{X}_h|g\rangle = |gh^{-1}\rangle.
\end{equation}
The choice of $Z$-type operators is much less clear.  A fairly natural extrapolation from the abelian case is to consider $Z$-type operators that act diagonally in the position basis via multiplication by the corresponding matrix coefficients:
\begin{equation}
\widehat{Z}_{mn}^\ell|g\rangle = Z_{mn}^\ell(g)|g\rangle. 
\end{equation}
The problem is that the $X$- and $Z$-type operators taken together do not form a Pauli \emph{group} (indeed, the $Z$-type operators alone do not even form a group), so we cannot use them to construct stabilizer codes outright.  Notwithstanding, these operators form an orthonormal basis for operators on $L^2(\G)$, so we can proceed analogously to the construction of GKP.

Codes on these spaces can again be constructed using the $\S \subset \L \subset \G$ technique, and will be called \bfit{generalized GKP codes}.  ACP focus primarily on the case $\G = SO(3)$.  This is the configuration space of a rigid body pinned to a point, which serves as the ball-and-stick model for the rotational degrees of freedom of a molecule.  The orientation and angular momentum of a rigid body, being conjugate variables, are subject to similar kinds of diffusive errors as $x$ and $p$ for an oscillator.  The corresponding ``molecular codes'' of ACP can correct any combination of sufficiently small rotational and angular momentum errors, subject to the tradeoff that increasing the magnitude of the correctable angles decreases the magnitude of the correctable angular momentum kicks (and vice versa).  While not strictly stabilizer codes, these codes can be described as the degenerate ground space of a Hamiltonian constructed from a sum of commuting stabilizer generators.

Details for molecular codes on $SO(3)$, as well as a framework for codes on nonabelian $\G$, can be found in \cite{PhysRevX.10.031050}.  While we will not need the explicit analysis here, a noteworthy point is that when $\G$ is a nonabelian group, the subgroups $\S$ and $\L$ may themselves be nonabelian.  Different choices of $\S$ and $\L$ allow for the construction of codewords that protect against a variety of different sets of errors.  Detectable position shifts correspond to elements of $\G/\L$, while logical shift errors correspond to elements of $\L/\S$.

\subsection{Continuous Codes: ACP on the Sphere} \label{ACPcodes}

GKP codes and generalized GKP codes are all formulated for configuration spaces that have a group structure.  ACP \cite{PhysRevX.10.031050} consider an even further generalization of GKP codes in which the configuration space is not a group, but rather a homogeneous manifold.  Like (gen\-er\-al\-ized) GKP codes, these codes protect against noise that simultaneously affects two conjugate variables describing the degrees of freedom of a continuous-variable system.

ACP consider the specific scenario where the configuration space is the two-sphere $S^2$, which describes a heterogeneous linear molecule fixed to a point (called a linear rotor) as well as a particle on a sphere.  Since $S^2$ can be regarded as the quotient space $SO(3)/SO(2)$, we can essentially take the analysis for $SO(3)$ as a starting point and branch each irrep of $SO(3)$ down to $SO(2)$.

The physical Hilbert space is $L^2(S^2)$.  We write $|v\rangle = |\theta, \varphi\rangle$ for position eigenstates on $S^2$, where $v$ is a unit vector on $S^2\subset \R^3$ and $(\theta, \varphi)$ are angular coordinates.  We write momentum eigenstates as $\sket{\ell}{m}$, where $\ell$ is a nonnegative integer and $|m|\leq \ell$.  The continuous position basis and the discrete angular momentum basis are connected via
\begin{equation}
|v\rangle = \sum_{\ell\geq 0}\sum_{|m|\leq \ell} Y_m^\ell(v)^\ast\sket{\ell}{m}, \qquad \sket{\ell}{m} = \int_{S^2} dv\, Y_m^\ell(v)|v\rangle,
\label{ACPpositions}
\end{equation}
where the $Y_m^\ell$ are (complex) spherical harmonics.  Note that $\langle v|\smqty*{\ell \\ m}\rangle = Y_m^\ell(v)$.

Only the left $X$-type operators from $SO(3)$ survive: a rotation $R\in SO(3)$ is represented by a unitary operator $X_R$, which acts as
\begin{equation}
X_R|v\rangle = |Rv\rangle.
\end{equation}
This action has fixed points, as $|v\rangle$ is invariant under any rotation about the axes $\pm v$.  In the angular momentum basis, we have
\begin{equation}
X_R\sket{\ell}{m} = \sum_{|m'|\leq \ell} D^\ell_{m'm}(R)\sket{\ell}{m'},
\label{rotangmom}
\end{equation}
where the $D^\ell_{m'm}$ are Wigner $D$-matrix elements.  Note that the $X_R$ operators do not disturb the irrep label $\ell$, only the internal label $m$.

The $Z$-type operators
\begin{equation}
\hat{Y}_m^\ell = \int_{S^2} dv\, Y_m^\ell(v)|v\rangle\langle v|,
\label{momentumkick}
\end{equation}
which are diagonal in the position basis, are called momentum kicks.  They act on position kets as $\hat{Y}_m^\ell|v\rangle = Y_m^\ell(v)|v\rangle$.  On the other hand,
\begin{equation}
\sbra{L}{M}\hat{Y}_m^\ell\sket{\ell'}{m' } = \int_{S^2} dv\, Y_M^L(v)^\ast Y_m^\ell(v)Y_{m'}^{l'}(v).
\label{selectionrules}
\end{equation}
This integral vanishes unless the angular momentum selection rules are satisfied: $|\ell - \ell'|\leq L\leq \ell + \ell'$ and $M = m + m'$.

The $X$-type operators $X_R$ together with the $Z$-type operators $\hat{Y}_m^\ell$ morally comprise the ``Pauli group'' on $L^2(S^2)$, despite not forming an actual \emph{group}.  They also form an \emph{overcomplete} frame, rather than an orthonormal basis, for operators on $L^2(S^2)$.  One can think of them as Pauli operators in the sense that the $X_R$ translate the position kets whereas the $\hat{Y}_m^\ell$ translate the momentum kets, analogously to their namesakes in finite quantum error correction.  Both also act ``locally'' in the sense that if $R$ is a small rotation, then $X_R|v\rangle$ is close to $|v\rangle$, and if $\ell, m$ are small, then $\hat{Y}_m^\ell\sket{\ell'}{m'}$ is close to $\sket{\ell'}{m'}$.

To construct an \bfit{ACP code}, we again choose a nested sequence of groups $\S \subset \L \subset \G$.  However, $\G$ is now not the configuration space itself, but rather the largest group of continuous ``translations'' (orientation-preserving isometries) of that space.  When the configuration space is a sphere, $\G = SO(3)$ and the corresponding codes are called ``linear rotor codes.''  Our focus in this work will be on constructing finite analogues of these linear rotor codes.

The subgroups of $SO(3)$ are well-classified, but to illustrate the simplest such codes, we restrict our attention to the cyclic subgroups $\S = \langle X_{R_z(2\pi/N)}\rangle\cong \Z_N$ and $\L = \langle X_{R_z(\pi/N)}\rangle\cong \Z_{2N}$.  Here, $N$ is a positive integer and $R_z(\theta)$ denotes a rotation about the $z$-axis by the angle $\theta$.  The quotient group $\L/\S$ contains two cosets.  If we act with these on a fiducial ket, say $|\pi/2, 0\rangle$, and take a uniform superposition, then we obtain the following logical states in the position basis:
\begin{equation}
|\overline{0}\rangle = \frac{1}{\sqrt{N}}\sum_{h\in \mathbb{Z}_N} \left|\frac{\pi}{2}, \frac{2\pi h}{N}\right\rangle, \qquad |\overline{1}\rangle = \frac{1}{\sqrt{N}}\sum_{h\in \mathbb{Z}_N} \left|\frac{\pi}{2}, \frac{2\pi h}{N} + \frac{\pi}{N}\right\rangle.
\label{linrotorcodewords}
\end{equation}
These codewords are equal superpositions of equatorial position eigenstates.  They span a two-dimensional subspace of $L^2(S^2)$.

The general logic of GKP tells us that a position shift (rotation) on $S^2$ is correctable as long as it keeps every constituent point of a codeword inside its corresponding Voronoi cell, which is the set of all points on $S^2$ that are closer to that point than to any other constituent point of a codeword.\footnote{We are deliberately oversimplifying the argument here.  In fact, to satisfy the quantum error correction conditions, it is important that the codewords $|\overline{0}\rangle$ and $|\overline{1}\rangle$ be related by a parity transformation, which acts as $X_P|v\rangle = |{-v}\rangle$.  This condition requires that the integer $N$ be odd.  In addition, the reliance of ACP \cite{PhysRevX.10.031050} on this parity argument restricts their discussion to qubit codes on the sphere ($\mathbb{Z}_N\subset \mathbb{Z}_{dN}\subset SO(3)$ codes with $d = 2$).  We defer a more careful discussion to Section \ref{LLLcodes}. \label{parityargument}}  From the explicit expressions \eqref{linrotorcodewords}, we see that these Voronoi cells are spherical lunes bounded by lines of longitude with angular separation $\pi/N$.  On the other hand, a short computation shows that in the angular momentum basis, we have
\begin{equation}
|\overline{r}\rangle = \sqrt{N}\sum_{\ell\geq 0}\sum_{|pN|\leq \ell} (-1)^{pr}Y^\ell_{pN}(\pi/2, 0)^\ast\sket{\ell}{pN}
\label{ACPangmombasis}
\end{equation}
for $r\in \{0, 1\}$.  Hence the codewords only have support on angular momentum eigenstates $\sket{\ell}{m}$ where $m$ is a multiple of $N$.  This property, together with the angular momentum selection rules \eqref{selectionrules}, implies that the value of $m$ in a momentum shift error $\hat{Y}^\ell_m$ can be measured modulo $N$.  The value of $m$ modulo $N$ determines $m$ if $|m| < N/2$.  Since $|m|\leq \ell$, a momentum shift is correctable if $\ell < N/2$.  In summary, increasing $N$ (analogous to the inverse of the parameter $\alpha$ in the GKP codes on $\R$) reduces the code's ability to correct position errors and increases its ability to correct momentum errors.

However, the fact that $X_R$ and $\hat{Y}_m^\ell$ errors are now overcomplete for operators on $L^2(S^2)$ turns out to imply that this code cannot correct \emph{combined} position and momentum shifts of the form $\rho\mapsto X_R\hat{Y}^\ell_m\rho\hat{Y}^{\ell\,\dag}_m X_R^\dag$.  It can correct sufficiently small position shifts or sufficiently small momentum shifts individually, but not products of both \cite{PhysRevX.10.031050}.  The extra symmetry of a linear rotor compared to that of a rigid rotor leads to a smaller configuration space, and hence less ``room'' for diagnosing errors.  Fortunately, for our eventual goal of constructing finite-dimensional linear rotor codes, only rotational errors will turn out to be relevant.

\section{Lowest Landau Level Codes} \label{LLLcodes}

We now motivate and present our main result: the lowest Landau level (LLL) codes.

\subsection{Motivation: Planar LLL}

The physical insight that motivates the present work is that the finite GKP code described in Section \ref{finiteGKP} can be realized by an electrically charged particle in a uniform magnetic field in two dimensions \cite{Gottesman:2000di}.  The Hamiltonian $H$ of such a particle commutes with \emph{magnetic} trans\-la\-tions rather than with ordinary translations (see Appendix \ref{LLs}).  The commutator of two magnetic translation operators $T_1$ and $T_2$ gives rise to an Aharonov-Bohm phase $2\pi k$, where $k$ is the number of magnetic flux quanta enclosed by the corresponding path:
\begin{equation}
T_1 T_2 T_1^{-1}T_2^{-1} = e^{2\pi ik}, \qquad k = \frac{\Phi}{\Phi_0}, \qquad \Phi_0 = \frac{2\pi\hbar}{q}.
\end{equation}
Here, $\Phi$ is the enclosed flux and $q$ is the particle's charge.  The two translations commute if $k$ is an integer.  Therefore, $T_1$ and $T_2$ form a maximally commuting set if the unit cell of the lattice that they generate encloses one flux quantum.  Suppose that this is the case: then the simultaneous eigenbasis of the operators $H$, $T_1$, $T_2$ furnishes a Landau level of $N$ degenerate states, where $N$ is the total number of flux quanta.  The operators
\begin{equation}
\overline{Z} = T_1^{1/K}, \qquad \overline{X} = T_2
\end{equation}
satisfy \eqref{logicalcommutator}.  These are the logical operators on a $K$-dimensional code space stabilized by the check operators
\begin{equation}
S_Z = \overline{Z}{}^K = T_1, \qquad S_X = \overline{X}{}^K = T_2^K.
\end{equation}
Suppose further that we impose periodic boundary conditions so that
\begin{equation}
T_1^{r_1} = (T_2^K)^{r_2} = 1.
\end{equation}
Then the number of flux quanta through the resulting torus is $N = Kr_1 r_2$, and the stabilizer generators $S_Z$ and $S_X$ tile the torus with $r_1 r_2$ cells containing $K$ flux quanta each.  The wavefunction of any state in the code subspace is periodic with respect to these cells.  The error operators
\begin{equation}
Z = T_1^{1/Kr_2}, \qquad X = T_2^{1/r_1}
\end{equation}
implement fractional shifts of these periodic wavefunctions on a scale smaller than the size of a single cell.

The fundamental principle at work in the above example is that in the strong-field limit, a system of charged particles in a background magnetic field is described by a Lagrangian that is first-order in time derivatives.  In this limit, the kinetic terms become negligible, the Hamiltonian vanishes, and the configuration space becomes an effective phase space.  If the configuration space (hence the effective phase space) is compact, then the associated Hilbert space is finite-dimensional.  In the case of the lowest Landau level on a torus $T^2$, magnetic translations generate a two-dimensional compact phase space.

\subsection{LLL Hilbert Space}

We have seen that taking an appropriate infinite-dimension limit of the qudit codes reviewed in Section \ref{finiteGKP} reproduces the continuous-variable GKP codes of Section \ref{GKPcodes} for the phase space $\mathbb{R}^2$ \cite{Gottesman:2000di}.  We now show that there likewise exist finite-dimensional incarnations of the continuous-variable ACP codes of Section \ref{ACPcodes}.

Additional complications arise when the geometry of the configuration space is curved.  Nonetheless, drawing inspiration from the preceding discussion, a reasonable strategy for ``finitizing'' the quantum codes of ACP \cite{PhysRevX.10.031050} is to start with a physical system whose configuration space $M$ admits a symplectic structure.  One might then hope that in some LLL limit, the noncompact phase space $T^\ast M$ goes to the compact phase space $M$ and the corresponding infinite-dimensional Hilbert space goes to a finite-dimensional one.  These considerations rule out the molecular codes based on $SO(3)$ and suggest that we focus on the linear rotor codes based on $S^2$.

The spherical Landau problem thus provides a natural physical setup for a finite-di\-men\-sion\-al ACP-inspired quantum code.  Consider a particle of electric charge $q$ constrained to move on a sphere $S^2$ of radius $R$ around a magnetic monopole of magnetic charge $BR^2$ with $qB > 0$.  The corresponding magnetic field and electromagnetic angular momentum are
\begin{equation}
\nabla\times\vec{A} = \frac{BR^2\hat{r}}{r^2}, \qquad \vec{L} = \vec{r}\times(-i\hbar\nabla - q\vec{A}) - qBR^2\hat{r},
\label{spheresetup}
\end{equation}
where $[L_i, L_j] = i\hbar\epsilon_{ijk}L^k$.  The Hamiltonian is diagonalized by the simultaneous eigenstates $\sket{\ell}{m}$ of $\vec{L}^2$ and $L_3$, which satisfy $\vec{L}^2\sket{\ell}{m} = \hbar^2\ell(\ell + 1)\sket{\ell}{m}$ and $L_3\sket{\ell}{m} = \hbar m\sket{\ell}{m}$.  We read off the spectrum by writing
\begin{equation}
H = \frac{1}{2m}(-i\hbar\nabla - q\vec{A})^2|_{r=R} = \frac{\vec{L}^2 - \hbar^2 j^2}{2mR^2} = \frac{\hbar^2(\ell(\ell + 1) - j^2)}{2mR^2},
\end{equation}
where the energy levels $\ell$ are bounded from below as
\begin{equation}
\ell\geq \boxed{j\equiv \frac{qBR^2}{\hbar}}\in \tfrac{1}{2}\mathbb{Z}.
\label{jbound}
\end{equation}
The boxed identification in \eqref{jbound} can be viewed as a ``correspondence principle'' that relates the classical and quantum Casimirs $qBR^2$ and $\hbar j$, respectively \cite{TedSphere}; the fact that $j$ is a half-integer follows from the Dirac quantization condition (see Appendix \ref{sphere} for details).  In a suitable longitudinal gauge that is regular at the north pole, the wavefunctions are monopole (or spin-weighted) spherical harmonics:
\begin{equation}
\langle\theta, \varphi|\smqty*{\ell \\ m}\rangle = \monopole(\theta, \varphi), \quad \ell\geq j, \quad -\ell\leq m\leq \ell.
\end{equation}
These functions are written explicitly in \eqref{Yjlm}.\footnote{More precisely, monopole spherical harmonics are sections of a complex line bundle over $S^2$, but we have made a choice of gauge that turns them into genuine functions.}  By taking a double-scaling limit of large $B$ and small $R$ with $j$ fixed, one can make the energy spacing between levels arbitrarily large while preserving the degeneracy of the \bfit{lowest Landau level} ($\ell = j$).  This projection to the LLL truncates the infinite-dimensional space of rotational states to a finite-dimensional subspace of $2j + 1$ degenerate states $\sket{j}{m}$, where $m = -j, \ldots, j$.  The resulting system effectively transforms in the spin-$j$ representation of $SU(2)$.

A noteworthy feature of this construction is that in the LLL limit, angular momentum kicks that take states in the LLL up to states of higher $\ell$ are energetically disfavored.  Thus quantum codes constructed in the LLL, unlike those of ACP \cite{PhysRevX.10.031050}, should only need to protect against rotational errors that go between the states with $\ell = j$.  Again, the fundamental reason for this simplification is that in the LLL limit of the spherical charge-monopole system, the phase space is reduced from the noncompact cotangent bundle $T^\ast S^2$ to $S^2$.  The LLL on $S^2$, as on any compact surface, then has a finite degeneracy.  We make some additional remarks on the quantization of this system in Appendix \ref{sphere}.

Can this finite-dimensional system realize an ACP code?  There are two apparent obstacles.  First, magnetic rotations satisfy a different algebra than magnetic translations due to the different topologies of the sphere and the torus.  Magnetic translations on the torus are exponentiated versions of the guiding center operators $R_x$ and $R_y$, which satisfy
\begin{equation}
[R_x, R_y] = -\frac{i\hbar}{qB}.
\label{canonicalcommutator}
\end{equation}
However, magnetic rotations on the sphere are exponentiated versions of the guiding center angular momenta in \eqref{spheresetup}, which satisfy
\begin{equation}
[L_i, L_j] = i\hbar\epsilon_{ijk}L^k.
\label{notcanonicalcommutator}
\end{equation}
While \eqref{canonicalcommutator} takes the form of a canonical commutation relation, \eqref{notcanonicalcommutator} does not.  In particular, unlike magnetic translations, magnetic rotations are kinematically identical to ordinary rotations in the absence of a background magnetic field.\footnote{The large-radius limit of the sphere leads to a contraction of the rotation algebra \eqref{notcanonicalcommutator}, as two generators suffice for translations on the plane.}  This fact would seem to stand in the way of extrapolating the framework of GKP to the spherical LLL, as there is no sense in which the operators in \eqref{notcanonicalcommutator} are canonically conjugate.  Second, the LLL projection does not preserve the would-be ACP codewords on $S^2$, which have support on all Landau levels.  Namely, suppose that we modify the linear rotor analysis of ACP to use monopole spherical harmonics instead of ordinary spherical harmonics, which comprise a different basis for $L^2(S^2)$.  Then the position eigenstates are
\begin{equation}
|\theta, \varphi\rangle = \sum_{\ell\geq j}\sum_{|m|\leq \ell} \monopole(\theta, \varphi)^\ast\sket{\ell}{m},
\label{allLLpositions}
\end{equation}
where---in contrast to \eqref{ACPpositions}---the $\sket{\ell}{m}$ are now eigenstates of the electromagnetic angular momenta in \eqref{spheresetup}.  A calculation similar to that producing \eqref{ACPangmombasis} shows that the codewords \eqref{linrotorcodewords} now take the following form in the angular momentum basis:
\begin{equation}
|\overline{r}\rangle = \sqrt{N}\sum_{\ell\geq j}\sum_{|pN - j|\leq \ell} (-1)^{pr}\monopoleY{j}{\ell}{pN - j}(\pi/2, 0)^\ast\sket{\ell}{pN - j}
\end{equation}
(see Appendix \ref{allLLs}).  Since these codewords involve states from all Landau levels, the LLL projection is na\"ively incompatible with the ACP construction.

Both of these problems can be surmounted by constructing codewords analogous to those of ACP directly in the finite-dimensional LLL Hilbert space.  But in attempting to do so, we immediately confront another obstacle: while the $2j + 1$ angular momentum eigenstates $\sket{j}{m}$ form a basis for the LLL Hilbert space, they fall far short of a basis for the space of $L^2$ functions on $S^2$.  Therefore, they cannot be used to construct the position eigenstates that form the backbone of GKP and ACP codewords.

This final problem has an elegant solution.  Consider the normalized \emph{spin coherent states} \cite{Radcliffe_1971, Arecchi:1972td}
\begin{equation}
|\theta, \varphi\rangle_j\equiv \frac{1}{(1 + |z|^2)^j}e^{zL_-/\hbar}\sket{j}{j},
\label{spincohdef}
\end{equation}
where $z$ is a complex parameter related to the angular coordinates $(\theta, \varphi)$ by stereographic projection: $z = \tan(\theta/2)e^{i\varphi}$.  We write these states with a $j$ subscript to distinguish them from position eigenstates on the sphere, and we alternatively denote them by $|\Omega\rangle_j$ or $|\vec{n}\rangle_j$, where $\Omega\equiv (\theta, \varphi)$ and $\vec{n}$ is the unit vector $(\sin\theta\cos\varphi, \sin\theta\sin\varphi, \cos\theta)$.  A more enlightening rewriting of \eqref{spincohdef} in terms of $(\theta, \varphi)$ variables permits a direct comparison to \eqref{ACPpositions} and \eqref{allLLpositions}:
\begin{equation}
|\theta, \varphi\rangle_j = \sum_{m=-j}^j \sqrt{\frac{(2j)!}{(j + m)!(j - m)!}}\cos^{j+m}(\theta/2)\sin^{j-m}(\theta/2)e^{i(j - m)\varphi}\sket{j}{m}.
\end{equation}
These spin coherent states are the best possible approximations to position eigenstates that can be constructed from the finite collection of states $\{\sket{j}{m}\}$.  They become more sharply peaked as the dimension of the Hilbert space grows.  Indeed, they form an overcomplete set for the LLL Hilbert space, and they approach orthogonality in the limit that $j\to\infty$:
\begin{equation}
|{}_j\langle\vec{n}_1|\vec{n}_2\rangle_j| = \left(\frac{1 + \vec{n}_1\cdot \vec{n}_2}{2}\right)^j.
\label{norminnerproduct}
\end{equation}
Moreover, they are eigenstates of the angular momentum operator in the $(\theta, \varphi)$ direction:
\begin{equation}
(L_1\sin\theta\cos\varphi + L_2\sin\theta\sin\varphi + L_3\cos\theta)|\theta, \varphi\rangle_j = \hbar j|\theta, \varphi\rangle_j.
\end{equation}
This is the sense in which spin coherent states comprise the ``position basis'' dual to the finite-di\-men\-sional angular momentum basis $\sket{j}{m}$.  We summarize many more of their properties in Appendix \ref{spincoherentstates}.

The spin coherent states $|\theta, \varphi\rangle_j$ form the foundation of our ACP-like codes.  Using these states, we construct approximate GKP codewords that become exact as $j\to\infty$.  The fact that these states are ``smeared'' relative to true position eigenstates necessitates \emph{approximate} quantum error correction (see Appendix \ref{approxQEC}).

Note that whereas the large-dimension limit of the GKP qudit codes reproduces the GKP oscillator codes, the corresponding limit of our codes is distinct from the ACP construction.  This is because the $j\to\infty$ limit of the LLL Hilbert space produces an infinite-dimensional Hilbert space in the ``horizontal'' direction rather than the ``vertical'' direction (i.e., a single infinitely degenerate level rather than an infinite sum of finitely degenerate levels).\footnote{In other words, the limit in which finite GKP codes approach continuous GKP codes is one in which the compact phase space $T^2$ becomes the noncompact phase space $\R^2$.  In the ACP setting, the analogous limit would take $S^2$ to $\R^2$ rather than to $T^\ast S^2$.}  For this reason, we refer to these ``finitized ACP codes'' as \bfit{LLL codes}.

\subsection{LLL Pauli Group} \label{errorbasis}

What is a natural error basis for our codes?  We saw in Section \ref{ACPcodes} that the ``Pauli group'' in the ACP setting is composed of $X$-type operators $X_R$ and $Z$-type operators $\hat{Y}_m^\ell$.  In the presence of a magnetic monopole, the latter are amended to ${}_j\hat{Y}_m^\ell$.  Despite not being a \emph{group} in the formal sense, it resembles the Pauli group in many ways (e.g., it spans the set of all operators on $L^2(S^2)$ and each operator acts locally).  The $X_R$ act within a single Landau level (see \eqref{rotangmom}), while the ${}_j \hat{Y}_m^\ell$ relate states in different Landau levels (see \eqref{selectionrules}).

Projecting this collection of operators down to the LLL leaves the $X$-type operators $X_R$ untouched but eliminates the $Z$-type operators ${}_j \hat{Y}_m^\ell$, as momentum kicks are energetically disfavored.  In this way, the Pauli group becomes an actual \emph{group} again; it is isomorphic to $SU(2)$ since the $X_R$ are projective representations of $SO(3)$.

In fact, the unitary rotation operators $X_R$ form an overcomplete set of linear operators within any level of fixed $\ell$, and in particular for the LLL Hilbert space, which is isomorphic to $\mathbb{C}^{2j + 1}$ with $j\equiv qBR^2/\hbar$ as in \eqref{jbound}.  Indeed, the $X_R$ are represented by Wigner $D$-matrices on the LLL Hilbert space.  Burnside's theorem \cite{lang2005algebra} implies that the complex linear span of the image of any $d$-dimensional unitary irrep $\rho : G\to U(d)$ of a compact group $G$ is the set of all $d\times d$ complex matrices.  Since the Wigner $D$-matrices comprise the image of the $(2j + 1)$-dimensional complex irrep of $SO(3)$, they are complete.\footnote{However, being non-orthogonal, the $X_R$ do not form a unitary error basis in the usual sense.  A unitary error basis for $\mathbb{C}^d$ is a set of $d^2$ unitary $d\times d$ matrices that are orthonormal with respect to the normalized Hilbert-Schmidt inner product $\langle A, B\rangle = \Tr(A^\dag B)/d$.  For example, the $d^2$ unitary operators $X^a Z^b$ ($a, b = 0, \linebreak[0] \ldots, d - 1$), where $X, Z$ are generalized Pauli matrices, form a unitary error basis.  The Weyl basis $e^{i(\alpha p + \beta x)/\hbar}$ is obtained from these in the infinite-$d$ limit.  The $Z$-type operator of GKP is simply the exponential of a multiple of the operator $L_3$ on the LLL Hilbert space.  The $X$-type operators of GKP are similar (but not identical) to $L_\pm\equiv L_1\pm iL_2$, where $[L_3, L_\pm] = \pm\hbar L_\pm$ and $[L_+, L_-] = 2\hbar L_3$.}

The $X_R$ can be expanded in the angular momentum basis as
\begin{equation}
X_R = e^{-i\alpha L_3/\hbar}e^{-i\beta L_2/\hbar}e^{-i\gamma L_3/\hbar} = \sum_{m, n = -j}^j D^j_{mn}(\alpha, \beta, \gamma)\sket{j}{m}\sbra{j}{n},
\end{equation}
where the rotation $R = R(\alpha, \beta, \gamma)\in SO(3)$ is given in terms of Euler angles.  Their action on angular momentum kets follows straightforwardly.  On the other hand, the $X_R$ have an action on spin coherent states that resembles an actual rotation, up to a phase factor.  This phase factor will be important for our analysis, so we pause to discuss it.

First note that spin coherent states themselves can be formulated in terms of rotations:
\begin{equation}
|\theta, \varphi\rangle_j = X_{R(\varphi, \theta, -\varphi)}\sket{j}{j} = \sum_{m=-j}^j D^j_{mj}(\varphi, \theta, -\varphi)\sket{j}{m}.
\label{spincohasrotation}
\end{equation}
Hence $|\theta, \varphi\rangle_j$ is simply a rotated version of the highest-weight state $\sket{j}{j}$.  The equivalence of \eqref{spincohasrotation} to \eqref{spincohdef} is shown in Appendix \ref{disentangling}.  The definition \eqref{spincohdef} or \eqref{spincohasrotation} fixes a phase convention for spin coherent states, as such states are in one-to-one correspondence with points on the sphere only up to a phase.  This is a consequence of the fact that the $SO(2)$ subgroup of $SO(3)$ that preserves a given point on the sphere acts on the corresponding spin coherent state by a phase.  To illustrate this point, let $R(\varphi, \theta, -\varphi)$ be the ``canonical'' rotation in \eqref{spincohasrotation} that takes the north pole to the point $(\theta, \varphi)$ on $S^2$.  If $R(\varphi_1, \varphi_2, \varphi_3)$ also takes the north pole to $(\theta, \varphi)$, then $\vec{n}(\varphi_1, \varphi_2) = \vec{n}(\theta, \varphi)$, which implies that $\varphi_2 = \theta$ and $\varphi_1 = \varphi$.  Hence the corresponding $D$-matrices indeed differ by a phase,
\begin{equation}
D^\ell_{mn}(\varphi_1, \varphi_2, \varphi_3) = e^{-i(\varphi_3 + \varphi)n}D^\ell_{mn}(\varphi, \theta, -\varphi),
\end{equation}
as do the corresponding states: $X_{R(\varphi_1, \varphi_2, \varphi_3)}\sket{j}{j} = e^{-ij(\varphi_3 + \varphi)}|\vec{n}\rangle_j$.  Letting $R(\alpha, \beta, \gamma)$ be an arbitrary rotation, we also check that
\begin{align}
X_{R(\alpha, \beta, \gamma)}|\vec{n}\rangle_j &= \sum_{m=-j}^j\left[\sum_{m'=-j}^j D^j_{mm'}(\alpha, \beta, \gamma)D^j_{m'j}(\varphi, \theta, -\varphi)\right]\sket{j}{m} \\
&= \sum_{m=-j}^j D^j_{mj}((\alpha, \beta, \gamma)\circ (\varphi, \theta, -\varphi))\sket{j}{m},
\end{align}
so rotations compose as expected.  In general,
\begin{equation}
R(\alpha, \beta, \gamma)R(\varphi, \theta, -\varphi)\neq R(\varphi', \theta', -\varphi')
\end{equation}
for some canonical rotation $R(\varphi', \theta', -\varphi')$, so a rotation takes a spin coherent state with a given orientation to the spin coherent state with the rotated orientation only up to a phase:
\begin{equation}
X_R|\vec{n}\rangle_j\propto |R\vec{n}\rangle_j.
\label{phaseforrots}
\end{equation}
With these considerations in mind, it will prove convenient to have a formula for the matrix element of an arbitrary rotation $R = R(\alpha, \beta, \gamma)$ between any two spin coherent states:
\begin{align}
{}_j\langle\Omega'|X_R|\Omega\rangle_j &= \bigg[\left(e^{-i(\alpha + \gamma)/2}\cos\frac{\theta'}{2}\cos\frac{\theta}{2} + e^{i(\alpha + \gamma)/2}e^{i(\varphi - \varphi')}\sin\frac{\theta'}{2}\sin\frac{\theta}{2}\right)\cos\frac{\beta}{2} \label{theamplitude} \\
&\hspace{1 cm} - \left(e^{-i(\alpha - \gamma)/2}e^{i\varphi}\cos\frac{\theta'}{2}\sin\frac{\theta}{2} - e^{i(\alpha - \gamma)/2}e^{-i\varphi'}\sin\frac{\theta'}{2}\cos\frac{\theta}{2}\right)\sin\frac{\beta}{2}\bigg]^{2j}, \nonumber
\end{align}
where $\Omega = (\theta, \varphi)$ and $\Omega' = (\theta', \varphi')$.  This formula is derived in Appendix \ref{matrixelements}.

We will put \eqref{theamplitude} to use when checking the quantum error correction conditions of Knill and Laflamme \cite{PhysRevA.55.900}.  Given a $d$-dimensional code spanned by the states $|\overline{0}\rangle, |\overline{1}\rangle, \ldots, |\overline{d - 1}\rangle$ within the LLL Hilbert space $\mathbb{C}^{2j+1}$, these conditions state that any two correctable rotation errors $R$ and $R'$ must satisfy
\begin{equation}
\langle\overline{0}|X_{R^{-1}R'}|\overline{0}\rangle = \langle\overline{1}|X_{R^{-1}R'}|\overline{1}\rangle = \cdots = \langle\overline{d - 1}|X_{R^{-1}R'}|\overline{d - 1}\rangle
\label{diagonalcond}
\end{equation}
(i.e., $\langle\overline{k}|X_{R^{-1}R'}|\overline{k}\rangle$ should be independent of $k$) as well as
\begin{equation}
\langle\overline{j}|X_{R^{-1}R'}|\overline{k}\rangle = 0, \qquad j\neq k.
\label{offdiagonalcond}
\end{equation}
We refer to \eqref{diagonalcond} as the ``diagonal'' quantum error correction condition and to \eqref{offdiagonalcond} as the ``off-diagonal'' one.  In writing these conditions, we have used $X_R^\dag X_{R'} = X_{R^{-1}}X_{R'} = X_{R^{-1}R'}$.  For our application, it will be useful to relax the exact conditions \eqref{diagonalcond} and \eqref{offdiagonalcond} to ap\-prox\-i\-mate ones: by demanding that
\begin{align}
|\langle\overline{j}|X_{R^{-1}R'}|\overline{j}\rangle - \langle\overline{k}|X_{R^{-1}R'}|\overline{k}\rangle| &< \delta, \label{AQECCdiag} \\
|\langle\overline{j}|X_{R^{-1}R'}|\overline{k}\rangle| &< \epsilon \label{AQECCoffdiag}
\end{align}
(for $j\neq k$) where $\delta, \epsilon$ are arbitrarily small, we can still recover encoded states to arbitrarily good accuracy (see Appendix \ref{KLconditions}).\footnote{To understand these conditions more geometrically, recall that for the linear rotor codes of ACP \cite{PhysRevX.10.031050}, one of the requisite conditions for exact quantum error correction is that a correctable rotation $R$ should keep any constituent point of a codeword inside its corresponding Voronoi cell.  This is precisely equivalent to the condition that for any two correctable rotations $R$ and $R'$, the combination $R^{-1}R'$ cannot map any constituent point of a codeword onto a constituent point of another codeword, and therefore that the matrix element of $R^{-1}R'$ between any two distinct codewords is zero, in accord with the condition \eqref{offdiagonalcond}.

For approximate quantum error correction with spin coherent states in place of infinitely peaked position eigenstates, a codeword can have nonzero overlap with the result of applying the rotation $R^{-1}R'$ to another codeword, even if all correctable rotations keep all constituent points of codewords inside their own Voronoi cells.  This is because the support of the spin coherent state $|\theta, \varphi\rangle_j$ is not limited to the point $(\theta, \varphi)$ on $S^2$.}

In summary, by projecting $L^2(S^2)$ down to the lowest-energy LLL subspace $\C^{2j+1}$, we obtain a finite-dimensional system for which the only physically relevant errors are position shifts (rotations).  We have thus converted a problem of \emph{quantum} error correction into one that seems much more \emph{classical}.  The projected Pauli group is $\langle X_R\rangle$.  Under the assumption of locality, large rotation errors are less likely to occur than small ones.  While a spin-$j$ Hilbert space admits no true ``position'' eigenstates, spin coherent states provide approximations thereof.  We now proceed to adapt the linear rotor codes of ACP to spin coherent states in this finite-dimensional Hilbert space.

\subsection{Spherical LLL Codes} \label{sphericalLLL}

To construct an ACP-like code in the LLL on the sphere, we again pick a CSS subgroup chain $\S \subset \L \subset \G$.  Let us first consider a qubit code based on $\Z_N \subset \Z_{2N} \subset SO(3)$ like we did in Section \ref{ACPcodes}.  For ACP, higher $N$ means better correction of momentum errors and worse correction of position errors.  However, our setup exhibits no similar tradeoff because momentum kicks between energy levels have already been suppressed by making the spectral gap arbitrarily large.  We are therefore motivated to take $N$ as small as possible: that is, we choose $N = 1$ so that each basis codeword is a \emph{single} coherent state.  In other words, the principle that guides the construction of our codes is that the support of the different codewords on $S^2$ should be as far apart as possible.  Thus our ideal codes are similar to classical error-correcting codes and spherical designs (equidistributed points on the sphere).

As motivated, the best (and simplest) LLL code is the qubit code $\C^2 \hookrightarrow \C^{2j+1}$ for which $\S \cong \Z_1$ and $\L \cong \Z_2$.  The codewords are spin coherent states localized to antipodal points on the sphere.  It is also arguably the most natural code in this context, since every rotation has two fixed points (furthermore, only in this situation are the resulting codewords \emph{exactly} orthogonal, by \eqref{norminnerproduct}).  If we choose our fiducial position ket to be oriented toward the north pole, then the corresponding codewords are
\begin{align}
|\overline{0}\rangle &\equiv |0, \varphi_0\rangle_j = \sket{j}{j}, \label{qubit0} \\
|\overline{1}\rangle &\equiv |\pi, \varphi_0\rangle_j = e^{2ij\varphi_0}\sket{j}{-j}. \label{qubit1}
\end{align}
For reasons that will become clear, we have fixed a fiducial azimuthal angle $\varphi_0$ whose only effect is on the phase of the $|\overline{1}\rangle$ state.  Regardless of its value, the points $(0, \varphi_0)$ and $(\pi, \varphi_0)$ correspond to $(0, 0, 1)$ and $(0, 0, -1)$ when the sphere is thought of as embedded in $\mathbb{R}^3$.

To determine which rotation errors this code can correct, we turn to the quantum error correction conditions \eqref{diagonalcond} and \eqref{offdiagonalcond}.  We write the composite rotation $T = R^{-1}R'$ in terms of Euler angles $\alpha, \beta, \gamma$.  As a special case of \eqref{theamplitude}, we have
\begin{equation}
{}_j\langle\Omega|X_T|\Omega\rangle_j = \left[\left(\cos\frac{\alpha + \gamma}{2} - i\sin\frac{\alpha + \gamma}{2}\cos\theta\right)\cos\frac{\beta}{2} - i\sin\left(\varphi - \frac{\alpha - \gamma}{2}\right)\sin\theta\sin\frac{\beta}{2}\right]^{2j}
\label{diagonalamplitude}
\end{equation}
and therefore
\begin{equation}
\langle\overline{0}|X_T|\overline{0}\rangle = e^{-ij(\alpha + \gamma)}\cos^{2j}\frac{\beta}{2}, \qquad \langle\overline{1}|X_T|\overline{1}\rangle = e^{ij(\alpha + \gamma)}\cos^{2j}\frac{\beta}{2},
\label{disturbing}
\end{equation}
irrespective of $\varphi_0$.  The inopportune phase discrepancy in \eqref{disturbing} reflects the property of spin coherent states that a rotation about their axis of orientation generally incurs a nontrivial phase (unlike for position eigenstates).  To satisfy the diagonal condition
\begin{equation}
\langle\overline{0}|X_T|\overline{0}\rangle = \langle\overline{1}|X_T|\overline{1}\rangle
\end{equation}
exactly, one can require that $\alpha + \gamma = 0$ for any correctable rotation or product thereof.  A natural way to do this is to choose our set of correctable rotations to be all rotations about any fixed equatorial axis, i.e., the $y$-axis conjugated by a fixed $z$-rotation.  Such rotations are clearly not complete for operators on the LLL Hilbert space.  Specifically, we fix a reference angle $\varphi_0$, which coincides with the angle $\varphi_0$ in \eqref{qubit0} and \eqref{qubit1}.  Then any two correctable rotations take the form
\begin{equation}
X_R = e^{-i\varphi_0 L_3/\hbar}e^{-i\theta L_2/\hbar}e^{i\varphi_0 L_3/\hbar}, \qquad X_{R'} = e^{-i\varphi_0 L_3/\hbar}e^{-i\theta'L_2/\hbar}e^{i\varphi_0 L_3/\hbar},
\end{equation}
for some angles $\theta, \theta'$.  The combined rotation
\begin{equation}
X_{R^{-1}}X_{R'} = e^{-i\varphi_0 L_3/\hbar}e^{-i(\theta' - \theta)L_2/\hbar}e^{i\varphi_0 L_3/\hbar}
\end{equation}
is likewise a $y$-rotation conjugated by a $z$-rotation by $\varphi_0$, whereupon \eqref{disturbing} gives
\begin{equation}
\langle\overline{0}|X_{R^{-1}}X_{R'}|\overline{0}\rangle = \langle\overline{1}|X_{R^{-1}}X_{R'}|\overline{1}\rangle = \cos^{2j}\left(\frac{\theta' - \theta}{2}\right),
\end{equation}
so that the diagonal condition \eqref{diagonalcond} is satisfied exactly.  As for the off-diagonal condition \eqref{offdiagonalcond}, suppose that we restrict the magnitude of the $y$-rotation angle in all elements of our set of correctable errors to satisfy $|\theta| < \theta_0$.  Then we have by \eqref{norminnerproduct} that
\begin{equation}
|\langle\overline{0}|X_{R^{-1}}X_{R'}|\overline{1}\rangle| < \left(\frac{1 - \cos 2\theta_0}{2}\right)^j,
\end{equation}
which can be chosen as small as desired by taking $\theta_0$ sufficiently small.  In other words,
\begin{equation}
|\langle\overline{0}|X_T|\overline{1}\rangle| = \left(\frac{1 - \cos\Theta}{2}\right)^j
\label{boundthis}
\end{equation}
where $\Theta$ is the angle between the north pole and its image under $T$; we can then do approximate error correction by imposing $|\langle\overline{0}|X_T|\overline{1}\rangle| < \epsilon$, or $|\Theta| < \arccos(1 - 2\epsilon^{1/j})$ (see \eqref{AQECCoffdiag}).

We now move on to more general qudit LLL codes based on the cyclic subgroups $\S\cong \Z_1$ and $\L\cong \Z_d$.  For ease of generalization, we change the fiducial ket to be the coherent state $|\pi/2, 0\rangle_j$, so that the codewords are supported along the equator.  The logical position kets for this equatorial qudit code are then
\begin{equation}
|\overline{k}\rangle = \left|\frac{\pi}{2}, \frac{2\pi k}{d}\right\rangle_j, \qquad k = 0, \ldots, d - 1.
\label{quditcodewords}
\end{equation}
This finite code is analogous to the $\mathbb{Z}_N\subset \mathbb{Z}_{dN}\subset SO(3)$ abelian linear rotor code of ACP, where we have chosen $N = 1$ to optimize for position shifts because momentum shifts play no role.

To examine the error correction conditions, we again write $T = R^{-1}R'$ in terms of Euler angles $\alpha, \beta, \gamma$.  From \eqref{diagonalamplitude}, we have
\begin{equation}
\langle\overline{k}|X_T|\overline{k}\rangle = \left[\cos\frac{\alpha + \gamma}{2}\cos\frac{\beta}{2} - i\sin\left(\frac{2\pi k}{d} - \frac{\alpha - \gamma}{2}\right)\sin\frac{\beta}{2}\right]^{2j}.
\end{equation}
The diagonal error correction conditions \eqref{diagonalcond} are clearly satisfied if $R, R'$ (and hence $T$) are equatorial rotations ($\beta = 0$), in which case
\begin{equation}
\langle\overline{k}|X_T|\overline{k}\rangle|_{\beta = 0} = \cos^{2j}\left(\frac{\alpha + \gamma}{2}\right).
\end{equation}
From our previous discussion of the antipodal qubit code, we would expect that when $d = 2$, we have more freedom to choose the set of correctable rotations.  Explicitly, we get
\begin{align}
\langle\overline{0}|X_T|\overline{0}\rangle|_{d=2} &= \left(\cos\frac{\alpha + \gamma}{2}\cos\frac{\beta}{2} + i\sin\frac{\alpha - \gamma}{2}\sin\frac{\beta}{2}\right)^{2j}, \label{d2qudit1} \\
\langle\overline{1}|X_T|\overline{1}\rangle|_{d=2} &= \left(\cos\frac{\alpha + \gamma}{2}\cos\frac{\beta}{2} - i\sin\frac{\alpha - \gamma}{2}\sin\frac{\beta}{2}\right)^{2j}. \label{d2qudit2}
\end{align}
In this case, aside from restricting to equatorial rotations ($\beta = 0$), we could also consider rotations such that the condition $\alpha = \gamma$ is preserved under composition.  It is not hard to see that such rotations are arbitrary $z$-rotations conjugated by a fixed $x$-rotation, where the $x$-axis connects the orientations of the codewords $|\overline{0}\rangle = |\pi/2, 0\rangle_j$ and $|\overline{1}\rangle = |\pi/2, \pi\rangle_j$ (with $\langle\overline{0}|\overline{0}\rangle = \langle\overline{1}|\overline{1}\rangle = 1$ and $\langle\overline{0}|\overline{1}\rangle = 0$).  Now consider the off-diagonal error correction conditions.  Specializing to equatorial rotations ($\beta = 0$, with $\Theta\equiv \alpha + \gamma$), we have from \eqref{norminnerproduct} that
\begin{equation}
|{}_j\langle\pi/2, \varphi'|X_T|\pi/2, \varphi\rangle_j| = \left(\frac{1 + \cos(\Theta + \varphi - \varphi')}{2}\right)^j.
\end{equation}
Since the smallest value of $|\varphi - \varphi'|$ for two distinct codewords is $2\pi/d$, to enforce
\begin{equation}
|\langle\overline{j}|X_T|\overline{k}\rangle| < \epsilon
\end{equation}
for $j\neq k$, it suffices to require that $|\Theta| < 2\pi/d - \arccos(2\epsilon^{1/j} - 1)$.  Since this is a restriction on $T = R^{-1}R'$, the rotation angle $\Theta$ for a correctable rotation must then take values in \emph{half} this range.  Further specializing to $d = 2$, for which $\varphi - \varphi' = \pm\pi$, gives back \eqref{boundthis}.

For truly approximate quantum error correction, it suffices to demand that the diagonal conditions \eqref{diagonalcond} (and not only the off-diagonal conditions \eqref{offdiagonalcond}) be satisfied approximately, i.e., that \eqref{AQECCdiag} hold for some $\delta$.  One can do so by allowing for rotations through a bounded angle about an axis that is perturbed slightly away from the $z$-axis.  A general rotation $T$ will rotate the states $|\overline{k}\rangle$ by different amounts, depending on how close they are to the axis of rotation (equatorial rotations have the property that all codewords $|\overline{k}\rangle$ are equidistant from the axis of rotation).  Even after relaxing the requirement that the rotations be exactly equatorial, we would not expect such rotations to be complete for the LLL Hilbert space.  Indeed, if a set of errors is correctable, then its linear span is also correctable.  To be able to correct arbitrary errors on the entire Hilbert space, we would need to be able to correct $(2j + 1)^2$ linearly independent rotations, but $j$ can be arbitrarily large.

The fact that the correctable rotations for our code are (approximately) equatorial makes it more akin to a planar rotor code \cite{PhysRevX.10.031050} than to a linear rotor code.  The primary advantage of the spherical LLL over the planar rotor, or a rigid rotor more generally, is that it eliminates the need to contend with momentum kicks.  The magnetic field is an adjustable parameter, unlike the moment of inertia of a molecule, so the spectral gap can in principle be made arbitrarily large.  However, to make the error model of equatorial rotations fully realistic within this setup may require an additional electric potential to confine the charged particle to the equator.  Moreover, maintaining the degeneracy of the LLL requires fine-tuning between the magnetic field strength and the size of the sphere.\footnote{One might wonder whether this code can be implemented in a simpler system, such as a particle on a ring in the presence of a background magnetic field, in which the equatorial error model is ``built in.''  This is not possible, however, because such a system exhibits no projection to a finite-dimensional space of states in the large-field limit.  Hearkening back to the beginning of this section, the fundamental reason is that $S^1$ is not a symplectic manifold.}

\subsection{Check Operators and Logical Operators}

Having demonstrated that the conditions for approximate quantum error correction are sat\-is\-fied (for the specified set of correctable rotations), we would like to formulate appropriate logical and check operators for these codes, which inform the error diagnosis and recovery procedure.  Whereas certain powers of logical $X$ and logical $Z$ furnish convenient commuting check operators for ACP codes (assuming that these two operators commute on the entire Hilbert space and not just on the code subspace), our codes require only one check operator.  The intuition is that a $Z$-type operator serves as a check operator for $X$-type errors.

Before jumping into the general scenario, let us examine what happens for the antipodal code described by \eqref{qubit0} and \eqref{qubit1}.  A reasonable choice for an $X$-type logical operator that swaps $|\overline{0}\rangle$ and $|\overline{1}\rangle$ is a rotation by $\pi$ about the chosen equatorial axis:
\begin{equation}
X_{R(\varphi_0, \pi, -\varphi_0)} = e^{-i\varphi_0 L_3/\hbar}e^{-i\pi L_2/\hbar}e^{i\varphi_0 L_3/\hbar}.
\end{equation}
It satisfies
\begin{equation}
X_{R(\varphi_0, \pi, -\varphi_0)}^2 = X_{R(\varphi_0, 2\pi, -\varphi_0)} = (-1)^{2L_3/\hbar} = (-1)^{2j},
\end{equation}
which can be seen from \eqref{rotationformula} or from the fact that the operators $e^{-2\pi iL_1/\hbar} = e^{-2\pi iL_2/\hbar} = e^{-2\pi iL_3/\hbar}$ are scalar matrices ($\pm 1$) that depend only on whether $j$ is an integer or half-integer.  As a consequence of $X_{R(\varphi_0, \pm 2\pi, -\varphi_0)} = (-1)^{2j}$, the rotation in the opposite direction satisfies
\begin{equation}
X_{R(\varphi_0, -\pi, -\varphi_0)} = (-1)^{2j}X_{R(\varphi_0, \pi, -\varphi_0)}.
\end{equation}
Therefore, when $j$ is an integer, we have
\begin{equation}
X_{R(\varphi_0, \pi, -\varphi_0)}\sket{j}{j} = e^{2ij\varphi_0}\sket{j}{-j}, \qquad X_{R(\varphi_0, \pi, -\varphi_0)}^2 = 1,
\end{equation}
so $X_{R(\varphi_0, \pi, -\varphi_0)}$ acts like logical $X$ on the states \eqref{qubit0} and \eqref{qubit1}.

However, rotations cannot be used to realize a $Z$-type logical operator.  Indeed, among rotations, the only candidate for such an operator would be a rotation through an angle $\alpha$ about the $z$-axis: $X_{R(\alpha, 0, 0)} = e^{-i\alpha L_3/\hbar}$.  We have
\begin{equation}
X_{R(\alpha, 0, 0)}|\overline{0}\rangle = e^{-ij\alpha}|\overline{0}\rangle, \qquad X_{R(\alpha, 0, 0)}|\overline{1}\rangle = e^{ij\alpha}|\overline{1}\rangle,
\end{equation}
so such a rotation would act proportionally to logical $Z$ when $e^{2ij\alpha} = -1$ (and as long as we take $\alpha$ to be a multiple of $2\pi/j$, such a rotation preserves both codewords).  But correctable errors take the codewords to states of the form $X_{R(\varphi_0, \theta, -\varphi_0)}\sket{j}{j}$, which lie along a great circle, and any such ``error state'' should be an eigenstate of the $Z$-type check operator derived from the desired $Z$-type logical operator.  A full $2\pi$ rotation about the $z$-axis preserves each point on the great circle of error states (in fact, it preserves every point on the sphere), but the operator $X_{R(2\pi, 0, 0)} = e^{-2\pi iL_3/\hbar}$ simply acts as $(-1)^{2j}$ on the entire Hilbert space.

To solve this problem, we move ahead to discussing the logical and check operators for the more general $\Z_1 \subset \Z_d$ cyclic subgroup codes.  It is helpful to review the corresponding operators for continuous ACP codes.  By analogy with the stabilizer formalism, the $X$-type and $Z$-type check operators of ACP (which do not generate a group) are chosen such that their mutual eigenspace with unit eigenvalue is precisely the code subspace.  For the abelian ACP codes, any two operators satisfying the following criteria are valid $X$-type and $Z$-type check operators:
\begin{itemize}
\item An $X$-type check operator should preserve each codeword.
\item A $Z$-type check operator should have unit eigenvalue at every constituent point of the codewords.
\item The two check operators should commute with each other to ensure that they can be measured simultaneously.
\end{itemize}
In practice, we construct the check operators from the logical operators, which ensures that these two sets of operators commute with each other.  The main ingredient missing from our discussion so far is the construction of a $Z$-type logical operator for our codes.

As a point of departure, ACP propose to use angular momentum kicks as $Z$-type logical operators.  Specifically, consider the $\mathbb{Z}_N\subset \mathbb{Z}_{2N}$ linear rotor code \eqref{linrotorcodewords}.  The spherical har\-monics $Y^\ell_m(\theta, \varphi)$ have $\varphi$-dependence $e^{im\varphi}$, so $\hat{Y}^\ell_{\pm N}$ for any $\ell\geq N$ (or more simply, $e^{\pm iN\hat{\varphi}}$) acts as logical $Z$.  Hence $\hat{Y}^\ell_{\pm 2N}$ for any $\ell\geq 2N$ (or more simply, $e^{\pm 2iN\hat{\varphi}}$) acts as a $Z$-type check operator that reduces to the identity on the code subspace.\footnote{The disadvantage of using momentum kicks as logical operators is that they are not unitary on the entire Hilbert space.  One option is to take Hermitian combinations of them instead.  For example \cite{PhysRevX.10.031050}, one can use
\begin{align}
S_Z &= \cos(2N\hat{\varphi})\sin^{2N}\hat{\theta}\propto \hat{Y}^{2N}_{2N} + \hat{Y}^{2N}_{-2N}, \\
S_X &= \cos(2\pi L_3/N\hbar)\propto e^{-i(2\pi/N)L_3/\hbar} + e^{i(2\pi/N)L_3/\hbar}
\end{align}
as Hermitian $Z$-type and $X$-type check operators.  More generally, we can construct $S_Z$ as a linear combination of operators of the form $(\hat{v}\cdot w)^p\equiv \int_{S^2} dv\, (v\cdot w)^p|v\rangle\langle v|$, where $\hat{v}\equiv (\sin\hat{\theta}\cos\hat{\varphi}, \sin\hat{\theta}\sin\hat{\varphi}, \cos\hat{\theta})$ is the position operator in spherical coordinates, $w$ is a (generally complex) vector, and $p$ is a nonnegative integer.}

The lesson to draw from the previous paragraph is that $Z$-type operators are naturally diagonal in the ``position basis.''  Since our codes in the LLL Hilbert space do not have access to position eigenstates or to a ``position operator,'' we can only aim to follow this guideline approximately.

Consider, then, the equatorial qudit code defined in Section \ref{sphericalLLL}.  Up to phases, we expect the codewords to be
\begin{equation}
|\pi/2, 2\pi k/d\rangle_j = X_{R(2\pi k/d, \pi/2, -2\pi k/d)}\sket{j}{j}, \qquad k = 0, \ldots, d - 1.
\end{equation}
As our $X$-type logical operator, we choose
\begin{equation}
\overline{X} = e^{-i(2\pi/d)L_3/\hbar} = X_{R(2\pi/d, 0, 0)} = X_{R(0, 0, 2\pi/d)}.
\end{equation}
For this operator to satisfy $\overline{X}{}^d = 1$, we must take $j$ to be an integer.  We find that
\begin{align}
\overline{X}|\pi/2, 2\pi k/d\rangle_j &= X_{R(2\pi(k + 1)/d, \pi/2, -2\pi k/d)}\sket{j}{j} \\
&= X_{R(2\pi(k + 1)/d, \pi/2, -2\pi(k + 1)/d)}e^{-i(2\pi/d)L_3/\hbar}\sket{j}{j} \\
&= e^{-2\pi ij/d}|\pi/2, 2\pi(k + 1)/d\rangle_j.
\end{align}
Therefore, for $\overline{X}$ to have the desired action on codewords, we have two simple options (up to redefining all of the codewords by an overall phase):
\begin{enumerate}
\item Take
\begin{equation}
|\overline{k}\rangle = e^{-2\pi ijk/d}|\pi/2, 2\pi k/d\rangle_j.
\end{equation}
\item Take $|\overline{k}\rangle = |\pi/2, 2\pi k/d\rangle_j$ and assume that $j$ is a multiple of $d$.
\end{enumerate}
Both options are compatible with the (approximate) quantum error correction conditions.  In either case, we have
\begin{equation}
\overline{X} : |\overline{k}\rangle\mapsto |\overline{\text{$k + 1$ (mod $d$)}}\rangle.
\end{equation}
Note that these phase conventions differ from those in the antipodal qubit code of Section \ref{sphericalLLL} because here, the reference point (north pole) used in defining spin coherent states no longer coincides with the orientation of the codeword $|\overline{0}\rangle$.

For our qudit code, the $d^\text{th}$ power of the $X$-type logical operator (or the would-be $X$-type check operator) is the identity.  Hence the only constraints on the $Z$-type check operator are that it should have the codewords as eigenstates with eigenvalue $+1$, and it should commute with the logical operators.

Perhaps the simplest candidate for a $Z$-type logical operator is
\begin{equation}
\overline{Z} = \int d\Omega\, e^{i\varphi}|\Omega\rangle_j\ds{}_j\langle\Omega|,
\end{equation}
which satisfies
\begin{equation}
\overline{Z}|\overline{k}\rangle\approx e^{2\pi ik/d}|\overline{k}\rangle, \qquad \overline{Z}{}^d\approx \int d\Omega\, e^{id\varphi}|\Omega\rangle_j\ds{}_j\langle\Omega|,
\end{equation}
so that $\overline{Z}{}^d$ approximates the identity on the code subspace (but \emph{not} on the entire Hilbert space, since it should serve as a check operator).  These relations are approximate because the states $|\Omega\rangle_j$ are not orthogonal.

Any operator on the LLL Hilbert space has a diagonal expansion of the form
\begin{equation}
\int d\Omega\, P(\Omega)|\Omega\rangle_j\ds{}_j\langle\Omega|.
\label{diagonalexpansion}
\end{equation}
For real $P(\Omega)$, such an operator is Hermitian.  Two operators
\begin{equation}
\int d\Omega\, P_1(\Omega)|\Omega\rangle_j\ds{}_j\langle\Omega|, \qquad \int d\Omega\, P_2(\Omega)|\Omega\rangle_j\ds{}_j\langle\Omega|
\end{equation}
that are ``diagonal in the position basis'' do not necessarily commute, as
\begin{equation}
\int d\Omega\, d\Omega'\, P_1(\Omega)P_2(\Omega')|\Omega\rangle_j\ds{}_j\langle\Omega|\Omega'\rangle_j\ds{}_j\langle\Omega'|\neq \int d\Omega\, d\Omega'\, P_2(\Omega)P_1(\Omega')|\Omega\rangle_j\ds{}_j\langle\Omega|\Omega'\rangle_j\ds{}_j\langle\Omega'|
\end{equation}
in general.  Only in the limit $j\to\infty$ do such operators approach diagonal matrices and thus mutually commute.  A ``diagonal'' operator \eqref{diagonalexpansion} has an expectation value
\begin{equation}
\int d\Omega'\, P(\Omega')|{}_j\langle\Omega'|\Omega\rangle_j|^2
\end{equation}
and a nontrivial variance in the state $|\Omega\rangle_j$, which approach $P(\Omega)$ and 0 as $j\to\infty$, respectively.  These quantities are computable in the saddle-point approximation for large $j$.

When does an operator of the form \eqref{diagonalexpansion} commute with the logical operators?  Since $X_R|\Omega\rangle_j$ equals $|R\Omega\rangle_j$ up to a phase, we have the exact equality
\begin{equation}
X_R|\Omega\rangle_j\ds{}_j\langle\Omega|X_R^\dag = |R\Omega\rangle_j\ds{}_j\langle R\Omega|.
\end{equation}
By rotational invariance of the measure $d\Omega$, we therefore have
\begin{equation}
X_R\int d\Omega\, P(\Omega)|\Omega\rangle_j\ds{}_j\langle\Omega| = \int d\Omega\, P(\Omega)|R\Omega\rangle_j\ds{}_j\langle R\Omega|X_R = \int d\Omega\, P(R^{-1}\Omega)|\Omega\rangle_j\ds{}_j\langle\Omega|X_R.
\end{equation}
It follows that
\begin{equation}
\overline{Z}\overline{X} = e^{2\pi i/d}\overline{X}\overline{Z},
\end{equation}
as desired.  In particular, $\overline{Z}{}^d$ commutes with both $\overline{X}$ and $\overline{Z}$.

Note that $\overline{Z}$ as defined above is only approximately a logical operator because it only approximately preserves the code subspace.  Therefore, there is nothing distinguished about using $\overline{Z}{}^d$ as an approximate check operator.  We could alternatively use
\begin{equation}
\overline{Z}_d\equiv \int d\Omega\, e^{id\varphi}|\Omega\rangle_j\ds{}_j\langle\Omega|
\end{equation}
directly as a $Z$-type check operator, which commutes with $\overline{X}$ but (unlike $\overline{Z}{}^d$) only approximately commutes with $\overline{Z}$.  Whichever $Z$-type check operator we choose should allow us to \emph{approximately} measure $\varphi$ modulo $2\pi/d$.

Unfortunately, the operators $\overline{Z}$, $\overline{Z}{}^d$, $\overline{Z}_d$ are only approximately unitary, which presents an obstacle to measuring them.  There is a workaround in the case that $d = 2$: we can replace them with operators that are exactly Hermitian.  For instance, one can use
\begin{equation}
\int d\Omega\, \cos\varphi|\Omega\rangle_j\ds{}_j\langle\Omega|, \qquad \int d\Omega\, \cos^2\varphi|\Omega\rangle_j\ds{}_j\langle\Omega|
\end{equation}
as $Z$-type logical and check operators, respectively.  However, this solution does not work for $d > 2$ because in those cases, $\overline{Z}$ must have complex (approximate) eigenvalues rather than merely $\pm 1$.

The fact that the operator $\int d\Omega\, e^{i\varphi}|\Omega\rangle_j\ds{}_j\langle\Omega|$ is approximately unitary means that it can be measured approximately for large $j$.  Equivalently, the Hermitian operators
\begin{equation}
\int d\Omega\, \cos\varphi|\Omega\rangle_j\ds{}_j\langle\Omega|, \qquad \int d\Omega\, \sin\varphi|\Omega\rangle_j\ds{}_j\langle\Omega|
\end{equation}
approximately commute at large $j$, and can be measured simultaneously.

Having formulated the check operators, in practice, we diagnose the error syndrome by transferring the error to an ancilla using a conditional rotation (this is the scheme, originally due to Steane \cite{Steane}, used by GKP and ACP \cite{Gottesman:2000di, PhysRevX.10.031050} for diagnosing rotation errors).  The check operators are a mathematical proxy for this procedure.  In Appendix \ref{ancillas}, we show that the approximation error in this procedure is exponentially suppressed as $j\to\infty$.

\section{Discussion} \label{discussion}

In this paper, we have proposed approximate quantum error-correcting codes in which quantum information is stored in the orientation of spin coherent states in the finite-dimensional Hilbert space of a single (large) spin.  Such codes satisfy approximate versions of the Knill-Laflamme conditions.\footnote{Somewhat different applications of spin coherent states to approximate quantum error correction can be found in \cite{PhysRevA.97.032346}.}  We have discussed logical operators that act as generalized Pauli $X$ and $Z$ within the code subspace, as well as check operators that can be measured to diagnose and recover from certain rotational errors.  Many open questions remain: for instance, one might hope to formulate a universal set of gates acting on the code subspace, or at least a set that generates the Clifford group.  Further work is also required to make such a scheme truly practical by enabling fault-tolerant quantum operations on the encoded information (in other words, coherent information processing of spin coherent states).

While our discussion has used the spherical Landau problem as a theoretical platform (as often done in studies of the quantum Hall effect \cite{Haldane:1983xm}), it may not provide the most practical implementation of our codes.  Indeed, our mathematical framework extends to any finite-di\-men\-sional quantum system on which rotations constitute the most physically natural set of error operations.  Such a system could be engineered as a collective spin built from a large number of small spins (see \cite{PhysRevX.10.031050}), in which case it would need to be checked whether rotational errors acting on the emergent large spin constitute a realistic error model.  More concretely, our codewords could be realized in atomic ensembles \cite{RevModPhys.90.035005}.  In this context, the totally symmetric sector of the Hilbert space of $2j$ qubits (spin-$1/2$ particles) has spin $j$, and a spin coherent state in this symmetric subspace is constructed by orienting each constituent spin in the same direction.

Our LLL codes resemble classical spherical codes that maximize the distance between points on a sphere.\footnote{Exact spherical codes are known for 2, 3, 4, 6, or 12 points, corresponding to placing the points at the vertices of a diameter, an equilateral triangle on a great circle, a regular tetrahedron, a regular octahedron, or a regular icosahedron, respectively.}  They are analogous to, but simpler than, ACP codes \cite{PhysRevX.10.031050} based on discrete subgroups of $SO(3)$.  However, we have only presented finite-dimensional realizations of the abelian linear rotor codes (cyclic subgroup codes) of ACP, which involve spin coherent states that are evenly distributed along a great circle.  More general configurations of spin coherent states may not satisfy the quantum error correction conditions (due to the phases that spin coherent states acquire under rotations), which presents an obstacle to constructing LLL analogues of the nonabelian subgroup codes.

A different method for constructing quantum error-correcting codes in the Hilbert space of a large spin has been put forward by Gross \cite{Gross}.  That method prioritizes the construction of a transversal gate set over the correction of all probable errors.  The code subspaces are irreps of discrete subgroups of $SU(2)$ that appear inside a given irrep of $SU(2)$, and the corresponding codewords are comparatively simple linear combinations of $\sket{j}{m}$ states.  It would be interesting to explore possible connections between our proposal and that of \cite{Gross}, and in particular to design codes that combine their relative advantages.

Finally, contemplating the generalization of our codes to configuration space geometries beyond $S^2$ raises many questions about their deeper mathematical structure.  The quantum Hall effect has been defined and studied on a variety of compact manifolds in two and higher dimensions, including the spheres $S^2$, $S^4$, and $S^8$ (via division algebras) \cite{Haldane:1983xm, Zhang:2001xs, Bernevig:2003yz}; complex projective spaces $\mathbb{CP}^n$ \cite{Karabali:2002im}; arbitrary even-dimensional spheres (via Clifford algebras) \cite{Hasebe:2010vp}; and most generally, coset spaces $G/H$ where $G$ and $H$ are compact Lie groups \cite{Dolan:2003bj}.\footnote{In those cases where the stated manifold is not symplectic, such as $S^{2n}$ for $n\geq 2$, one must account for internal (isospin) degrees of freedom to obtain the full configuration space \cite{Bernevig:2002eq}.}  In all of these cases, close connections to generalized coherent states \cite{Perelomov:1971bd, Onofri1975, Perelomovbook, Dunne:1991cs} and to noncommutative geometry \cite{Douglas:2001ba} abound.\footnote{The algebra of magnetic translations on a compact space naturally describes a noncommutative version of that space.  The Landau problem can also be studied on a configuration space that is itself noncommutative \cite{Nair:2000ii, Morariu:2001dv}.}  Moreover, the existence of parallels between GKP codes and Landau levels is no accident.  On one hand, the GKP approach to quantum error correction takes as its guiding principle the noncommutative geometry of phase space.  On the other hand, background magnetic fields induce noncommutativity between position coordinates, which turns into noncommutativity of phase space coordinates in the LLL limit.  All of these considerations hint at richer connections between GKP-like codes and noncommutative geometry waiting to be explored.

\section*{Acknowledgements}

We thank Victor Albert, Phuc Nguyen, Jess Riedel, and Ian Teixeira for helpful discussions and correspondence, as well as Ted Jacobson for facilitating this collaboration.  YF thanks the UT quantum information group for inspiration and the Aspen Center for Physics (sup\-port\-ed by NSF grant PHY-1607611) for hospitality during a portion of this work.  The work of YF and WF was supported in part by the NSF grant PHY-1914679.  The work of EK was supported in part by the NSF QLCI grant OMA-2120757.

\newpage

\appendix

\section{Landau Levels} \label{LLs}

Here, we describe the physical setup and set our conventions for Landau levels.  The material is standard; some accessible references include \cite{Tong:2016kpv, murayamaLL, murayamaspin}.  We make $\hbar$ and other parameters explicit.  We use boldface for differential forms.

Consider a spinless particle of electric charge $q$ and mass $m$ in a background magnetic field $\vec{B} = \nabla\times \vec{A}$ (in practice, $q = -e$).  The classical Lagrangian and Hamiltonian are
\begin{equation}
L = \frac{1}{2}m\dot{\vec{x}}^2 + q\dot{\vec{x}}\cdot \vec{A}, \qquad H = \frac{1}{2m}(\vec{p} - q\vec{A})^2.
\label{classlagham}
\end{equation}
The kinetic momentum $m\dot{\vec{x}}$ is gauge-invariant, while the canonical momentum $\vec{p} = m\dot{\vec{x}} + q\vec{A}$ is not.  We have the classical Poisson brackets
\begin{equation}
\{x_i, p_j\} = \delta_{ij}, \qquad \{x_i, x_j\} = \{p_i, p_j\} = 0, \qquad \{m\dot{x}_i, m\dot{x}_j\} = q\epsilon_{ijk}B^k.
\end{equation}
The quantum Hamiltonian is
\begin{equation}
H = \frac{1}{2m}\vec{\pi}^2, \qquad \vec{\pi}\equiv \vec{p} - q\vec{A},
\end{equation}
now in terms of operators satisfying the canonical commutation relations
\begin{equation}
[x_i, p_j] = i\hbar\delta_{ij}, \qquad [x_i, x_j] = [p_i, p_j] = 0, \qquad [\pi_i, \pi_j] = i\hbar q\epsilon_{ijk}B^k.
\end{equation}
The distinction between classical phase space coordinates and quantum operators is implicit.

\subsection{Plane}

Consider planar motion in a uniform, perpendicular magnetic field.  We write $\vec{x} = (x, y, 0)$ and $\vec{B} = (0, 0, B)$, taking $qB > 0$.  We define the cyclotron frequency and magnetic length
\begin{equation}
\omega_B\equiv \frac{qB}{m}, \qquad \ell_B\equiv \left(\frac{\hbar}{qB}\right)^{1/2}.
\end{equation}
It is convenient to work in terms of kinetic momenta and guiding center coordinates rather than canonical momenta and canonical coordinates.  We have $[\pi_x, \pi_y] = i\hbar qB$, allowing us to define the Landau level ladder operators
\begin{equation}
a\equiv \frac{1}{\sqrt{2\hbar qB}}(\pi_x + i\pi_y), \qquad a^\dag\equiv \frac{1}{\sqrt{2\hbar qB}}(\pi_x - i\pi_y)
\end{equation}
satisfying $[a, a^\dag] = 1$ and in terms of which
\begin{equation}
H = \hbar\omega_B\left(a^\dag a + \frac{1}{2}\right).
\end{equation}
The energy levels are now manifest.  We can argue for the degeneracies as follows.  We define the guiding center operators
\begin{equation}
R_x\equiv x + \frac{\pi_y}{m\omega_B}, \qquad R_y\equiv y - \frac{\pi_x}{m\omega_B},
\end{equation}
which classically correspond to the coordinates of the center of the cyclotron orbit.  They commute with the kinetic momenta and are therefore constants of motion:
\begin{equation}
[\pi_i, R_j] = 0 \implies [H, R_x] = [H, R_y] = 0.
\end{equation}
We compute that
\begin{equation}
[R_x, R_y] = -i\ell_B^2,
\label{guidingcentercomm}
\end{equation}
so the guiding center ladder operators
\begin{equation}
b\equiv -\frac{i}{\sqrt{2}\ell_B}(R_x - iR_y), \qquad b^\dag\equiv \frac{i}{\sqrt{2}\ell_B}(R_x + iR_y)
\end{equation}
satisfy $[b, b^\dag] = 1$ and commute with $H$.  The classical intuition for the noncommutativity of the guiding center coordinates is that the cyclotron orbits prevent us from localizing states in both spatial directions simultaneously.  The corresponding minimum uncertainty is
\begin{equation}
\Delta R_x\Delta R_y\sim 2\pi\ell_B^2,
\end{equation}
so a semiclassical estimate of the number of states in each Landau level within an area $A$ is
\begin{equation}
\frac{A}{\Delta R_x\Delta R_y}\sim \frac{qBA}{2\pi\hbar}.
\end{equation}
The above considerations are independent of gauge.

The unitary magnetic translation operators are defined as
\begin{equation}
T_x(r) = e^{irR_y/\ell_B^2}, \qquad T_y(r) = e^{-irR_x/\ell_B^2}.
\end{equation}
They shift the guiding center operators,
\begin{equation}
T_i(r)^\dag R_i T_i(r) = R_i + r, \qquad i = x, y,
\end{equation}
and they satisfy
\begin{equation}
T_x(r_x)T_y(r_y) = e^{ir_x r_y/\ell_B^2}T_y(r_y)T_x(r_x) = e^{2\pi iBr_x r_y/\Phi_0}T_y(r_y)T_x(r_x).
\end{equation}
(Recall: if $[X, Y]$ is a $c$-number, then $e^{X + Y} = e^{X}e^{Y}e^{-\frac{1}{2}[X, Y]}$ and $e^Y Xe^{-Y} = X + [Y, X]$.)

To write the Landau level wavefunctions in a specific basis, it is convenient to work in either the Landau (translationally invariant) gauge, where $[H, p_x] = 0$ or $[H, p_y] = 0$, or the symmetric (rotationally invariant) gauge, where $[H, J_z] = 0$.  In either case, one can argue semiclassically for the degeneracy in a finite area of the plane.  We consider the symmetric gauge $\vec{A} = \frac{B}{2}(-y, x, 0)$ because it allows us to introduce complex coordinates.  In this gauge, the (canonical, not kinetic) angular momentum operator $J_z\equiv xp_y - yp_x$ can be written as
\begin{equation}
J_z = \hbar(b^\dag b - a^\dag a).
\end{equation}
Hence the Hilbert space is spanned by the Fock states
\begin{equation}
|N, M\rangle\equiv \frac{(a^\dag)^N(b^\dag)^M}{\sqrt{N!M!}}|0, 0\rangle, \qquad H|N, M\rangle = \hbar\omega_B\left(N + \frac{1}{2}\right)|N, M\rangle,
\end{equation}
where $a|0, 0\rangle = b|0, 0\rangle = 0$ and $N, M\geq 0$, which are simultaneously eigenstates of $J_z$:
\begin{equation}
J_z|N, M\rangle = \hbar m_z|N, M\rangle, \qquad m_z\equiv M - N.
\end{equation}
In particular, $m_z\geq -N$ in the $N^\text{th}$ Landau level.  In terms of the coordinate $z = x + iy$, we have in symmetric gauge that
\begin{alignat}{2}
a &= -i\sqrt{2}\left(\ell_B\bar{\partial} + \frac{z}{4\ell_B}\right), \qquad & a^\dag &= -i\sqrt{2}\left(\ell_B\partial - \frac{\bar{z}}{4\ell_B}\right), \\
b &= -i\sqrt{2}\left(\ell_B\partial + \frac{\bar{z}}{4\ell_B}\right), \qquad & b^\dag &= -i\sqrt{2}\left(\ell_B\bar{\partial} - \frac{z}{4\ell_B}\right),
\end{alignat}
and $J_z = \hbar(z\partial - \bar{z}\bar{\partial})$.  The wavefunctions of the LLL states (those annihilated by $a$) therefore take the form
\begin{equation}
f(z)e^{-|z|^2/4\ell_B^2}
\end{equation}
for holomorphic $f$ (up to normalization).  Specifically,
\begin{equation}
\langle z, \bar{z}|0, M\rangle\sim \left(\frac{z}{\ell_B}\right)^M e^{-|z|^2/4\ell_B^2}.
\end{equation}
One can construct the overlaps $\langle z, \bar{z}|N, M\rangle$ similarly.  Coherent states centered at $z = z_0$, obtained by acting on the ground state with exponentials of $a^\dag$, are eigenstates of $a$ with eigenvalue $\propto z_0$; they are linear combinations of states in all Landau levels, and their time evolution describes cyclotron orbits.

By extending the structure group from $U(1)$ to $GL(1, \mathbb{C})$, we can regard the gauge field as a connection on a complex line bundle over $\mathbb{C}$.  In symmetric (real) gauge, we have
\begin{equation}
\mathbf{A} = \frac{B}{2}(-y\, dx + x\, dy) = -\frac{iB}{4}(\bar{z}\, dz - z\, d\bar{z}).
\end{equation}
We pass to holomorphic gauge via a complexified gauge transformation:
\begin{equation}
\mathbf{A}' = \mathbf{A} + d\Lambda = -\frac{iB}{2}\bar{z}\, dz, \qquad \Lambda\equiv -\frac{iB}{4}|z|^2.
\end{equation}
Since $\psi\to e^{iq\Lambda/\hbar}\psi$, this changes the inner product as follows:
\begin{equation}
\int d^2 z\, e^{-|z|^2/2\ell_B^2}\psi_1(z)^\ast\psi_2(z)
\end{equation}
(the measure factor compensates).  In holomorphic gauge, we have
\begin{alignat}{2}
a &= -i\sqrt{2}\ell_B\bar{\partial}, \qquad & a^\dag &= -i\sqrt{2}\left(\ell_B\partial - \frac{\bar{z}}{2\ell_B}\right), \\
b &= -i\sqrt{2}\ell_B\partial, \qquad & b^\dag &= -i\sqrt{2}\left(\ell_B\bar{\partial} - \frac{z}{2\ell_B}\right).
\end{alignat}
We see that the ground-state wavefunctions are precisely the holomorphic functions.  The advantage of this rewriting is that when passing to a compact Riemann surface, the exact ground-state degeneracy can be computed as the number of holomorphic sections of a certain holomorphic line bundle.  On the plane, we have $\mathbf{A} = A_z\, dz + A_{\bar{z}}\, d\bar{z}$ and
\begin{equation}
\mathbf{B} = d\mathbf{A} = (\partial_z A_{\bar{z}} - \partial_{\bar{z}}A_z)\, dz\wedge d\bar{z} = B\, dz\wedge d\bar{z}.
\end{equation}
On a compact surface, we take $\mathbf{B}$ to be a multiple of the volume form, keeping in mind that the total magnetic flux must be quantized in units of the magnetic flux quantum $\Phi_0 = 2\pi\hbar/q$, or the amount of flux within an area $2\pi\ell_B^2$.

Finally, the lowest Landau level is a prototype for coordinate noncommutativity.  When $m = 0$, the classical Lagrangian and Hamiltonian are $L = q\dot{\vec{x}}\cdot \vec{A}$ and $H = 0$.  Integrating by parts gives $L = -qBy\dot{x}$ in any gauge.  So we have a reduced phase space with one canonical coordinate $x$ and its conjugate momentum $p = -qBy$, giving $\{x, p\} = 1$ and therefore
\begin{equation}
[x, y] = -\frac{i\hbar}{qB}.
\end{equation}
Note that the guiding center coordinates are noncommutative in exactly the same way as the ordinary coordinates in the LLL, by \eqref{guidingcentercomm}.  In this sense, noncommutativity is not specific to the LLL.

\subsection{Torus}

We parametrize the torus $T^2$ by the \emph{dimensionless} complex coordinate $z$ where $z\sim z + 1$ and $z\sim z + \tau$ with $\operatorname{Im}\tau > 0$.  Equivalently, $z = \tilde{x} + \tau\tilde{y}$ with $\tilde{x}\sim \tilde{x} + 1$ and $\tilde{y}\sim \tilde{y} + 1$.  Here,
\begin{equation}
\tilde{x}\equiv \frac{x}{L_x}, \qquad \tilde{y}\equiv \frac{y}{L_y},
\end{equation}
where $x\sim x + L_x$ and $y\sim y + L_y$.  The volume form is
\begin{equation}
\omega = dx\wedge dy = L_x L_y\, d\tilde{x}\wedge d\tilde{y} = \frac{L_x L_y}{\bar{\tau} - \tau}\, dz\wedge d\bar{z}, \qquad \int_{T^2} \omega = L_x L_y.
\end{equation}
We take
\begin{equation}
\mathbf{B} = B\omega = \frac{N\Phi_0}{\bar{\tau} - \tau}\, dz\wedge d\bar{z},
\end{equation}
where $BL_x L_y = N\Phi_0$ and $N$ is the number of flux quanta.  In holomorphic gauge ($\mathbf{A} = A_z\, dz$ and $\mathbf{B} = -\partial_{\bar{z}}A_z\, dz\wedge d\bar{z}$), invariance under $z\to z + 1$ fixes
\begin{equation}
\mathbf{A} = -\frac{N\Phi_0(\bar{z} - z)}{\bar{\tau} - \tau}\, dz.
\end{equation}
Under $z\to z + \tau$, $\mathbf{A}$ undergoes the following gauge transformation:
\begin{equation}
\mathbf{A}\to \mathbf{A} - N\Phi_0\, dz = \mathbf{A} - d(N\Phi_0 z + c\hbar/q),
\end{equation}
where $c$ is a dimensionless constant.  Correspondingly, wavefunctions must transform by a factor of $e^{-i(2\pi Nz + c)}$ under $z\to z + \tau$.  The level-$N$ theta functions defined by\footnote{Alternatively, $\Theta_{m, N}(z) = \vartheta\left[\begin{smallmatrix} m/N \\ 0 \end{smallmatrix}\right](Nz, N\tau)$ where the generalized elliptic theta functions are
\begin{equation}
\vartheta\left[\begin{array}{c} a \\ b \end{array}\right](z, \tau) = \sum_{n\in \mathbb{Z}} e^{\pi i(n + a)^2\tau}e^{2\pi i(n + a)(z + b)}.
\end{equation}}
\begin{equation}
\Theta_{m, N}(z) = \sum_{n\in \mathbb{Z} + \frac{m}{N}} e^{N(\pi in^2\tau + 2\pi inz)}, \qquad m = 0, \ldots, N - 1
\end{equation}
are periodic under $z\to z + 1$ and quasiperiodic in the desired manner under $z\to z + \tau$:
\begin{align}
\Theta_{m, N}(z + 1) &= \Theta_{m, N}(z), \\
\Theta_{m, N}(z + \tau) &= e^{-N(2\pi iz + \pi i\tau)}\Theta_{m, N}(z).
\end{align}
They hence furnish a basis of LLL wavefunctions on the torus.  In real gauge ($\mathbf{A} = A_z\, dz + \overline{A_z}\, d\bar{z}$ and $\mathbf{B} = (\partial_z\overline{A_z} - \partial_{\bar{z}}A_z)\, dz\wedge d\bar{z}$), invariance under $z\to z + 1$ fixes
\begin{equation}
\mathbf{A} = -\frac{N\Phi_0(\bar{z} - z)}{2(\bar{\tau} - \tau)}(dz + d\bar{z}).
\end{equation}
We have the relation
\begin{equation}
\mathbf{A}_\text{hol} = \mathbf{A}_\text{real} + \frac{N\Phi_0}{4(\bar{\tau} - \tau)}d(\bar{z} - z)^2,
\end{equation}
so the holomorphic inner product is given by
\begin{equation}
\int \omega e^{-N\pi i(\bar{z} - z)^2/(\bar{\tau} - \tau)}\psi_1(z)^\ast\psi_2(z).
\end{equation}
This is invariant under both $z\to z + 1$ and $z\to z + \tau$.

We can obtain the algebra of generalized Pauli operators ($N$-dimensional $X, Z$) from the action of magnetic translations on these explicit LLL wavefunctions.  In holomorphic gauge, we compute the guiding center operators to be
\begin{equation}
R_x = L_x z - \frac{i\ell_B^2}{L_y}(\tau\partial_z + \bar{\tau}\partial_{\bar{z}}), \qquad R_y = \frac{i\ell_B^2}{L_x}(\partial_z + \partial_{\bar{z}}).
\end{equation}
Following GKP \cite{Gottesman:2000di}, we set
\begin{align}
T_1 &\equiv T_x(L_x/r_1) = e^{-(1/r_1)(\partial_z + \partial_{\bar{z}})}, \\
T_2 &\equiv T_y(L_y/Kr_2) = e^{-(1/Kr_2)(iL_x L_y z/\ell_B^2 + \tau\partial_z + \bar{\tau}\partial_{\bar{z}})}.
\end{align}
The error operators $Z = T_1^{1/Kr_2}$ and $X = T_2^{1/r_1}$ clearly satisfy
\begin{equation}
ZX = e^{(1/Kr_1 r_2)^2(iL_x L_y/\ell_B^2)}XZ = e^{2\pi i/N}XZ\equiv \omega XZ,
\end{equation}
where $N = Kr_1 r_2$.  The operators $T_1$ and $T_2$ themselves commute.  In the LLL, we can set $\partial_{\bar{z}} = 0$ and write
\begin{equation}
Z = e^{-(1/N)\partial_z}, \qquad X = e^{-2\pi iz - (\tau/N)\partial_z}.
\end{equation}
With respect to the basis $\Theta_{m, N}(z)$ for $m = 0, \ldots, N - 1$, we compute that
\begin{align}
Z\Theta_{m, N}(z) &= \omega^{-m}\Theta_{m, N}(z), \\
X\Theta_{m, N}(z) &= e^{-\pi i\tau/N}\Theta_{m - 1, N}(z).
\end{align}
So the codewords are the (unnormalized) superpositions
\begin{equation}
\sum_{n=0}^{r_2 - 1} \Theta_{(Kn + j)r_1, N}(z), \qquad j = 0, \ldots, K - 1,
\end{equation}
as in Section \ref{finiteGKP}.

\subsection{Sphere} \label{sphere}

\subsubsection{Complex Coordinates}

For a sphere $S^2$ of radius $R$, stereographic projection from the south pole yields the map
\begin{equation}
z = 2R\tan(\theta/2)e^{i\varphi}
\end{equation}
between spherical coordinates $(\theta, \varphi)$ and complex coordinates $(z, \bar{z})$ and hence the volume form
\begin{equation}
\omega = R^2\sin\theta\, d\varphi\wedge d\theta = -\frac{i}{2}\frac{dz\wedge d\bar{z}}{(1 + |z|^2/4R^2)^2}.
\end{equation}
We take $\mathbf{B} = B\omega$ where $4\pi R^2 B = N\Phi_0$.  A suitable holomorphic gauge (regular at $z = 0$) is
\begin{equation}
\mathbf{A} = -\frac{iB\bar{z}}{2(1 + |z|^2/4R^2)}\, dz.
\end{equation}
Under the coordinate change $w = 4R^2/z$, $\omega$ is invariant:
\begin{equation}
\omega = -\frac{i}{2}\frac{dw\wedge d\bar{w}}{(1 + |w|^2/4R^2)^2}.
\end{equation}
On the other hand, we have
\begin{equation}
\mathbf{A} = \frac{iB}{2(w/4R^2)(1 + |w|^2/4R^2)}\, dw.
\end{equation}
Via a gauge transformation
\begin{equation}
\mathbf{A}\to \mathbf{A} - 2iBR^2\, d\log(w/2R) = -\frac{iB\bar{w}}{2(1 + |w|^2/4R^2)}\, dw,
\end{equation}
$\mathbf{A}$ is rendered regular at $w = 0$.  Wavefunctions transform as $\psi\to (w/2R)^N\psi$.  We require LLL wavefunctions to be holomorphic, so that $\psi(z)$ and $(w/2R)^N\psi(4R^2/w)$ are regular at $z = 0$ and $w = 0$, respectively.  Hence a basis of LLL wavefunctions is given by
\begin{equation}
1, z, \ldots, z^N.
\end{equation}
A suitable real gauge (regular at $z = 0$) is given by
\begin{equation}
\mathbf{A} = -\frac{iB}{4(1 + |z|^2/4R^2)}(\bar{z}\, dz - z\, d\bar{z}).
\label{realgaugesouth}
\end{equation}
In terms of $w = 4R^2/z$, we have
\begin{equation}
\mathbf{A} = \frac{iBR^2}{1 + |w|^2/4R^2}\left(\frac{dw}{w} - \frac{d\bar{w}}{\bar{w}}\right).
\end{equation}
Via a gauge transformation
\begin{equation}
\mathbf{A}\to \mathbf{A} - iBR^2\, d\log\left(\frac{w}{\bar{w}}\right) = -\frac{iB}{4(1 + |w|^2/4R^2)}(\bar{w}\, dw - w\, d\bar{w}),
\label{gaugesouthtonorth}
\end{equation}
$\mathbf{A}$ is rendered regular at $w = 0$.  In terms of $z$, we have the relation
\begin{equation}
\mathbf{A}_\text{hol} = \mathbf{A}_\text{real} - iBR^2\, d\log(1 + |z|^2/4R^2),
\end{equation}
so the holomorphic inner product is given by
\begin{equation}
\int \frac{\omega}{(1 + |z|^2/4R^2)^N}\psi_1(z)^\ast\psi_2(z).
\end{equation}
This is equivalent to
\begin{equation}
\int \frac{\omega}{(1 + |w|^2/4R^2)^N}(w/2R)^N\psi_1(4R^2/w)^\ast(\bar{w}/2R)^N\psi_2(4R^2/w),
\end{equation}
so the inner product is invariant under a change of patch.

\subsubsection{Spherical Coordinates}

To treat the full Landau problem, we work in spherical coordinates.  We consider a particle of electric charge $q$ constrained to move on a sphere $S^2$ of radius $R$ around a monopole of magnetic charge $BR^2$ \cite{Shnir:2005vvi}.  Its Hamiltonian is
\begin{equation}
H = \frac{1}{2m}(-i\hbar\nabla - q\vec{A})^2|_{r=R},
\end{equation}
where we may choose a longitudinal gauge that is regular at the south or the north pole:
\begin{equation}
\vec{A} = -\frac{BR^2(\varepsilon + \cos\theta)}{r\sin\theta}\hat{\varphi}, \qquad \varepsilon = \begin{cases} +1 & \text{regular at south pole}, \\ -1 & \text{regular at north pole}. \end{cases}
\end{equation}
This vector potential is divergenceless and singular along the positive or the negative $z$-axis, depending on whether $\varepsilon = +1$ or $\varepsilon = -1$ ($z\equiv x_3$ meaning the Cartesian coordinate, not the complex coordinate).  Away from the singularity, we have
\begin{equation}
\nabla\times\vec{A} = \frac{BR^2\hat{r}}{r^2}.
\end{equation}
Equivalently, we have
\begin{equation}
\mathbf{A}_\text{real} = -BR^2(\varepsilon + \cos\theta)\, d\varphi, \qquad \mathbf{B} = BR^2\sin\theta\, d\theta\wedge d\varphi.
\label{Areal}
\end{equation}
This real gauge coincides with \eqref{realgaugesouth} for $\varepsilon = -1$.  The two signs $\varepsilon = \pm 1$ are related by a gauge transformation, as in \eqref{gaugesouthtonorth}:
\begin{equation}
\mathbf{A}_\text{real}|_{\varepsilon = +1} = \mathbf{A}_\text{real}|_{\varepsilon = -1} - 2BR^2\, d\varphi.
\end{equation}
We compute that
\begin{equation}
H = \frac{1}{2m}\left[-\frac{\hbar^2}{R^2}\Delta_{S^2} + \frac{q^2 B^2 R^2(\varepsilon + \cos\theta)^2}{\sin^2\theta} - \frac{2i\hbar qB(\varepsilon + \cos\theta)}{\sin^2\theta}\partial_\varphi\right]
\end{equation}
where the Laplace-Beltrami operator on $S^2$ is
\begin{equation}
\Delta_{S^2} = \frac{1}{\sin^2\theta}\partial_\varphi^2 + \frac{1}{\sin\theta}\partial_\theta(\sin\theta\partial_\theta),
\end{equation}
whose eigenvalues are $-l(l + 1)$ with degeneracies $2l + 1$ for $l\geq 0$.  The standard angular momentum operators are
\begin{equation}
\vec{D} = -i\hbar(-\sin\varphi\partial_\theta - \cos\varphi\cot\theta\partial_\varphi, \cos\varphi\partial_\theta - \sin\varphi\cot\theta\partial_\varphi, \partial_\varphi).
\label{dee}
\end{equation}
They satisfy $[D_i, D_j] = i\hbar\epsilon_{ijk}D^k$ and do not commute with $H$.  The ``good'' (guiding center) angular momenta, accounting for the contribution from the electromagnetic field, are
\begin{align}
\vec{L} &= \vec{r}\times(-i\hbar\nabla - q\vec{A}) - qBR^2\hat{r} \\
&= \vec{D} - \varepsilon qBR^2\left(\frac{(\varepsilon + \cos\theta)\cos\varphi}{\sin\theta}, \frac{(\varepsilon + \cos\theta)\sin\varphi}{\sin\theta}, -1\right), \label{ell}
\end{align}
where $\hat{r} = \vec{r}/r$.  They satisfy $[L_i, L_j] = i\hbar\epsilon_{ijk}L^k$ and commute with $H$.\footnote{Indeed \cite{Greiter}, they satisfy $[L_i, X_j] = i\hbar\epsilon_{ijk}X^k$ for $X = \hat{r}, \Lambda, L$, where we have defined the position operator $\hat{r}\equiv (\cos\varphi\sin\theta, \sin\varphi\sin\theta, \cos\theta)$ and the dynamical angular momenta $\vec{\Lambda}\equiv \vec{r}\times(-i\hbar\nabla - q\vec{A})$.  The commutator with $\hat{r}$ follows from $[L_i, \hat{r}_j] = [D_i, \hat{r}_j]$.  The commutator with $\Lambda$ then follows from writing $\vec{L} = \vec{\Lambda} - qBR^2\hat{r}$.}  In fact, we have
\begin{equation}
H = \frac{\vec{L}^2 - q^2 B^2 R^4}{2mR^2} = \frac{\hbar^2\ell(\ell + 1) - q^2 B^2 R^4}{2mR^2},
\end{equation}
where the levels $\ell$ are constrained by an $L_3$ selection rule to satisfy
\begin{equation}
\ell\geq j\equiv \frac{qBR^2}{\hbar}
\end{equation}
(recall that $qB > 0$ by assumption).  Hence $\ell$ is either an integer or a half-integer according to the quantized value of $j$.  In the limit $m\to 0$, all states except for those with $\ell = j$ decouple.  Note that the gap between the LLL and the first excited level is
\begin{equation}
\frac{\hbar}{m}\left(qB + \frac{\hbar}{R^2}\right) = \hbar\omega_B\left(1 + \frac{1}{j}\right),
\end{equation}
so assuming that $q$ and $m$ are fixed, we can alternatively project to the LLL by taking $B$ large while also taking $R$ small to keep the LLL degeneracy $j$ fixed.

To obtain the explicit wavefunctions, we write
\begin{equation}
\vec{L}^2 = -\hbar^2\left[(1 - x^2)\partial_x^2 - 2x\partial_x - \frac{1}{1 - x^2}(i\partial_\varphi - j(\varepsilon + x))^2 - j^2\right], \quad H = \frac{\vec{L}^2 - \hbar^2 j^2}{2mR^2},
\end{equation}
where $x = \cos\theta$.  The properly normalized monopole spherical harmonics are given by
\begin{align}
\monopole(\theta, \varphi) = 2^m\sqrt{\frac{(2\ell + 1)(\ell - m)!(\ell + m)!}{4\pi(\ell - j)!(\ell + j)!}}&(1 - x)^{-(m + j)/2}(1 + x)^{-(m - j)/2} \nonumber \\[-2 mm]
&\times P_{\ell + m}^{(-(m + j), -(m - j))}(x)e^{i(m + j)\varphi}
\label{Yjlm}
\end{align}
for $\ell\geq j$ and $-\ell\leq m\leq \ell$, where the Jacobi polynomials are defined as
\begin{equation}
P_n^{(\alpha, \beta)}(x) = \frac{(-1)^n}{2^n n!}(1 - x)^{-\alpha}(1 + x)^{-\beta}\frac{d^n}{dx^n}((1 - x)^{n + \alpha}(1 + x)^{n + \beta}).
\end{equation}
They satisfy
\begin{equation}
-\left[(1 - x^2)\partial_x^2 - 2x\partial_x - \frac{1}{1 - x^2}(i\partial_\varphi + j(1 - x))^2 - j^2\right]\monopole(\theta, \varphi) = \ell(\ell + 1)\monopole(\theta, \varphi).
\end{equation}
For $j = 0$, the monopole spherical harmonics reduce to ordinary spherical harmonics:
\begin{equation}
\monopoleY{0}{\ell}{m}(\theta, \varphi) = Y^\ell_m(\theta, \varphi) = \sqrt{\frac{(2\ell + 1)(\ell - m)!}{4\pi(\ell + m)!}}P_\ell^m(x)e^{im\varphi}
\label{sphericalharmonics}
\end{equation}
for $\ell\geq 0$ and $-\ell\leq m\leq \ell$, where the associated Legendre polynomials are given by
\begin{equation}
P_\ell^m(x) = \frac{(-1)^{\ell + m}}{2^\ell\ell!}(1 - x^2)^{m/2}\frac{d^{\ell + m}}{dx^{\ell + m}}(1 - x^2)^\ell.
\end{equation}
Using $P_n^{(\beta, \alpha)}(-x) = (-1)^n P_n^{(\alpha, \beta)}(x)$, we obtain
\begin{equation}
\monopoleY{-j}{\ell}{m}(\pi - \theta, \varphi) = (-1)^{\ell + m}e^{-2ij\varphi}\monopole(\theta, \varphi)
\end{equation}
for $\ell\geq |j|$ and $-\ell\leq m\leq \ell$ (note that $\ell + m\in \mathbb{Z}$).  Hence the simultaneous eigenfunctions of $\vec{L}^2$ and $L_3 = -i\hbar\partial_\varphi + \varepsilon\hbar j$ are
\begin{equation}
\langle\theta, \varphi|\smqty*{\ell \\ m}\rangle\equiv \monopole(\theta, \varphi)\times \begin{cases} (-1)^{\ell + m}e^{-2ij\varphi} & \varepsilon = +1, \\ 1 & \varepsilon = -1, \end{cases}
\label{angularwavefunctions}
\end{equation}
with eigenvalues $\hbar^2\ell(\ell + 1)$ for $\vec{L}^2$ and $\hbar m$ for $L_3$.  In particular, we have
\begin{equation}
\monopoleY{j}{j}{m}(\theta, \varphi) = \frac{(-1)^{j + m}}{2^j}\sqrt{\frac{(2j + 1)(2j)!}{4\pi(j - m)!(j + m)!}}(1 - x)^{(j + m)/2}(1 + x)^{(j - m)/2}e^{i(j + m)\varphi},
\end{equation}
which gives a basis of LLL wavefunctions.  From \eqref{dee} and \eqref{ell}, we have the raising and lowering operators
\begin{equation}
L_\pm = L_1\pm iL_2 = \hbar e^{\pm i\varphi}\left[\pm\partial_\theta + i\cot\theta\partial_\varphi - \frac{j(1 + \varepsilon\cos\theta)}{\sin\theta}\right].
\end{equation}
For properly normalized wavefunctions, it follows from the algebra that
\begin{equation}
L_\pm\langle\theta, \varphi|\smqty*{\ell \\ m}\rangle = \hbar\sqrt{\ell(\ell + 1) - m(m\pm 1)}\langle\theta, \varphi|\smqty*{\ell \\ m\pm 1}\rangle.
\end{equation}
The Hilbert space splits into irreps of $SU(2)$ labeled by $\ell$.  We now revisit how an $L_3$ selection rule determines the allowed levels $\ell$.  Note that $P_n^{(\alpha, \beta)}(1) = \binom{n + \alpha}{n}$, so that
\begin{equation}
\monopole(0, \varphi) = \begin{cases} 0 & m\neq -j, \\ \sqrt{\frac{2\ell + 1}{4\pi}} & m = -j. \end{cases}
\end{equation}
Note also that $P_n^{(\alpha, \beta)}(-1) = (-1)^n\binom{n + \beta}{n}$, so that
\begin{equation}
\monopole(\pi, \varphi) = \begin{cases} 0 & m\neq j, \\ (-1)^{\ell + j}\sqrt{\frac{2\ell + 1}{4\pi}}e^{2ij\varphi} & m = j. \end{cases}
\end{equation}
Hence we see explicitly that
\begin{equation}
\langle\theta = 0|\smqty*{\ell \\ m}\rangle = \sqrt{\frac{2\ell + 1}{4\pi}}\delta_{m, -j}\times \begin{cases} (-1)^{\ell - j}e^{-2ij\varphi} & \varepsilon = +1, \\ 1 & \varepsilon = -1, \end{cases}
\label{explicitzero}
\end{equation}
and in particular that
\begin{equation}
\text{$\langle\theta = 0|\smqty*{\ell \\ m}\rangle = 0$ for $m\neq -j$}.
\end{equation}
This implies that $\langle\theta, \varphi|\smqty*{\ell \\ m}\rangle = 0$ unless $\ell\geq j$.  These are the allowed levels, each occurring once.  From the fact that
\begin{equation}
\langle\theta = 0|L_3|\smqty*{\ell \\ m}\rangle = -\hbar j\langle\theta = 0|\smqty*{\ell \\ m}\rangle = \hbar m\langle\theta = 0|\smqty*{\ell \\ m}\rangle,
\end{equation}
regardless of gauge (as we check using \eqref{explicitzero}), we deduce that
\begin{equation}
L_3|\theta = 0\rangle = -\hbar j|\theta = 0\rangle.
\end{equation}
This $L_3$ matrix element is the origin of the selection rule.  Note that the overlap $\langle\theta = 0|\smqty*{\ell \\ m}\rangle$ (and hence the state $|\theta = 0\rangle$) can have $\varphi$-dependence, depending on the gauge.

It is useful to have some alternative expressions for the $\monopole$.  Using Euler angles in the $z$-$y$-$z$ convention ($\alpha$ and $\gamma$ range from 0 to $2\pi$, while $\beta$ ranges from 0 to $\pi$), we define the Wigner $D$-matrices and $d$-matrices by
\begin{align}
D^\ell_{mn}(\alpha, \beta, \gamma) &= \sbra{\ell}{m}e^{-i\alpha L_3/\hbar}e^{-i\beta L_2/\hbar}e^{-i\gamma L_3/\hbar}\sket{\ell}{n} = e^{-i\alpha m}d^\ell_{mn}(\beta)e^{-i\gamma n}, \label{wignerD} \\
d^\ell_{mn}(\beta) &= \sbra{\ell}{m}e^{-i\beta L_2/\hbar}\sket{\ell}{n} = D^\ell_{mn}(0, \beta, 0).
\end{align}
In the $z$-$y$-$z$ convention, the $d^\ell_{mn}$ are real, so
\begin{equation}
d^\ell_{mn}(-\theta) = d^\ell_{nm}(\theta)^\ast = d^\ell_{nm}(\theta).
\end{equation}
Therefore, writing $|\theta, \varphi\rangle = e^{-iL_3\varphi/\hbar}e^{-iL_2\theta/\hbar}|\theta = 0\rangle$, we have
\begin{equation}
\langle\theta, \varphi|\smqty*{\ell \\ m}\rangle = \sum_n e^{in\varphi}d^\ell_{mn}(\theta)\langle\theta = 0|\smqty*{\ell \\ n}\rangle.
\end{equation}
For nonnegative $m - n$ and $m + n$ (i.e., $m\geq |n|$), we have the relation \cite{Edmonds}
\begin{equation}
d_{mn}^\ell(\beta) = (-1)^{m - n}\sqrt{\frac{(\ell - m)!(\ell + m)!}{(\ell - n)!(\ell + n)!}}\left(\cos\frac{\beta}{2}\right)^{m + n}\left(\sin\frac{\beta}{2}\right)^{m - n}P_{\ell - m}^{(m - n, m + n)}(\cos\beta).
\label{wignerd}
\end{equation}
We also have the symmetry relations
\begin{equation}
d^\ell_{mn}(\beta) = (-1)^{m - n}d^\ell_{nm}(\beta) = (-1)^{m - n}d^\ell_{-m, -n}(\beta).
\label{drelations}
\end{equation}
On the other hand, the Jacobi polynomials satisfy \cite{biedenharn1981angular}
\begin{align}
P_n^{(\alpha, \beta)}(x) &= \frac{(n + \alpha)!(n + \beta)!}{n!(n + \alpha + \beta)!}\left(\frac{x + 1}{2}\right)^{-\beta}P_{n + \beta}^{(\alpha, -\beta)}(x), \\
P_n^{(\alpha, \beta)}(x) &= \frac{(n + \alpha)!(n + \beta)!}{n!(n + \alpha + \beta)!}\left(\frac{x - 1}{2}\right)^{-\alpha}P_{n + \alpha}^{(-\alpha, \beta)}(x), \\
P_n^{(\alpha, \beta)}(x) &= \left(\frac{x - 1}{2}\right)^{-\alpha}\left(\frac{x + 1}{2}\right)^{-\beta}P_{n + \alpha + \beta}^{(-\alpha, -\beta)}(x).
\end{align}
The latter two of these relations give
\begin{align*}
P_{\ell - m}^{(m - n, m + n)}(\cos\beta) &= (-1)^{m - n}\frac{(\ell - n)!(\ell + n)!}{(\ell - m)!(\ell + m)!}\left(\sin\frac{\beta}{2}\right)^{-2(m - n)}P_{\ell - n}^{(-(m - n), m + n)}(\cos\beta), \\
P_{\ell - m}^{(m - n, m + n)}(\cos\beta) &= (-1)^{m - n}\left(\sin\frac{\beta}{2}\right)^{-2(m - n)}\left(\cos\frac{\beta}{2}\right)^{-2(m + n)}P_{\ell + m}^{(-(m - n), -(m + n))}(\cos\beta),
\end{align*}
respectively, from which we see that the right side of \eqref{wignerd} is consistent with the relations \eqref{drelations} for swapping $m, n$ and negating $m, n$.  Hence the identification \eqref{wignerd} in fact holds for all $-\ell\leq m, n\leq \ell$.  Comparing \eqref{Yjlm} and \eqref{wignerd} shows that
\begin{equation}
\monopole(\theta, \varphi) = \sqrt{\frac{2\ell + 1}{4\pi}}e^{i(m + j)\varphi}d_{j, -m}^\ell(\theta) = \sqrt{\frac{2\ell + 1}{4\pi}}D^\ell_{j, -m}(-\varphi, \theta, \varphi).
\end{equation}
Using $d^\ell_{mn}(\pi - \beta) = (-1)^{\ell + m}d^\ell_{m, -n}(\beta)$, this is equivalent to
\begin{equation}
\monopole(\theta, \varphi) = (-1)^{\ell + j}\sqrt{\frac{2\ell + 1}{4\pi}}e^{i(m + j)\varphi}d_{jm}^\ell(\pi - \theta).
\end{equation}
In the $\varepsilon = -1$ gauge, this is simply $\langle\theta, \varphi|\smqty*{\ell \\ m}\rangle$ (via \eqref{angularwavefunctions}).

Finally, how does the LLL Hilbert space arise from the quantization of a classical phase space?  The classical Lagrangian and Hamiltonian of the spherical Landau problem take the form \eqref{classlagham}:
\begin{align}
L &= \frac{1}{2}mR^2(\dot{\theta}^2 + \sin^2\theta\dot{\varphi}^2) - \hbar j(\varepsilon + \cos\theta)\dot{\varphi}, \label{classicalL} \\
H &= \frac{1}{2mR^2}\left[\left(\frac{\pi_\varphi + \hbar j(\varepsilon + \cos\theta)}{\sin\theta}\right)^2 + \pi_\theta^2\right],
\end{align}
where we have defined the canonical momenta
\begin{equation}
\pi_\varphi = mR^2\sin^2\theta\dot{\varphi} - \hbar j(\varepsilon + \cos\theta), \qquad \pi_\theta = mR^2\dot{\theta}.
\end{equation}
The classical versions of the standard and guiding center angular momenta are
\begin{align}
\vec{D}_\text{cl} &= (-\sin\varphi\pi_\theta - \cos\varphi\cot\theta\pi_\varphi, \cos\varphi\pi_\theta - \sin\varphi\cot\theta\pi_\varphi, \pi_\varphi), \\
\vec{L}_\text{cl} &= \vec{D}_\text{cl} - \varepsilon\hbar j\left(\frac{(\varepsilon + \cos\theta)\cos\varphi}{\sin\theta}, \frac{(\varepsilon + \cos\theta)\sin\varphi}{\sin\theta}, -1\right).
\end{align}
For finite $m$, the phase space is $(2 + 2)$-dimensional, and we obtain the Poisson brackets
\begin{equation}
[(D_i)_\text{cl}, (D_j)_\text{cl}]_\text{PB} = \epsilon_{ijk}(D^k)_\text{cl}, \qquad [(L_i)_\text{cl}, (L_j)_\text{cl}]_\text{PB} = \epsilon_{ijk}(L^k)_\text{cl}
\end{equation}
with respect to $(\theta, \pi_\theta, \varphi, \pi_\varphi)$.  For $m = 0$, the phase space is $(1 + 1)$-dimensional, and there exists a distinguished polarization in which $\varphi$ is the canonical coordinate.  We have
\begin{equation}
\pi_\varphi = -\hbar j(\varepsilon + \cos\theta), \qquad \pi_\theta = 0,
\end{equation}
and the classical guiding center angular momenta reduce to
\begin{align}
\vec{L}_\text{cl} &= -\hbar j(\cos\varphi\sin\theta, \sin\varphi\sin\theta, \cos\theta) \\
&= \textstyle \left(-\cos\varphi\sqrt{(\hbar j)^2 - (L_3)_\text{cl}^2}, -\sin\varphi\sqrt{(\hbar j)^2 - (L_3)_\text{cl}^2}, \pi_\varphi + \varepsilon\hbar j\right).
\end{align}
The Poisson brackets on the reduced phase space $(\varphi, \pi_\varphi)$ take the same form as for $m > 0$.  The corresponding quantum operators satisfying $[L_3, L_\pm] = \pm\hbar L_\pm$ and $[L_+, L_-] = 2\hbar L_3$ are
\begin{equation}
L_\pm = -\sqrt{\hbar j\pm L_3}e^{\pm i\varphi}\sqrt{\hbar j\mp L_3}, \qquad L_3 = -i\hbar\partial_\varphi + \varepsilon\hbar j,
\end{equation}
where $L_\pm\equiv L_1\pm iL_2$.  We have $\vec{L}^2 = \hbar^2 j(j + 1)$.  The normalized wavefunctions
\begin{equation}
\langle\varphi|\smqty*{j \\ m}\rangle = \frac{1}{\sqrt{2\pi}}(-1)^m e^{i(m - \varepsilon j)\varphi}
\end{equation}
are eigenfunctions of $L_3$ and satisfy
\begin{equation}
L_\pm\langle\varphi|\smqty*{j \\ m}\rangle = \hbar\sqrt{j(j + 1) - m(m\pm 1)}\langle\varphi|\smqty*{j \\ m\pm 1}\rangle.
\end{equation}
(For a planar rotor, by contrast, the wavefunctions are $\frac{1}{\sqrt{2\pi}}e^{im\varphi}$ for $m\in \mathbb{Z}$.) Note that these wavefunctions are $2\pi$-periodic in $\varphi$.

In summary, setting $m = 0$ in the Lagrangian \eqref{classicalL} results in the action $q\int \mathbf{A}_\text{real}$ with $\mathbf{A}_\text{real}$ as in \eqref{Areal}.  The Dirac quantization condition for a closed trajectory in the compact phase space $S^2$ then implies that $e^{iq\int_{S^2} \mathbf{B}/\hbar} = 1$, which is equivalent to $j\in \frac{1}{2}\mathbb{Z}$.  Quantizing this compact phase space yields a finite-dimensional Hilbert space with $2j + 1$ states $\sket{j}{m}$.

\section{Spin Coherent States} \label{spincoherentstates}

\subsection{Basic Properties}

Given a spin-$j$ multiplet $\{\sket{j}{m}\}$, one conventional definition of \emph{unnormalized} spin coherent states \cite{Radcliffe_1971, Arecchi:1972td} is the following:
\begin{equation}
|z\rangle\equiv e^{zL_-/\hbar}\sket{j}{j} = \sum_{m=-j}^j z^{j-m}\sqrt{\frac{(2j)!}{(j + m)!(j - m)!}}\sket{j}{m},
\label{spincoherent}
\end{equation}
where $z\in \mathbb{C}$.  The BCH formula $e^Y Xe^{-Y} = X + [Y, X] + \frac{1}{2!}[Y, [Y, X]] + \cdots$ implies that
\begin{equation}
e^{-zL_-/\hbar}(L_3, L_-, L_+)e^{zL_-/\hbar} = (L_3 - zL_-, L_-, L_+ + 2zL_3 - z^2 L_-).
\end{equation}
Consequently, we find that
\begin{equation}
\frac{\bar{z}L_+ + zL_- + (1 - |z|^2)L_3}{1 + |z|^2}|z\rangle = \hbar j|z\rangle.
\label{zeigenvalueequation}
\end{equation}
For an $S^2$ of unit diameter, stereographic projection from the south pole yields the map
\begin{equation}
z = \tan(\theta/2)e^{i\varphi}
\label{unitstereographic}
\end{equation}
between spherical coordinates $(\theta, \varphi)$ and complex coordinates $(z, \bar{z})$.  Hence \eqref{zeigenvalueequation} becomes
\begin{equation}
(\vec{L}\cdot \vec{n})|z\rangle = \hbar j|z\rangle,
\end{equation}
where $\vec{L} = (L_1, L_2, L_3)$ and $\vec{n} = (\sin\theta\cos\varphi, \sin\theta\sin\varphi, \cos\theta)$ is the unit vector in the $(\theta, \varphi)$ direction determined by $z$. (Note that for spin coherent states obtained by rotating $|j, -j\rangle$, the corresponding map would require stereographic projection through the \emph{north} pole.) From \eqref{spincoherent}, these states have the overlaps
\begin{equation}
\langle z_1|z_2\rangle = (1 + \bar{z}_1 z_2)^{2j}.
\label{overlaps}
\end{equation}
The resolution of the identity takes the form
\begin{equation}
\frac{2j + 1}{\pi}\int \frac{d^2 z}{(1 + |z|^2)^{2j + 2}}|z\rangle\langle z| = \sum_{m=-j}^j \sket{j}{m}\sbra{j}{m} = 1.
\end{equation}
These states are overcomplete for the spin-$j$ Hilbert space.

The geometrical meaning of these states becomes more apparent if we define their normalized counterparts
\begin{equation}
|\Omega\rangle_j\equiv |\theta, \varphi\rangle_j\equiv |\vec{n}\rangle_j\equiv \frac{|z\rangle}{(1 + |z|^2)^j}.
\end{equation}
Then we can write
\begin{equation}
|{}_j\langle\vec{n}_1|\vec{n}_2\rangle_j| = \left[\frac{|1 + \bar{z}_1 z_2|^2}{(1 + |z_1|^2)(1 + |z_2|^2)}\right]^j = \left(\frac{1 + \vec{n}_1\cdot \vec{n}_2}{2}\right)^j.
\end{equation}
These states are not orthonormal, but they approach orthonormality (become more sharply peaked) as $j\to\infty$.  Using these states, we can construct ``approximate'' ACP codewords \cite{PhysRevX.10.031050} for finite $j$.  In terms of the normalized states
\begin{equation}
|\Omega\rangle_j = \sum_{m=-j}^j y^j_m(\theta, \varphi)^\ast\sket{j}{m}
\end{equation}
where
\begin{equation}
y^j_m(\theta, \varphi)^\ast\equiv \sqrt{\frac{(2j)!}{(j + m)!(j - m)!}}\cos^{j+m}(\theta/2)\sin^{j-m}(\theta/2)e^{i(j - m)\varphi},
\end{equation}
the inner product becomes
\begin{equation}
{}_j\langle\Omega'|\Omega\rangle_j = \left(\cos\frac{\theta}{2}\cos\frac{\theta'}{2} + e^{i(\varphi - \varphi')}\sin\frac{\theta}{2}\sin\frac{\theta'}{2}\right)^{2j},
\label{innerproduct}
\end{equation}
and the completeness relation becomes
\begin{equation}
\frac{2j + 1}{4\pi}\int d\Omega\, |\Omega\rangle_j\ds{}_j\langle\Omega| = 1, \qquad d\Omega\equiv \sin\theta\, d\theta\wedge d\varphi.
\end{equation}
To invert the expansion of $|\Omega\rangle_j$ in terms of $\sket{j}{m}$, we use
\begin{equation}
\frac{2j + 1}{4\pi}\int d\Omega\, y^j_m(\theta, \varphi)^\ast y^j_{m'}(\theta, \varphi) = \delta_{mm'}
\label{ycompleteness}
\end{equation}
and therefore
\begin{equation}
\sket{j}{m} = \frac{2j + 1}{4\pi}\int d\Omega\, y^j_m(\theta, \varphi)|\Omega\rangle_j.
\end{equation}
The completeness relation \eqref{ycompleteness} follows from the beta function integral
\begin{equation}
\int_0^\pi d\theta\cos^a(\theta/2)\sin^b(\theta/2) = \frac{\Gamma(\frac{a + 1}{2})\Gamma(\frac{b + 1}{2})}{\Gamma(\frac{a + b + 2}{2})}.
\label{betaintegral}
\end{equation}
The $y^j_m$ are essentially the ``lowest'' monopole spherical harmonics $\monopole$ for $\ell = j$:
\begin{equation}
\monopoleY{j}{j}{-m}(\theta, \varphi) = (-1)^{j - m}\sqrt{\frac{2j + 1}{4\pi}}y^j_m(\theta, \varphi)^\ast.
\end{equation}
In terms of Wigner $D$-matrices, we have simply
\begin{equation}
y^j_m(\theta, \varphi)^\ast = (-1)^{j - m}D^j_{jm}(-\varphi, \theta, \varphi) = D^j_{mj}(\varphi, \theta, -\varphi),
\end{equation}
where the second equality follows from the symmetry property
\begin{equation}
D^\ell_{nm}(\gamma, \beta, \alpha) = (-1)^{m-n}D^\ell_{mn}(\alpha, \beta, \gamma).
\end{equation}
Therefore, we have
\begin{equation}
|\Omega\rangle_j = \sum_{m=-j}^j D^j_{mj}(\varphi, \theta, -\varphi)\sket{j}{m}.
\end{equation}
So we see that spin coherent states are simply rotated versions of $\sket{j}{j}$:
\begin{equation}
|\Omega\rangle_j = X_{R(\varphi, \theta, -\varphi)}\sket{j}{j} = e^{-i\varphi L_3/\hbar}e^{-i\theta L_2/\hbar}e^{i\varphi L_3/\hbar}\sket{j}{j} = e^{ij\varphi}e^{-i\varphi L_3/\hbar}e^{-i\theta L_2/\hbar}\sket{j}{j}.
\label{spincoherentrotated}
\end{equation}
This conclusion also follows from the disentangling formulas of Appendix \ref{disentangling}.

Finally, given any linear operator $\hat{O}$ on the spin-$j$ Hilbert space, the functions $P(\Omega)$ and $Q(\Omega)$ defined by
\begin{equation}
\hat{O} = \frac{2j + 1}{4\pi}\int d\Omega\, P(\Omega)|\Omega\rangle_j\ds{}_j\langle\Omega|, \qquad Q(\Omega) = {}_j\langle\Omega|\hat{O}|\Omega\rangle_j
\label{symbols}
\end{equation}
are called the \emph{upper} and \emph{lower symbols}.  Any $\hat{O}$ has an expansion of the form on the left of \eqref{symbols} \cite{Perelomovbook}.  For example, in the context of Section \ref{errorbasis}, one could define the would-be ``momentum kick'' operators
\begin{equation}
\hat{y}^j_m\equiv \frac{2j + 1}{4\pi}\int d\Omega\, y^j_m(\Omega)|\Omega\rangle_j\ds{}_j\langle\Omega|.
\label{momkick}
\end{equation}
Unlike the momentum kick operators of ACP \cite{PhysRevX.10.031050}, these act \emph{within} the LLL Hilbert space rather than taking states up to higher Landau levels.

\subsection{Matrix Elements} \label{matrixelements}

It is convenient to have an expression for the amplitude ${}_j\langle\Omega'|X_R|\Omega\rangle_j$ \cite{Radcliffe_1971}.  Given a rotation $R = R(\alpha, \beta, \gamma)$, we compute that
\begin{align}
{}_j\langle\Omega'|X_R|\Omega\rangle_j &= \textstyle \sum_{m', m = -j}^j y^j_{m'}(\theta', \varphi')\sbra{j}{\smash{m'}}X_R\sket{j}{m}y^j_m(\theta, \varphi)^\ast \\
&= \textstyle \sum_{m', m = -j}^j D^j_{jm'}(\varphi', -\theta', -\varphi')D^j_{m'm}(\alpha, \beta, \gamma)D^j_{mj}(\varphi, \theta, -\varphi) \\
&= D^j_{jj}((\varphi', -\theta', -\varphi')\circ (\alpha, \beta, \gamma)\circ (\varphi, \theta, -\varphi)),
\end{align}
where we have used
\begin{equation}
D^\ell_{mn}(\alpha, \beta, \gamma)^\ast = D^\ell_{nm}(-\gamma, -\beta, -\alpha) \implies y^j_m(\theta, \varphi) = D^j_{jm}(\varphi, -\theta, -\varphi).
\end{equation}
Writing $R(\alpha', \beta', \gamma') = R(\varphi', -\theta', -\varphi')R(\alpha, \beta, \gamma)R(\varphi, \theta, -\varphi)$, we have
\begin{equation}
D^j_{jj}(\alpha', \beta', \gamma') = e^{-ij(\alpha' + \gamma')}d^j_{jj}(\beta') = e^{-ij(\alpha' + \gamma')}\left(\frac{1 + \cos\beta'}{2}\right)^j.
\end{equation}
To obtain a more explicit formula, we can write
\begin{align}
{}_j\langle\Omega'|X_R|\Omega\rangle_j &= \sum_{m', m = -j}^j y^j_{m'}(\theta', \varphi')D^j_{m'm}(\alpha, \beta, \gamma)y^j_m(\theta, \varphi)^\ast \\
&= (2j)!\sum_{m', m = -j}^j \frac{\cos^{j+m'}(\theta'/2)\sin^{j-m'}(\theta'/2)\cos^{j+m}(\theta/2)\sin^{j-m}(\theta/2)}{\sqrt{(j + m')!(j - m')!(j + m)!(j - m)!}} \nonumber \\
&\hspace{3 cm} \times e^{-i(j - m')\varphi'}e^{i(j - m)\varphi}e^{-i\alpha m'}e^{-i\gamma m}d^j_{m'm}(\beta),
\end{align}
where $D^\ell_{mn}(\alpha, \beta, \gamma) = e^{-i\alpha m}d^\ell_{mn}(\beta)e^{-i\gamma n}$.  We then use the following expansion of Wigner $d$-matrix elements (which follows from an identity for Jacobi polynomials):
\begin{align}
d^\ell_{mn}(\beta) &= (-1)^{m - n}\sqrt{(\ell + m)!(\ell - m)!(\ell + n)!(\ell - n)!} \nonumber \\
&\phantom{==} \times \sum_{s = s_\text{min}}^{s_\text{max}} \frac{(-1)^s\left(\cos\frac{\beta}{2}\right)^{2\ell + n - m - 2s}\left(\sin\frac{\beta}{2}\right)^{m - n + 2s}}{s!(\ell + n - s)!(m - n + s)!(\ell - m - s)!},
\end{align}
where the sum extends over all integer values of $s$ for which the arguments of the factorials are nonnegative: $s_\text{min} = \operatorname{max}(0, n - m)$ and $s_\text{max} = \operatorname{min}(\ell + n, \ell - m)$.  This gives
\begin{align}
&{}_j\langle\Omega'|X_R|\Omega\rangle_j = (2j)!\sum_{m', m = -j}^j (-1)^{m' - m}\cos^{j+m'}(\theta'/2)\sin^{j-m'}(\theta'/2)\cos^{j+m}(\theta/2)\sin^{j-m}(\theta/2) \nonumber \\
&\times e^{-i(j - m')\varphi'}e^{i(j - m)\varphi}e^{-i\alpha m'}e^{-i\gamma m}\sum_{s = \operatorname{max}(0, m - m')}^{\operatorname{min}(j + m, j - m')} \frac{(-1)^s\left(\cos\frac{\beta}{2}\right)^{2j + m - m' - 2s}\left(\sin\frac{\beta}{2}\right)^{m' - m + 2s}}{s!(j + m - s)!(m' - m + s)!(j - m' - s)!}.
\end{align}
After some rearranging, we get
\begin{align}
{}_j\langle\Omega'|X_R|\Omega\rangle_j &= (2j)!\left(\cos\frac{\theta'}{2}\cos\frac{\theta}{2}\cos\frac{\beta}{2}\right)^{2j}e^{-ij(\alpha + \gamma)}\sum_{m', m = -j}^j\sum_{s = \operatorname{max}(0, m - m')}^{\operatorname{min}(j + m, j - m')} \nonumber \\
&\frac{(-1)^s\left(-e^{i(\alpha - \varphi')}\tan\frac{\theta'}{2}\right)^{j-m'}\left(-e^{i(\varphi + \gamma)}\tan\frac{\theta}{2}\right)^{j-m}\left(\tan\frac{\beta}{2}\right)^{m' - m + 2s}}{s!(j + m - s)!(m' - m + s)!(j - m' - s)!}.
\end{align}
Letting $x\equiv -e^{i(\alpha - \varphi')}\tan\frac{\theta'}{2}$, $y\equiv -e^{i(\varphi + \gamma)}\tan\frac{\theta}{2}$, and $z\equiv \tan\frac{\beta}{2}$, we can write this as
\begin{equation}
{}_j\langle\Omega'|X_R|\Omega\rangle_j = \left(\cos\frac{\theta'}{2}\cos\frac{\theta}{2}\cos\frac{\beta}{2}\right)^{2j}e^{-ij(\alpha + \gamma)}S(x, y, z)
\end{equation}
where
\begin{equation}
S(x, y, z)\equiv (2j)!\sum_{m', m = -j}^j\sum_{s = \operatorname{max}(0, m - m')}^{\operatorname{min}(j + m, j - m')} \frac{(-1)^s x^{j-m'}y^{j-m}z^{m' - m + 2s}}{s!(j + m - s)!(m' - m + s)!(j - m' - s)!}.
\end{equation}
To evaluate the sum, we first switch the order of summation by writing the inequalities that ensure that the arguments of the factorials are nonnegative as
\begin{equation}
j + m\geq s\geq 0, \qquad j - s\geq m'\geq m - s.
\end{equation}
Then, by successive applications of the binomial theorem, we obtain:
\begin{align}
S(x, y, z) &= (2j)!\sum_{m = -j}^j\sum_{s = 0}^{j + m}\sum_{m' = m - s}^{j - s} \frac{(-1)^s x^{j-m'}y^{j-m}z^{m' - m + 2s}}{s!(j + m - s)!(m' - m + s)!(j - m' - s)!} \\
&= (2j)!\sum_{m = -j}^j\sum_{s = 0}^{j + m} \frac{(-xz)^s(y(x + z))^{j - m}}{s!(j + m - s)!(j - m)!} \\
&= (2j)!\sum_{m = -j}^j \frac{(1 - xz)^{j + m}(y(x + z))^{j - m}}{(j + m)!(j - m)!} \\
&= (1 + xy + yz - xz)^{2j}.
\end{align}
For instance, to evaluate the innermost sum, we used
\begin{equation}
\sum_{m' = m - s}^{j - s} \frac{x^{j-m'}z^{m' - m + 2s}}{(m' - m + s)!(j - m' - s)!} = \frac{(xz)^s(x + z)^{j - m}}{(j - m)!}.
\end{equation}
Altogether, we find that
\begin{align}
{}_j\langle\Omega'|X_R|\Omega\rangle_j &= e^{-ij(\alpha + \gamma)}\bigg[\left(\cos\frac{\theta'}{2}\cos\frac{\theta}{2} + e^{i(\alpha + \gamma)}e^{i(\varphi - \varphi')}\sin\frac{\theta'}{2}\sin\frac{\theta}{2}\right)\cos\frac{\beta}{2} \nonumber \\
&\hspace{2.5 cm} - \left(e^{i(\gamma + \varphi)}\cos\frac{\theta'}{2}\sin\frac{\theta}{2} - e^{i(\alpha - \varphi')}\sin\frac{\theta'}{2}\cos\frac{\theta}{2}\right)\sin\frac{\beta}{2}\bigg]^{2j}. \label{amplitude}
\end{align}
This expression is equivalent to \eqref{theamplitude}.  It reduces to the inner product \eqref{innerproduct} when $\alpha = \beta = \gamma = 0$.  The above logic follows that of \cite{Radcliffe_1971}, but differs in some details.

We can check explicitly that
\begin{equation}
|{}_j\langle\Omega'|X_R|\Omega\rangle_j| = |{}_j\langle\vec{n}'|R\vec{n}\rangle_j| = \left(\frac{1 + \vec{n}'\cdot R\vec{n}}{2}\right)^j,
\end{equation}
where, as a rotation acting on vectors in $\mathbb{R}^3$,
\begin{equation}
R(\alpha, \beta, \gamma) = \left(\begin{array}{ccc}
\cos\alpha\cos\beta\cos\gamma - \sin\alpha\sin\gamma & -\cos\gamma\sin\alpha - \cos\alpha\cos\beta\sin\gamma & \cos\alpha\sin\beta \\
\cos\alpha\sin\gamma + \cos\beta\cos\gamma\sin\alpha & \cos\alpha\cos\gamma - \cos\beta\sin\alpha\sin\gamma & \sin\alpha\sin\beta \\
-\cos\gamma\sin\beta & \sin\beta\sin\gamma & \cos\beta
\end{array}\right).
\end{equation}
Hence ${}_j\langle\Omega'|X_R|\Omega\rangle_j$ coincides with ${}_j\langle\vec{n}'|R\vec{n}\rangle_j$ up to a phase.  This phase becomes important when considering superpositions of coherent states or codes constructed therefrom.

\subsection{Disentangling Formulas} \label{disentangling}

The connection between the definition \eqref{spincoherent} of spin coherent states via lowering operators and the expression \eqref{spincoherentrotated} in terms of rotations can be seen as follows.  A standard trick is that if we use a faithful representation of the Lie algebra to derive a group identity, i.e., a relation between exponentials (as opposed to an identity in the universal enveloping algebra), then it is valid irrespective of representation.  This is essentially the content of the BCH-like ``disentangling theorem'' of \cite{Arecchi:1972td}.  For $SU(2)$, the simplest representation is the spin-$1/2$ representation in terms of Pauli matrices: $\vec{L} = \frac{\hbar}{2}\vec{\sigma}$.  In our case,
\begin{align}
X_{R(\varphi, \theta, -\varphi)} &= e^{-i\varphi L_3/\hbar}e^{-i\theta L_2/\hbar}e^{i\varphi L_3/\hbar} \\
&\leftrightarrow e^{-i\varphi\sigma_3/2}e^{-i\theta\sigma_2/2}e^{i\varphi\sigma_3/2} = \left(\begin{array}{cc} \cos\frac{\theta}{2} & -e^{-i\varphi}\sin\frac{\theta}{2} \\ e^{i\varphi}\sin\frac{\theta}{2} & \cos\frac{\theta}{2} \end{array}\right),
\end{align}
where ``$\leftrightarrow$'' indicates ``passing to the spin-$1/2$ representation.''  On the other hand, we have
\begin{align}
e^{x_- L_-/\hbar}e^{x_3 L_3/\hbar}e^{x_+ L_+/\hbar} &\leftrightarrow e^{x_-(\sigma_1 - i\sigma_2)/2}e^{x_3\sigma_3/2}e^{x_+(\sigma_1 + i\sigma_2)/2} \\
&= \left(\begin{array}{cc} e^{x_3/2} & x_+ e^{x_3/2} \\ x_- e^{x_3/2} & e^{-x_3/2} + x_- x_+ e^{x_3/2} \end{array}\right).
\end{align}
Choosing
\begin{equation}
e^{x_3/2} = \cos\frac{\theta}{2}, \qquad x_- = e^{i\varphi}\tan\frac{\theta}{2}, \qquad x_+ = -e^{-i\varphi}\tan\frac{\theta}{2}
\end{equation}
shows that
\begin{equation}
X_{R(\varphi, \theta, -\varphi)} = e^{zL_-/\hbar}\left(\cos^{2L_3/\hbar}\frac{\theta}{2}\right)e^{-\bar{z}L_+/\hbar},
\label{rotationformula}
\end{equation}
with $z$ as in \eqref{unitstereographic}.  This gives
\begin{equation}
\frac{|z\rangle}{(1 + |z|^2)^j} = \left(\cos^{2j}\frac{\theta}{2}\right)e^{zL_-/\hbar}\sket{j}{j} = X_{R(\varphi, \theta, -\varphi)}\sket{j}{j} = |\Omega\rangle_j,
\end{equation}
as claimed. (A canonical rotation has many alternative presentations, such as $X_{R(\varphi, \theta, -\varphi)} = e^{i\theta(\sin\varphi L_1 - \cos\varphi L_2)/\hbar}$.)

As another application, these disentangling formulas can be used to determine the set of correctable rotations for the $d = 2$ case of the equatorial qudit code (see \eqref{d2qudit1} and \eqref{d2qudit2}).  To see that a $z$-rotation conjugated by an $x$-rotation has Euler angles $\alpha = \gamma$, we note that
\begin{align}
e^{-i\alpha_1 L_1/\hbar}e^{-i\alpha_3 L_3/\hbar}e^{i\alpha_1 L_1/\hbar} &\leftrightarrow e^{-i\alpha_1\sigma_1/2}e^{-i\alpha_3\sigma_3/2}e^{i\alpha_1\sigma_1/2} \\
&= \left(\begin{array}{cc} \cos\frac{\alpha_3}{2} - i\cos\alpha_1\sin\frac{\alpha_3}{2} & \sin\alpha_1\sin\frac{\alpha_3}{2} \\ -\sin\alpha_1\sin\frac{\alpha_3}{2} & \cos\frac{\alpha_3}{2} + i\cos\alpha_1\sin\frac{\alpha_3}{2} \end{array}\right),
\end{align}
and to write this in the form of a generic rotation
\begin{align}
e^{-i\alpha L_3/\hbar}e^{-i\beta L_2/\hbar}e^{-i\gamma L_3/\hbar} &\leftrightarrow e^{-i\alpha\sigma_3/2}e^{-i\beta\sigma_2/2}e^{-i\gamma\sigma_3/2} \\
&= \left(\begin{array}{cc} e^{-i(\alpha + \gamma)/2}\cos\frac{\beta}{2} & -e^{-i(\alpha - \gamma)/2}\sin\frac{\beta}{2} \\ e^{i(\alpha - \gamma)/2}\sin\frac{\beta}{2} & e^{i(\alpha + \gamma)/2}\cos\frac{\beta}{2} \end{array}\right)
\end{align}
requires taking $\alpha = \gamma$.

\section{Approximate Quantum Error Correction} \label{approxQEC}

\subsection{Knill-Laflamme Conditions} \label{KLconditions}

Let us revisit the quantum error correction conditions for a qubit code, which is a choice of two-dimensional subspace $\mathcal{C} = \operatorname{span}(|\overline{0}\rangle, |\overline{1}\rangle)$ in a Hilbert space $\mathcal{H}$ (the discussion generalizes easily to qudits).  The effect of an error $E$ is to take the code subspace $\mathcal{C}$ to another subspace $E\mathcal{C}$.  Certain errors won't distort $\mathcal{C}$ beyond repair; these are called \emph{correctable}.

The exact quantum error correction conditions of Knill-Laflamme \cite{PhysRevA.55.900} state that for any two correctable errors $E, F$, we have
\begin{equation}
\langle\overline{1}|F^\dag E|\overline{0}\rangle = 0, \qquad \langle\overline{0}|F^\dag E|\overline{0}\rangle = \langle\overline{1}|F^\dag E|\overline{1}\rangle.
\label{errorcorrectexact}
\end{equation}
These conditions say two things:
\begin{enumerate}
\item Correctable errors act by isometries in the projective Hilbert space $\mathbb{P}\mathcal{H}$:
\begin{equation}
\langle\overline{1}|E^\dag E|\overline{0}\rangle = 0, \qquad \langle\overline{0}|E^\dag E|\overline{0}\rangle = \langle\overline{1}|E^\dag E|\overline{1}\rangle.
\label{firstcondition}
\end{equation}
In other words, correctable errors keep orthogonal vectors orthogonal and either shrink or expand $\mathcal{C}$ in a uniform way.
\item Different but indistinguishable errors act in the same way on the code subspace:
\begin{equation}
\langle\overline{1}|F^\dag E|\overline{0}\rangle = 0, \qquad \langle\overline{0}|F^\dag E|\overline{0}\rangle = \langle\overline{1}|F^\dag E|\overline{1}\rangle
\label{secondcondition}
\end{equation}
for $E\neq F$.  Note that if we further have $\langle\overline{0}|F^\dag E|\overline{0}\rangle = \langle\overline{1}|F^\dag E|\overline{1}\rangle = 0$, then $E\mathcal{C}\perp F\mathcal{C}$ for all $E\neq F$.  In this case, we can unambiguously measure the error syndrome and the code is called \emph{nondegenerate}.  Otherwise, different error spaces have nontrivial overlap and the code is called \emph{degenerate}.  In light of \eqref{firstcondition}, \eqref{secondcondition} says that the projector $P_F$ onto $F\mathcal{C}$ maps the basis $E|\overline{i}\rangle$ to $F|\overline{i}\rangle$ in $\mathbb{P}\mathcal{H}$, so that an $E$ error can be inverted regardless of whether the syndrome measurement projects onto $E\mathcal{C}$ or $F\mathcal{C}$.
\end{enumerate}
To formulate the approximate error correction conditions in a convenient way, we assume only unitary errors, so that the first condition \eqref{firstcondition} is automatically satisfied: $E^\dag E = 1$.  To enforce the second condition \eqref{secondcondition} approximately, we define the projection operator
\begin{equation}
P_F = F|\overline{0}\rangle\langle\overline{0}|F^\dag + F|\overline{1}\rangle\langle\overline{1}|F^\dag
\end{equation}
onto $F\mathcal{C}$ and demand that $P_F$ map $E|\overline{i}\rangle$ to $F|\overline{i}\rangle$ up to a constant $c$, but only approximately:
\begin{align}
P_F E|\overline{0}\rangle &= cF|\overline{0}\rangle + \epsilon_{00}F|\overline{0}\rangle + \epsilon_{01}F|\overline{1}\rangle, \label{proj1} \\
P_F E|\overline{1}\rangle &= cF|\overline{1}\rangle + \epsilon_{10}F|\overline{0}\rangle + \epsilon_{11}F|\overline{1}\rangle, \label{proj2}
\end{align}
where $c$ depends only on $E$ and $F$.  We can then invert the error approximately by applying $F^\dag$; the quality of the inversion is quantified by the fidelity (for pure states, the fidelity is simply $|\langle\psi|\psi'\rangle|^2$ where $|\psi\rangle$ is the original state and $|\psi'\rangle$ is the recovered state).  By solving \eqref{proj1}--\eqref{proj2}, we get
\begin{align}
\langle\overline{0}|F^\dag E|\overline{0}\rangle &= c + \epsilon_{00}, \\
\langle\overline{1}|F^\dag E|\overline{1}\rangle &= c + \epsilon_{11}, \\
\langle\overline{1}|F^\dag E|\overline{0}\rangle &= \epsilon_{01}, \\
\langle\overline{0}|F^\dag E|\overline{1}\rangle &= \epsilon_{10}.
\end{align}
So for approximate quantum error correction, we demand both
\begin{align}
|\langle\overline{0}|F^\dag E|\overline{0}\rangle - \langle\overline{1}|F^\dag E|\overline{1}\rangle| &< \delta, \\
|\langle\overline{1}|F^\dag E|\overline{0}\rangle| &< \epsilon,
\end{align}
for $\delta, \epsilon$ small.  More comprehensive discussions of approximate quantum error correction can be found in \cite{PhysRevA.56.2567, Schumacher2002-vr, 10.1007/11426639_17, Beny, Beny-Oreshkov, PhysRevA.81.062342, PhysRevA.86.012335}.

\subsection{Measurement with Ancillas} \label{ancillas}

GKP \cite{Gottesman:2000di} and ACP \cite{PhysRevX.10.031050} construct infinitely squeezed codewords that must be approximated by normalizable states.  Our error-correction scheme, by contrast, is inherently approximate.  We now estimate the error incurred by the finite spread of spin coherent states.  GKP and ACP compute the leakage error from the projections of their approximate codewords onto the complementary Voronoi cells.  However, we lack a sharply defined position operator, so we must take a different approach.

To measure the error syndrome for the equatorial qudit codes of Section \ref{sphericalLLL}, which is a $U(1)$ rotation modulo $\mathbb{Z}_d$, we use an ancilla system.  Let us first consider an idealized ancilla that admits an orthonormal set of position states parametrized by $U(1)/\mathbb{Z}_d$.  Measuring such an ancilla allows us to perfectly resolve the error syndrome in one shot.

We first initialize the ancilla in a $\mathbb{Z}_d$-invariant state, e.g., the uniform superposition of position eigenstates
\begin{equation}
|\overline{0}\rangle_X = \frac{1}{\sqrt{d}}\sum_{k=0}^{d-1} |\pi/2, 2\pi k/d\rangle.
\end{equation}
This is a logical-$X$ eigenstate with eigenvalue 1.  To map the syndrome onto the ancilla, we use the (approximate) controlled rotation (CROT) operator\footnote{Note that the CROT gate can also be used for initialization \cite{PhysRevX.10.031050}.}
\begin{equation}
\int d\Omega'\, (|\Omega'\rangle_j\ds{}_j\langle\Omega'|\otimes e^{-i\varphi'L_3/\hbar}),
\end{equation}
where $\Omega' = (\theta', \varphi')$.  We omit the overall normalization.  This operator is only approximately unitary because
\begin{equation}
\int d\Omega'\, d\Omega\, (|\Omega'\rangle_j\ds{}_j\langle\Omega'|\Omega\rangle_j\ds{}_j\langle\Omega|\otimes e^{-i(\varphi' - \varphi)L_3/\hbar})
\end{equation}
is not proportional to the identity (${}_j\langle\Omega'|\Omega\rangle_j$ is not a delta function).  Let $|\Omega\rangle_j = e^{-i\delta\varphi L_3/\hbar}|\overline{k}\rangle$ (with $|\overline{k}\rangle$ as in \eqref{quditcodewords}) be a noisy logical state, where an equatorial rotation acts on a general spin coherent state as
\begin{equation}
e^{-i\delta\varphi L_3/\hbar}|\theta, \varphi\rangle_j = X_{R(\varphi + \delta\varphi, \theta, -\varphi)}\sket{j}{j} = X_{R(\varphi + \delta\varphi, \theta, -\varphi - \delta\varphi)}e^{-i\delta\varphi L_3/\hbar}\sket{j}{j} = e^{-ij\delta\varphi}|\theta, \varphi + \delta\varphi\rangle_j.
\end{equation}
Then we act with the approximate CROT gate on $|\Omega\rangle_j\otimes |\overline{0}\rangle_X$, giving the state
\begin{equation}
\int d\Omega'\, {}_j\langle\Omega'|\Omega\rangle_j\left(|\Omega'\rangle_j\otimes \sum_{k=0}^{d-1} |\pi/2, 2\pi k/d + \varphi'\rangle\right),
\end{equation}
where we have dropped the overall normalization.  If the coherent states had no spread, then this would simply be a finite superposition:
\begin{equation}
|\Omega\rangle_j\otimes \sum_{k=0}^{d-1} |\pi/2, 2\pi k/d + \varphi\rangle.
\end{equation}
In our case, however, this is a continuous superposition.  Now we measure the ancilla in the position ($|\Omega\rangle$, no $j$ subscript) basis.  To determine the unnormalized probability density of measuring the state $|\pi/2, \varphi_m\rangle$, we first apply the partial projector
\begin{equation}
I\otimes |\pi/2, \varphi_m\rangle\langle\pi/2, \varphi_m|
\end{equation}
to the state, giving
\begin{equation}
\int d\Omega'\, {}_j\langle\Omega'|\Omega\rangle_j\left(|\Omega'\rangle_j\otimes \sum_{k=0}^{d-1} \delta(\varphi_m - \varphi' - 2\pi k/d)|\pi/2, \varphi_m\rangle\right) = |u\rangle\otimes |\pi/2, \varphi_m\rangle
\label{postmeasurement}
\end{equation}
where
\begin{equation}
|u\rangle\equiv \sum_{k=0}^{d-1} \int d\theta'\sin\theta'{}_j\langle\theta', \varphi_m - 2\pi k/d|\Omega\rangle_j|\theta', \varphi_m - 2\pi k/d\rangle_j.
\end{equation}
Up to normalization, \eqref{postmeasurement} is the post-measurement state.  The unnormalized probability density (with respect to $d\varphi_m$) of obtaining this state is
\begin{align}
P(\varphi_m) = |\langle u|u\rangle|^2 = \sum_{k, \ell = 0}^{d-1} \int d\theta'\sin\theta'\int d\theta''\sin\theta'' &{}_j\langle\theta', \varphi_m - 2\pi k/d|\Omega\rangle_j\ds{}_j\langle\Omega|\theta'', \varphi_m - 2\pi\ell/d\rangle_j \nonumber \\
&\times {}_j\langle\theta'', \varphi_m - 2\pi\ell/d|\theta', \varphi_m - 2\pi k/d\rangle_j.
\end{align}
As a consistency check, if there were no spread, then $P(\varphi_m)$ would indeed be nonzero only when $\varphi_m = 2\pi k/d + \varphi$ for some $k$.  Using that the state $|\Omega\rangle_j$ has $\theta = \pi/2$, the above integrals localize to $\theta' = \theta'' = \pi/2$ as $j\to\infty$ (in the saddle-point approximation).  Therefore, again dropping normalization, this becomes
\begin{align}
P(\varphi_m) &\sim \sum_{k, \ell = 0}^{d-1} \textstyle {}_j\langle\pi/2, \varphi_m - \frac{2\pi k}{d}|\Omega\rangle_j\ds{}_j\langle\Omega|\pi/2, \varphi_m - \frac{2\pi\ell}{d}\rangle_j\ds{}_j\langle\pi/2, \varphi_m - \frac{2\pi\ell}{d}|\pi/2, \varphi_m - \frac{2\pi k}{d}\rangle_j \nonumber \\
&= \sum_{k, \ell = 0}^{d-1} \left(\frac{1 + e^{-i(\varphi_m - 2\pi k/d - \varphi)}}{2}\right)^{2j}\left(\frac{1 + e^{i(\varphi_m - 2\pi\ell/d - \varphi)}}{2}\right)^{2j}\left(\frac{1 + e^{-2\pi i(k - \ell)/d}}{2}\right)^{2j}.
\end{align}
As a further approximation, we keep only the leading $k = \ell$ terms to get
\begin{equation}
P(\varphi_m)\sim \sum_{k=0}^{d-1} \left(\frac{1 + \cos(\varphi_m - 2\pi k/d - \varphi)}{2}\right)^{2j}.
\label{furtherapprox}
\end{equation}
The other terms are exponentially suppressed in $j$.  We can make a rough estimate of the error involved in resolving the error syndrome using Laplace's method.

First recall Laplace's method: consider the one-dimensional integral
\begin{equation}
I\equiv \int_a^b dx\, e^{tf(x)}g(x),
\end{equation}
where $f(x)$ has a single critical point at $x_0\in (a, b)$ with $f''(x_0) < 0$.  Taylor expanding gives
\begin{align}
I &= e^{tf(x_0)}\int_a^b dx\, e^{\frac{1}{2}tf''(x_0)(x - x_0)^2 + O(t(x - x_0)^3)}(g(x_0) + O(x - x_0)) \\
&= e^{tf(x_0)}\sqrt{\frac{2}{t|f''(x_0)|}}\int_{\sqrt{\frac{t|f''(x_0)|}{2}}(a - x_0)}^{\sqrt{\frac{t|f''(x_0)|}{2}}(b - x_0)} dy\, e^{-y^2 + O(y^3/\sqrt{t})}(g(x_0) + O(y/\sqrt{t})) \\
&\approx e^{tf(x_0)}\sqrt{\frac{2}{t|f''(x_0)|}}\int_{-\infty}^\infty dy\, e^{-y^2 + O(y^3/\sqrt{t})}(g(x_0) + O(y/\sqrt{t})) \\
&= e^{tf(x_0)}g(x_0)\sqrt{\frac{2\pi}{t|f''(x_0)|}}\left(1 + O(1/\sqrt{t})\right).
\end{align}
Specifically, say we take $a = x_0 - \epsilon$ and $b = x_0 + \epsilon$.  Then we have
\begin{equation}
I = e^{tf(x_0)}\sqrt{\frac{2}{t|f''(x_0)|}}\int_{-\sqrt{\frac{t|f''(x_0)|}{2}}\epsilon}^{\sqrt{\frac{t|f''(x_0)|}{2}}\epsilon} dy\, e^{-y^2 + O(y^3/\sqrt{t})}(g(x_0) + O(y/\sqrt{t})).
\end{equation}
The asymptotic expansion of the complementary error function is
\begin{equation}
\int_x^\infty dt\, e^{-t^2} = \frac{e^{-x^2}}{2x}\left(1 - O(1/x^2)\right).
\end{equation}
So we have
\begin{align}
\int_{\sqrt{\frac{t|f''(x_0)|}{2}}\epsilon}^\infty dy\, e^{-y^2}(1 + O(1/\sqrt{t})) &= \int_{-\infty}^{-\sqrt{\frac{t|f''(x_0)|}{2}}\epsilon} dy\, e^{-y^2}(1 + O(1/\sqrt{t})) \\
&= \frac{e^{-\frac{t|f''(x_0)|}{2}\epsilon^2}}{\sqrt{2t|f''(x_0)|}\epsilon}\left(1 - O\left(\frac{1}{t\epsilon^2}\right)\right)(1 + O(1/\sqrt{t})).
\end{align}
If we define
\begin{equation}
I_\infty\equiv e^{tf(x_0)}\sqrt{\frac{2}{t|f''(x_0)|}}\int_{-\infty}^\infty dy\, e^{-y^2 + O(y^3/\sqrt{t})}(g(x_0) + O(y/\sqrt{t})),
\end{equation}
then we have, to leading order in $1/t$,
\begin{equation}
\frac{I_\infty - I}{I_\infty}\approx \sqrt{\frac{2}{\pi t|f''(x_0)|}}\frac{e^{-\frac{t|f''(x_0)|}{2}\epsilon^2}}{\epsilon}
\end{equation}
(we assume that $\epsilon$ is constant and does not scale with $t$).  This is the error from dropping the tails of the Gaussians.  It is exponentially suppressed as $t\to\infty$.

Returning to \eqref{furtherapprox}, we focus on a single critical point $\varphi_0$:
\begin{equation}
P(\varphi_m)\sim e^{2jf(\varphi_m)}, \qquad f(\varphi_m)\equiv \log\left(\frac{1 + \cos(\varphi_m - \varphi_0)}{2}\right).
\end{equation}
We have
\begin{equation}
\int_0^{2\pi} d\varphi_m\, P(\varphi_m)\approx \sqrt{\frac{2\pi}{j}}\left(1 + O(1/\sqrt{j})\right),
\end{equation}
where the only critical point with a nontrivial contribution is $\varphi_m = \varphi_0$ with $f''(\varphi_0) = -1/2$.  Hence
\begin{equation}
\frac{\int_0^{2\pi} d\varphi_m\, P(\varphi_m) - \int_{\varphi_0 - \epsilon}^{\varphi_0 + \epsilon} d\varphi_m\, P(\varphi_m)}{\int_0^{2\pi} d\varphi_m\, P(\varphi_m)}\approx \sqrt{\frac{2}{\pi j}}\frac{e^{-j\epsilon^2/2}}{\epsilon}.
\end{equation}
This is the failure probability as $j\to\infty$, for a window of radius $\epsilon$ about $\varphi_0$.

A more realistic ancilla can itself be constructed using spin coherent states rather than position eigenstates \cite{PhysRevX.10.031050}.  We can again use the approximate CROT gate to map the error syndrome onto the spin-$j_\text{anc}$ ancilla, which is then measured in an overcomplete set of spin coherent states parametrized by $U(1)/\mathbb{Z}_d$.  This introduces an additional error in the measurement that becomes negligible as $j_\text{anc}\to\infty$.

\section{Generalizations} \label{morecodes}

In this appendix, we present three different generalizations of our LLL codes.  In Appendix \ref{allLLs}, we study ACP codes in the physical setting of a charged particle on a sphere in the presence of a magnetic monopole field, \emph{without} projecting to the LLL.  In Appendix \ref{LLLcyclic}, we formally show that the LLL analogues of the $\mathbb{Z}_N\subset \mathbb{Z}_{2N}$ cyclic subgroup codes of ACP perform optimally when $N = 1$.  In Appendix \ref{generalgroups}, we speculate on how our codes could be adapted to coherent states for arbitrary Lie groups.

\subsection{ACP Codes for the Spherical Landau Problem} \label{allLLs}

The direct Landau level analogues of the abelian linear rotor codes of ACP, for which both position and momentum errors are relevant, involve states from all Landau levels.  To construct such codes, we work in the full Hilbert space of the spherical Landau problem.  This amounts to modifying the linear rotor analysis of ACP to use monopole spherical harmonics in the place of ordinary spherical harmonics.\footnote{Given a particle on the sphere $S^2\cong SO(3)/SO(2)$, ACP implicitly branched $SO(3)\downarrow SO(2)$ as $\ell\downarrow A_0$ where $A_0$ is the trivial irrep of $SO(2)$.  Indeed, by the Peter-Weyl theorem, $L^2(SO(3))\cong \bigoplus_{\ell\geq 0} V_\ell\otimes V_\ell^\ast$ where $V_\ell$ is the spin-$\ell$ irrep of $SO(3)$, and by selecting the trivial irrep of $SO(2)$ inside each $V_\ell^\ast$, we obtain $L^2(S^2)\cong \bigoplus_{\ell\geq 0} V_\ell$.  In the presence of a magnetic monopole, we must instead branch $SO(3)\downarrow SO(2)$ using $\ell\downarrow A_j$ where $A_j$ is the irrep of $SO(2)$ given as $\smash{e^{ij\theta}}$.  This results in the Hilbert space $\bigoplus_{\ell\geq j} W_\ell$, where the $W_\ell$ are spanned by monopole spherical harmonics of spin weight $j$.}

Our $\monopole$ in \eqref{Yjlm} form a basis for $L^2(S^2)$.  With our normalization, they satisfy
\begin{equation}
\int_0^\pi \sin\theta\, d\theta\int_0^{2\pi} d\varphi\, \monopole(\theta, \varphi)^\ast\monopoleY{j}{\ell'}{m'}(\theta, \varphi) = \delta_{\ell\ell'}\delta_{mm'}.
\end{equation}
Again, we denote position eigenstates on $S^2$ by $|v\rangle$.  The continuous position basis is dual to the discrete angular momentum basis:
\begin{equation}
\sket{\ell}{m} = \int_{S^2} dv\, \monopole(v)|v\rangle, \qquad |v\rangle = \sum_{\ell\geq |j|}\sum_{|m|\leq \ell} \monopole(v)^\ast\sket{\ell}{m},
\label{continuousdiscretebasis}
\end{equation}
where (by our normalization)
\begin{align}
\langle{\smqty*{\ell \\ m}}|{\smqty*{\smash{\ell'}\vphantom{\ell} \\ \smash{m'}\vphantom{m}}}\rangle &= \int_{S^2} dv\, \monopole(v)^\ast\monopoleY{j}{\ell'}{m'}(v) = \delta_{\ell\ell'}\delta_{mm'}, \\
\langle v|v'\rangle &= \sum_{\ell\geq |j|}\sum_{|m|\leq \ell} \monopole(v)^\ast\monopole(v') = \delta^{S^2}(v - v').
\end{align}
We are working in the $\varepsilon = -1$ gauge of Appendix \ref{sphere}, where\footnote{Our $\monopole$ are precisely the $(Y_{j\ell m})_a$ of \cite{Wu:1976ge, Wu:1977qk}.  However, the $(Y_{j\ell m})_b = e^{-2ij\varphi}(Y_{j\ell m})_a$ of \cite{Wu:1976ge, Wu:1977qk} differ by signs from our $\varepsilon = +1$ wavefunctions in \eqref{angularwavefunctions}.}
\begin{equation}
\langle v|\smqty*{\ell \\ m}\rangle\equiv \monopole(v).
\end{equation}
We also have the following resolution of the identity:
\begin{equation}
\int_{S^2} dv\, |v\rangle\langle v| = \sum_{\ell\geq |j|}\sum_{|m|\leq \ell} \sket{\ell}{m}\sbra{\ell}{m} = 1_{S^2}.
\end{equation}
We represent rotations $R\in SO(3)$ by unitary operators $X_R$ acting as
\begin{equation}
X_R|v\rangle = |Rv\rangle, \qquad X_R\sket{\ell}{m} = \sum_{|m'|\leq \ell} D^\ell_{m'm}(R)\sket{\ell}{m'},
\end{equation}
where $D^\ell_{mn}(R) = \sbra{\ell}{m}X_R\sket{\ell}{n}$.

We consider the $\mathbb{Z}_N\subset \mathbb{Z}_{2N}$ qubit code with codewords \eqref{linrotorcodewords}:
\begin{equation}
|\overline{0}\rangle = \frac{1}{\sqrt{N}}\sum_{h\in \mathbb{Z}_N} \left|\frac{\pi}{2}, \frac{2\pi h}{N}\right\rangle, \qquad |\overline{1}\rangle = \frac{1}{\sqrt{N}}\sum_{h\in \mathbb{Z}_N} \left|\frac{\pi}{2}, \frac{2\pi h}{N} + \frac{\pi}{N}\right\rangle.
\end{equation}
Using $\monopole(\theta, \varphi) = \monopole(\theta, 0)e^{i(m + j)\varphi}$ and the expansion for $|\theta, \varphi\rangle$ in \eqref{continuousdiscretebasis}, we get
\begin{align}
\sum_{h\in \mathbb{Z}_N} |\theta, 2\pi h/N + \varphi_0\rangle &= \sum_{\ell\geq |j|}\sum_{|m|\leq \ell}\sum_{h\in \mathbb{Z}_N} \monopole(\theta, 2\pi h/N + \varphi_0)^\ast\sket{\ell}{m} \\
&= \sum_{\ell\geq |j|}\sum_{|m|\leq \ell} \monopole(\theta, 0)^\ast e^{-i(m + j)\varphi_0}\sum_{h\in \mathbb{Z}_N} e^{-2\pi i(m + j)h/N}\sket{\ell}{m} \\
&= N\sum_{\ell\geq |j|}\sum_{|pN - j|\leq \ell} \monopoleY{j}{\ell}{pN - j}(\theta, 0)^\ast e^{-ipN\varphi_0}\sket{\ell}{pN - j}.
\end{align}
Hence in the angular momentum basis, we have
\begin{equation}
|\overline{r}\rangle = \sqrt{N}\sum_{\ell\geq |j|}\sum_{|pN - j|\leq \ell} (-1)^{pr}\monopoleY{j}{\ell}{pN - j}(\pi/2, 0)^\ast\sket{\ell}{pN - j}
\end{equation}
for $r\in \{0, 1\}$.

The analysis of position shifts is the same as on $S^2$ without a magnetic monopole: the correctable position shifts are determined by the Voronoi cells of the codewords' constituent points, as described in Section \ref{ACPcodes}.  As for momentum shifts, the appropriate momentum kick operator for the spherical Landau problem is
\begin{equation}
{}_j\hat{Y}^\ell_m = \int_{S^2} dv\, \monopole(v)|v\rangle\langle v|.
\end{equation}
Its action on the codewords is given by
\begin{align}
{}_j\hat{Y}^\ell_m|\overline{0}\rangle &= \frac{\monopole(\pi/2, 0)}{\sqrt{N}}\sum_{h\in \mathbb{Z}_N} e^{2\pi i(m + j)h/N}\left|\frac{\pi}{2}, \frac{2\pi h}{N}\right\rangle, \\
{}_j\hat{Y}^\ell_m|\overline{1}\rangle &= \frac{\monopole(\pi/2, 0)e^{\pi i(m + j)/N}}{\sqrt{N}}\sum_{h\in \mathbb{Z}_N} e^{2\pi i(m + j)h/N}\left|\frac{\pi}{2}, \frac{2\pi h}{N} + \frac{\pi}{N}\right\rangle.
\end{align}
Since the codewords in the angular momentum basis have support only on states such that $m + j$ is a multiple of $N$, we can determine the value of $m + j$ in an error ${}_j\hat{Y}^\ell_m$ modulo $N$.  The value of $m + j$ modulo $N$ determines $m + j$ if $|m + j| < N/2$.  Since $|m|\leq \ell$, we need $\ell + j < N/2$ for the shift to be correctable, which is more restrictive for larger $j$ (we take $j$ positive for simplicity).  Indeed, since $\ell\geq j$, we need $N > 4j$ for any shifts to be correctable.  So these ``full spherical Landau'' codes seem to suffer from diminished performance compared to ordinary ACP linear rotor codes.

\subsection{More Cyclic Subgroup Codes in the LLL} \label{LLLcyclic}

In the main text, we argued that the antipodal ($N = 1$) case of the qubit code is optimal in the LLL.  Here, we show for illustrative purposes that LLL codes for larger cyclic subgroups do not perform as well.

In what follows, it will be convenient to have a formula for matrix elements of equatorial rotations $X_T\equiv e^{-i\Theta L_3/\hbar}$ between equatorial spin coherent states:
\begin{equation}
{}_j\langle\pi/2, \varphi'|X_T|\pi/2, \varphi\rangle_j = \left(\frac{e^{-i\Theta/2} + e^{i\Theta/2}e^{i(\varphi - \varphi')}}{2}\right)^{2j}.
\label{equatorialmatrixelement}
\end{equation}
For convenience, we also record here the formula
\begin{equation}
\sum_{h=0}^{N-1} (A + Be^{2\pi ih/N})^{2j} = A^{2j}\sum_{k=0}^{2j} \binom{2j}{k}(B/A)^k\sum_{h=0}^{N-1} e^{2\pi ihk/N} = NA^{2j}\sum_{k=0}^{\lfloor 2j/N\rfloor} \binom{2j}{kN}(B/A)^{kN}.
\label{ABsum}
\end{equation}
This expression is invariant under $B\to Be^{2\pi i/N}$.

Consider the abelian $\mathbb{Z}_N\subset \mathbb{Z}_{2N}\subset SO(3)$ (qubit) code, with
\begin{equation}
|\overline{0}\rangle = \frac{1}{\sqrt{\mathcal{N}}}\sum_{h\in \mathbb{Z}_N} \left|\frac{\pi}{2}, \frac{2\pi h}{N}\right\rangle_j, \qquad |\overline{1}\rangle = \frac{1}{\sqrt{\mathcal{N}}}\sum_{h\in \mathbb{Z}_N} \left|\frac{\pi}{2}, \frac{2\pi h}{N} + \frac{\pi}{N}\right\rangle_j.
\end{equation}
Using the $\Theta = 0$ case of \eqref{equatorialmatrixelement} as well as \eqref{ABsum}, we deduce the normalization factor $\mathcal{N}$ for which $\langle\overline{0}|\overline{0}\rangle = \langle\overline{1}|\overline{1}\rangle = 1$:
\begin{equation}
\mathcal{N} = \sum_{h, h' = 0}^{N-1} \left(\frac{1 + e^{2\pi i(h - h')/N}}{2}\right)^{2j} = N\sum_{h=0}^{N-1} \left(\frac{1 + e^{2\pi ih/N}}{2}\right)^{2j} = \frac{N^2}{2^{2j}}\sum_{k=0}^{\lfloor 2j/N\rfloor} \binom{2j}{kN}.
\end{equation}
As $j\to\infty$, the states $|\theta, \varphi\rangle_j$ approach orthonormality and $\mathcal{N}\to N$.

The diagonal error correction condition $\langle\overline{0}|X_T|\overline{0}\rangle = \langle\overline{1}|X_T|\overline{1}\rangle$ holds for equatorial rotations $X_T$.\footnote{The ``parity argument'' of ACP no longer implies that it holds for \emph{all} rotations when $N$ is odd because spin coherent states transform differently than position eigenstates under rotations and inversions, incurring possible phases.  See Footnote \ref{parityargument}.}  To satisfy the off-diagonal condition approximately, we further wish to impose $|\langle\overline{0}|X_T|\overline{1}\rangle| < \epsilon$.  Again specializing to equatorial rotations $X_T$, we use \eqref{equatorialmatrixelement} and \eqref{ABsum} to compute that
\begin{align}
\langle\overline{0}|X_T|\overline{1}\rangle &= \frac{e^{-ij\Theta}}{\mathcal{N}}\sum_{h, h' = 0}^{N-1} \left(\frac{1 + e^{i(\Theta + 2\pi(h - h')/N + \pi/N)}}{2}\right)^{2j} \\
&= \frac{Ne^{-ij\Theta}}{\mathcal{N}}\sum_{h=0}^{N-1} \left(\frac{1 + e^{i(\Theta + 2\pi h/N + \pi/N)}}{2}\right)^{2j} \label{sumoverh} \\
&= \frac{N^2 e^{-ij\Theta}}{2^{2j}\mathcal{N}}\sum_{k=0}^{\lfloor 2j/N\rfloor} (-1)^k e^{ikN\Theta}\binom{2j}{kN},
\end{align}
which is (up to an overall phase) invariant under $\Theta\to \Theta + 2\pi/N$.  We have
\begin{equation}
\lim_{j\to\infty} |\langle\overline{0}|X_T|\overline{1}\rangle| = \begin{cases} 0 & \text{if $\Theta + \pi/N\not\equiv 0\text{ (mod $2\pi/N$)}$}, \\ 1 & \text{if $\Theta + \pi/N\equiv 0\text{ (mod $2\pi/N$)}$}. \end{cases}
\end{equation}
This is easiest to see from the sum over $h$ in \eqref{sumoverh}.  From the above analysis, we see that smaller values of $N$ lead to fewer values of $\Theta$ for which $|\langle\overline{0}|X_T|\overline{1}\rangle|$ is large as $j\to\infty$, in accord with the error correction conditions.  Thus taking $N > 1$ results in codes that are neither as simple nor as effective as those proposed in the main text.

\subsection{General Lie Groups} \label{generalgroups}

We now briefly comment on the possibility of formulating similar codes involving generalized coherent states for a Lie group $G$ other than $SU(2)$ \cite{Perelomov:1971bd, Onofri1975}; for an overview, see \cite{Perelomovbook}.  Such states are parametrized by coset spaces $G/H$.  The following discussion holds for an arbitrary representation of $G$, or choice of abstract ``LLL Hilbert space.''  For the precise LLL degeneracies on the compact coset space $G/H$ and a discussion of the Landau problem thereon, see \cite{Dolan:2003bj}.

Suppose that errors are parametrized by a Lie group $G$.  Let $T : G\to U(\mathcal{H})$ be a unitary irreducible representation of $G$ on the Hilbert space $\mathcal{H}$.  Let $H$ be the isotropy subgroup of a fiducial state $|\psi_0\rangle\in \mathcal{H}$, i.e., the maximal subgroup of $G$ that acts on $|\psi_0\rangle$ by a phase:
\begin{equation}
T(h)|\psi_0\rangle = e^{i\alpha(h)}|\psi_0\rangle.
\end{equation}
Then we have a system of coherent states $|x(g)\rangle$ given by
\begin{equation}
T(g)|\psi_0\rangle = e^{i\alpha(g)}|x(g)\rangle,
\label{generalcoherent}
\end{equation}
where $x\in X\equiv G/H$.  Replacing $g$ in \eqref{generalcoherent} by $gh$ shows that
\begin{equation}
\alpha(gh) = \alpha(g) + \alpha(h).
\label{addphase}
\end{equation}
Furthermore, we have
\begin{equation}
T(g')|x(g)\rangle = e^{-i\alpha(g)}T(g'g)|\psi_0\rangle = e^{i\beta(g', g)}|x(g'g)\rangle
\label{generalcoherentact}
\end{equation}
where
\begin{equation}
\beta(g', g)\equiv \alpha(g'g) - \alpha(g).
\label{compositionphase}
\end{equation}
By \eqref{addphase}, the phase \eqref{compositionphase} depends only on the equivalence class $x(g)$ of $g$ in $G/H$.

Since the representation is unitary, we have by the Cauchy-Schwarz inequality that
\begin{equation}
|\langle x(g_1)|x(g_2)\rangle|\leq 1.
\end{equation}
We readily see that
\begin{align}
\langle x(g_1)|x(g_2)\rangle &= e^{i(\alpha(g_1) - \alpha(g_2))}\langle\psi_0|T(g_1^{-1}g_2)|\psi_0\rangle, \\
\langle x(gg_1)|x(gg_2)\rangle &= e^{i(\beta(g, g_1) - \beta(g, g_2))}\langle x(g_1)|x(g_2)\rangle.
\end{align}
In particular, we have
\begin{equation}
\langle x(g)|T(r)|x(g)\rangle = \langle\psi_0|T(g^{-1}rg)|\psi_0\rangle.
\label{inparticular}
\end{equation}
Now suppose we wish to construct a code in $\mathcal{H}$ for which the codewords are generalized coherent states.  For any $g, g'$ corresponding to distinct codewords, the (approximate) quantum error correction conditions require that
\begin{align}
\langle x(g)|T(r)|x(g)\rangle &= \langle x(g')|T(r)|x(g')\rangle, \\
\langle x(g')|T(r)|x(g)\rangle &\approx 0,
\end{align}
where the group operation $r$ is a combination of correctable errors.  In view of \eqref{inparticular}, one simple way to satisfy the first condition is to take $g, g'$ to be elements of the centralizer of the set of correctable errors.  We may then choose the $g, g'$ within the centralizer so that the second condition is satisfied as well as possible.  It may be the case that all of the $r$'s belong to some subgroup of $G$.

In our case, both the correctable rotations and the codeword orientations belong to an abelian subgroup of $G = SU(2)$ (really $SO(3)$, since we restrict to integer $j$).  From \eqref{amplitude}, we have when $\beta = 0$ that
\begin{equation}
{}_j\langle\Omega|X_T|\Omega\rangle_j = \left(\cos\frac{\alpha + \gamma}{2} - i\sin\frac{\alpha + \gamma}{2}\cos\theta\right)^{2j},
\end{equation}
so this matrix element is independent of $\varphi$ (i.e., it is the same for any states related by an equatorial rotation, regardless of whether those states lie on the equator).  An equatorial rotation is one that preserves the north pole, which is the reference point.  The key property is that conjugating an equatorial rotation by another equatorial rotation has no effect.

\newpage

\nocite{*}
\bibliographystyle{utphys}
\bibliography{\jobname}

\end{document}